\documentclass{SciPost}

\usepackage{color,bm,cite}
\usepackage{amsmath,amssymb}
\usepackage{epsfig}
\usepackage{verbatim}
\usepackage[T1]{fontenc}
\usepackage{array}

\usepackage{url,hyperref,cite}
\hypersetup{
    colorlinks=true, 
    linktoc=all,     
    linkcolor=blue,
    citecolor=magenta
}     

\newcommand{\la}{\langle}
\newcommand{\ra}{\rangle}
\newcommand{\dg}{^\dagger}
\newcommand{\p}{\partial}
\newcommand{\rd}{{\rm d}}

\makeatletter
\newsavebox{\@brx}
\newcommand{\llangle}[1][]{\savebox{\@brx}{\(\m@th{#1\langle}\)}%
  \mathopen{\copy\@brx\mkern2mu\kern-0.9\wd\@brx\usebox{\@brx}}}
\newcommand{\rrangle}[1][]{\savebox{\@brx}{\(\m@th{#1\rangle}\)}%
  \mathclose{\copy\@brx\mkern2mu\kern-0.9\wd\@brx\usebox{\@brx}}}
\makeatother

\begin{document}

\begin{center}{
    \Large
    \textbf{
      On computing non-equilibrium dynamics following a quench 
      %Computing time evolution following quenches in the Lieb-Liniger model 
    }
  }
\end{center}

\begin{center}
Neil J. Robinson,$^\ast$ Albertus J. J. M. de Klerk,$^\circ$ and Jean-S\'ebastien
Caux$^\dagger$
\end{center}

\begin{center}
  Institute for Theoretical Physics, University of Amsterdam,\\
  Postbus 94485, 1090 GL Amsterdam, The Netherlands
\end{center}

\begin{center}
  $^\ast$n.j.robinson@uva.nl\qquad
  $^\circ$a.j.j.m.deklerk@uva.nl\qquad
  $^\dagger$j.s.caux@uva.nl \\
  
  \vspace{3mm}\today
\end{center}

\section*{Abstract} {\bf Computing the non-equilibrium dynamics that follows a
  quantum quench is difficult, even in exactly solvable models. Results are
  often predicated on the ability to compute overlaps between the initial state
  and eigenstates of the Hamiltonian that governs time evolution.  Except for a
  handful of known cases, it is generically not possible to find these overlaps
  analytically. Here we develop a numerical approach to preferentially generate
  the states with high overlaps for a quantum quench starting from the ground
  state or an excited state of an initial Hamiltonian.  We use these
  preferentially generated states, in combination with a ``high overlap states
  truncation scheme'' and a modification of the numerical renormalization group,
  to compute non-equilibrium dynamics following a quench in the Lieb-Liniger
  model. The method is non-perturbative, works for reasonable numbers of
  particles, and applies to both continuum and lattice systems. It can also be
  easily extended to more complicated scenarios, including those with
  integrability breaking. }

\vspace{5pt}
\noindent\rule{\textwidth}{1pt}
\tableofcontents\thispagestyle{fancy}
\noindent\rule{\textwidth}{1pt}

%%%%%%%%%%%%%%%%%%%%%%%%
\section{Introduction}
%%%%%%%%%%%%%%%%%%%%%%%%
Non-equilibrium strongly correlated systems have been the subject of intense
study over the last
decade~\cite{essler2016quench,calabrese2016quantum,cazalilla2016quantum,bernard2016conformal,caux2016quench,vidmar2016generalized,ilievski2016quasilocal,langen2016prethermalization,vasseur2016nonequilibrium,deluca2016equilbration}.
Spurred on by experiments in ultra-cold atomic
gases~\cite{kinoshita2006quantum,polkovnikov2011colloquium,langen2015ultracold},
questions of a fundamental nature have taken the center stage: What general
principles govern properties of a non-equilibrium system?  Are there
non-equilibrium phases that have no equilibrium analogue? How does an isolated
quantum system equilibrate and thermalize when time evolution is unitary? In the
process of addressing such questions, it was realized that conservation laws
play a central role in the description of non-equilibrium physics, strongly
restricting the dynamics that can occur and the processes that govern
equilibration~\cite{rigol2007relaxation,rigol2008thermalization,calabrese2011quantum,gring2012relaxation,caux2012constructing,caux2013time,ilievski2015complete,langen2015experimental,vidmar2016generalized}.
Nowhere is this more evident than in integrable models, where the presence of an
extensive number of local conservation laws leads to an absence of
thermalization~\cite{rigol2007relaxation,rigol2008thermalization}. 

Integrable quantum many-body systems may, at first glance, appear to be little
more than an academic curiosity. One may imagine that for an  extensive number
of local conservation laws to exist, there must be an extreme fine-tuning of a
many-body Hamiltonian, and hence there is little chance of them being realized
in an experimental system. Fortunately, this is not the case.  Perhaps the
simplest non-trivial example is the Lieb-Liniger
model~\cite{lieb1963exact1,lieb1963exact2,korepin1997quantum} of delta-function
interacting bosons confined to a single spatial dimension, which is almost
perfectly realized in many cold atomic gas experiments (see, e.g.,
Ref.~\cite{kinoshita2006quantum,hofferberth2007nonequilibrium,bloch2008manybody,cazalilla2011one}),
including those that probe non-equilibrium dynamics. Thus integrability, and the
influence of conservation laws, can be directly examined in
experiment.\footnote{This is despite the fact that integrability is broken,
albeit weakly, in experiments. The timescales for observing integrability
breaking can be anomalously long, with the system exhibiting so-called
`prethermalization'~\cite{moeckel2008interaction,marcuzzi2013prethermalization,essler2014quench,bertini2015prethermalization,langen2016prethermalization},
where the proximity to an integrable point still strongly restricts the
dynamics. At very long times the system is still expected to
thermalize~\cite{bertini2015prethermalization,bertini2016thermalization,biebl2017thermalization}.}

As a result, non-equilibrium dynamics of the Lieb-Liniger model has received a
significant amount of theoretical
attention~\cite{kormos2013interaction,panfil2014metastable,kormos2014analytic,denardis2014solution,collura2014stationary,denardis2014analytical,denardis2015relaxation,zill2015relaxation,piroli2016multiparticle,piroli2016quantum,castroalvaredo2016emergent,zill2016coordinate,denardis2017probing,atas2017exact,palmai2017quasilocal,doyon2017largescale,doyon2018soliton,doyon2018exact,doyon2018geometric,denardis2018edge,bastianello2018exact,bastianello2018from,caux2019hydrodynamics,schemmer2019generalized,perfetto2019quench}.
In many of these studies the system is driven out of equilibrium via a quantum
quench~\cite{calabrese2006time} of the interaction strength. Analytical studies
have focused on cases where the initial states are eigenstates of the
Lieb-Liniger model with either $c=0$ or $c=\infty$, due to simplifications (both
cases being `non-interacting' in nature) that allow one to explicitly compute
overlaps between the initial state and eigenstates of the Hamiltonian governing
time evolution. With these overlaps at hand, expectation values of local
operators in the long-time limit can be computed via, for example, the quench
action method~\cite{caux2013time,caux2016quench}. Generally, accessing the
real-time dynamics of observables is still an outstanding challenge -- away from
the mentioned special initial states it is not known how to proceed.
Analytically, one does not know how to compute the overlaps, a crucial
ingredient for existing approaches, whilst numerically it is tough to deal with
continuum models in a rigorous and well-controlled manner. Brute force
computations, using the coordinate Bethe ansatz, are limited to very small
numbers of particles, $N\sim5$, and scale super-exponentially as $\propto
(N!)^2$ without additional
approximations~\cite{zill2015relaxation,zill2016coordinate}. 

In this work, we develop a novel numerical approach, motivated by the truncated
spectrum approach~\cite{james2017nonperturbative}, that allows one to compute
overlaps. Our algorithm allows us to efficiently express the initial state in
terms of the most important eigenstates of the Hamiltonian governing time
evolution. With these overlaps at hand, we can then study the structure of the
overlaps away from analytically tractable limits, compute real-time dynamics,
and access the long time limit via the diagonal
ensemble~\cite{rigol2008thermalization}. Here we present proof-of-principle
computational results for interaction quenches in the Lieb-Liniger model for a
reasonably small number of particles (although, we note, well beyond the reaches
of `brute force' coordinate Bethe ansatz
computations~\cite{zill2015relaxation,zill2016coordinate}).

The truncated spectrum approach has recently been used to compute overlaps and
non-equilibrium dynamics in both the Ising field
theory~\cite{rakovszky2016hamiltonian,kormos2017realtime,hodsagi2018quench,james2018nonthermal,robinson2018signatures}
and the sine-Gordon
model~\cite{horvath2017overlaps,kukuljan2018correlation,horvath2018overlap,horvath2018nonequilibrium}.
There are some fundamental differences between the approach taken here and those
previously considered. We will explicitly work with a \textit{strongly
correlated computational basis} formed from eigenstates of an
\textit{interacting} integrable quantum system. Strong correlations are then
inherently built into the basis, in contrast to the non-interacting bases of the
Ising and sine-Gordon models. Furthermore, our approach abandons the
conventional, energy-ordered, Hilbert space truncation metric and instead we are
able to \textit{preferentially target} the states contributing the largest
overlaps. Our approach reduces the computational cost of calculations by orders
of magnitude, as we will see below. 

%%%%
\subsection{Layout}
%%%%

In Sec.~\ref{Sec:LiebLiniger} we introduce the system that we study, the
Lieb-Liniger model, and its exact Bethe ansatz solution. We also discuss our
quench protocol and formulation of the quench problem in terms of a perturbed
Hamiltonian. Following this, in Sec.~\ref{Sec:Overlaps} we first describe the
``ideal'' numerical solution to the quench problem, and then discuss the
development of the high overlap states truncation scheme (HOSTS) -- an attempt
to construct precisely this. To do so, we describe how the basic truncated
spectrum approach works and apply it to the problem. This reveals that: (i) the
traditional truncated spectrum approach is not well-suited to the problem; (ii)
numerical renormalization group extensions of this method are also not
well-suited to the problem. The numerical renormalization group results give us
inspiration for an alternative algorithm, based upon a better ``ordering
metric'' for the Hilbert space truncation. We explore this and, putting all
these results together, can construct the initial state to reasonable accuracy
at some (not insignificant) burden.  

With this high overlap states truncation scheme in place, we then explore how to
\textit{preferentially generate} high overlap states for the truncation scheme.
This is discussed in Sec.~\ref{Sec:Abacus} and we illustrate its application in
efficiently constructing a given initial state to high precision for a
non-perturbative quench. This is not easily achievable within the conventional
truncated spectrum approach. We also provide a number of additional convergence
checks of our initial state in this section. With this algorithm at hand, we are
able to compute real time non-equilibrium dynamics following a quench, as
discussed in Sec.~\ref{Sec:RealTime}, and access the long-time limit via the
diagonal ensemble. 

In Sec.~\ref{Sec:MERG} we study strongly non-perturbative quenches, where
numerical renormalization group approaches within the high overlap truncation
scheme need some modification. A modified algorithm, the matrix element
renormalization group (MERG), is detailed in this section. We illustrate
problems of the HOSTS algorithm and the success of MERG in computing
non-equilibrium dynamics following strongly non-perturbative quenches.
Furthermore, we introduce a general version of the matrix element
renormalization group algorithm able to deal with excited states as well as
ground states.  We conclude in Sec.~\ref{Sec:Conclusions}, where we also suggest
a number of future directions for studies. 

%%%%%%%%%%%%%%%%%%%%%%%%
\section{The Lieb-Liniger model}
\label{Sec:LiebLiniger}
%%%%%%%%%%%%%%%%%%%%%%%%

The Lieb-Liniger model describes indistinguishable bosons confined to move in a
single spatial dimension, which are coupled via an ultra-local density-density
interaction. On a ring of circumference $R$ the Hamiltonian
reads~\cite{lieb1963exact1,lieb1963exact2}
\begin{align}
H(c) = \int_0^R \rd x \bigg( \frac{\hbar^2}{2m} \p_x \Psi\dg(x) \p_x \Psi(x) + c \Psi\dg(x) \Psi\dg(x) \Psi(x) \Psi(x) \bigg). \label{H}
\end{align}
Here $m$ is the mass of boson and $c$ is the interaction strength. Here we will
focus on the case of repulsive interactions, $c>0$, and henceforth we set
$2m=\hbar=1$ to define our units. We will consider the case of unit density
$N/R=1$ herein. 

%%%%%%%%%%%%%%%
\subsection{Bethe Ansatz Solution}
%%%%%%%%%%%%%%%

The Lieb-Liniger model is integrable and exactly
solvable~\cite{lieb1963exact1,lieb1963exact2,korepin1997quantum}; $N$-particle
eigenstates $|\{\lambda\}_N\ra$ are characterized by a set of $N$ real
rapidities $\{\lambda\}_N = \{ \lambda_1,\ldots,\lambda_N\}$ that satisfy the
Bethe equations
\begin{align}
e^{-\mathrm{i}\lambda_j R} = \prod_{\substack{l=1 \\ l\neq j}}^N \frac{ \lambda_l - \lambda_j + \mathrm{i}c}{\lambda_l - \lambda_j - \mathrm{i}c}. \label{betheEqs}
\end{align}
These states have momentum $P(\{\lambda\}_N)$ and energy $E(\{\lambda\}_N)$
given by 
\begin{align}
P\big(\{\lambda\}_N\big) = \sum_{j=1}^N \lambda_j, \qquad E\big(\{\lambda\}_N\big) = \sum_{j=1}^N \lambda_j^2. \label{def:PE}
\end{align}
Integrability of the model is realized through an infinite family of conserved
quantities, whose eigenvalues take the form
\begin{align}
Q_n\big(\{\lambda\}_N\big) = \sum_{j=1}^N \lambda_j^n, \quad n = 1,2,\ldots, \label{defQ}
\end{align}
where $Q_1 = P$, $Q_2=E$. We work with eigenstates $|\{\lambda\}_N\ra$ that are
normalized as~\cite{korepin1997quantum, gaudin1983}:
\begin{align}
\la \{\lambda\}_N | \{\lambda\}_N\ra = c^N \prod_{j < l} \frac{ (\lambda_j - \lambda_l)^2 + c^2}{(\lambda_j - \lambda_l)^2}\, \text{det}\ {\cal N}, \label{eq:defNorm}
\end{align}
where ${\cal N}$ is the $N\times N$ ``Gaudin matrix'', with elements
\begin{align}
{\cal N}_{jl}  = \delta_{jl} \bigg( R + \sum_{k=1}^N K(\lambda_j,\lambda_k) \bigg) - K(\lambda_j,\lambda_l), 
\label{eq:defGaudin}
\end{align}
and
\begin{align}
  K(\lambda,\mu) = \frac{2c}{c^2 + (\lambda-\mu)^2}.
  \label{eq:defK}
\end{align}

%%%%
\subsubsection{Characterizing eigenstates via integers: Logarithmic Bethe equations}
%%%%
The $N$-particle eigenstates, $|\{\lambda\}_N\rangle$, can be characterized via
sets of unique sets of quantum numbers, $\{I\}$, which are integer or half-odd
integer (depending on the parity of the particle number $N$). There is a
one-to-one correspondence between sets of quantum number and sets of rapidities,
defined via the Logarithmic Bethe equations
\begin{align}
  \lambda_j R = 2\pi I_j - 2 \sum_{l=1}^N \arctan\left(\frac{\lambda_j - \lambda_l}{c}\right),
  \label{eq:logbae}
\end{align}
where the quantum numbers satisfy
\begin{align}
  I_j \in \left\{
  \begin{array}{lll}
    \mathbb{Z}+\frac12 & \quad & \text{for}~N~\text{even},\\
    \mathbb{Z}         & \quad & \text{for}~N~\text{odd},
  \end{array}\right.
  \label{def:betheintegers}
\end{align}
and a Pauli principle, $I_j \neq I_l$ for $j\neq l$.

The mapping between quantum numbers and rapidities satisfies $\lambda_j >
\lambda_l$ if $I_j > I_l$ (due to the monotonic nature of the second term on the
right of Eq.~\eqref{eq:logbae}). From the definition of the energy in terms of
the rapidities, Eq.~\eqref{def:PE}, it follows then that  the ground state
configuration of quantum numbers is a ``Fermi sea'' of quantum numbers that are
symmetrically distributed about the origin. In the large $c$ limit, the
rapidities crystallize on to $\lambda_j \to (2\pi/R)I_j$, as if noninteracting
fermions. 

%%%%
\subsubsection{Equilibrium and non-equilibrium properties}
%%%%

The Lieb-Liniger model is perhaps the simplest non-trivial integrable model,
with many of its equilibrium properties being well understood. This includes
both thermodynamic properties and correlation functions of local
operators~\cite{korepin1997quantum, caux_dynamical_2006, calabrese_dynamics_2007,
panfil_finite-temperature_2014}. There are also known expressions for scalar
products~\cite{slavnov1989calculation}, as well as determinant representations
of matrix elements of local operators in the
eigenbasis~\cite{slavnov1989calculation,slavnov1990nonequaltime,kojima1997determinant,korepin1999form,pozsgay2011local,piroli2015exact},
some of which are detailed in the appendix (and will be used further in this
work). Recently, exact results for the full counting statistics and local
correlation functions have been obtained~\cite{bastianello2018exact}.

Non-equilibrium properties of the model following a quantum quench are much less
well understood, with important studies only emerging over the past six
years~\cite{kormos2013interaction,panfil2014metastable,kormos2014analytic,denardis2014solution,collura2014stationary,denardis2014analytical,denardis2015relaxation,piroli2016multiparticle,piroli2016quantum,atas2017exact,palmai2017quasilocal}.
Such studies have been rather restricted, relying on knowledge of the
\textit{overlaps} of eigenstates of the Hamiltonian at different interaction
strengths. These can, in some special limits, be extracted from integrability of
the model and simplifications which occur in those limits. Away from these
cases, such studies of non-equilibrium properties are hampered by lack of
knowledge of the overlaps and a dearth of techniques for calculating them. 

Recently, there have been a number of works that study the emergence of
non-equilibrium steady states in the Lieb-Liniger, in the context of dynamics
starting from inhomogeneous initial
states~\cite{castroalvaredo2016emergent,doyon2017largescale,doyon2018geometric,doyon2018soliton,denardis2018edge,bastianello2018exact,bastianello2018from,caux2019hydrodynamics,schemmer2019generalized,perfetto2019quench}.
These studies have been enabled by the generalized hydrodynamics
framework~\cite{castroalvaredo2016emergent,bertini2016transport}, an adaptation
of hydrodynamics to the case of integrable systems. This framework has also
allowed the computation of the Drude weight in the Lieb-Liniger
model~\cite{doyon2017drude}. To be clear, we will be considering only cases with
translational invariance here, i.e. global quantum quenches. 

%%%%%%%%%%
\subsection{The Quench Protocol}
%%%%%%%%%
We consider the following problem. The system is initialized in the ground state
of the Lieb-Liniger model~\eqref{H} at interaction strength $c_i>0$. At time
$t=0$ the interaction strength is instantaneously changed $c_i \to c_f > 0$ and
the system subsequently evolves in time according to $H(c_f)$. Of interest to us
is how to compute the time evolution and long-time limit of expectation values
of observables for \textit{generic values} of the initial and final interaction
strengths, $c_i$ and $c_f$.

For approaches such as the quench action~\cite{caux2013time,caux2016quench} a
crucial role is played by \textit{the overlaps}. The overlaps describe how an
initial state $|\Psi_i\ra$ is projected onto the eigenstates
$|\{\lambda\}^{(n)}_N\ra$ of $H(c_f)$, the Hamiltonian governing time evolution
\begin{align}
|\Psi_i\ra = \sum_{n=0}^\infty |\{\lambda\}^{(n)}_N\ra \underbrace{\la \{\lambda\}^{(n)}_N | \Psi_i\ra}_{\text{the overlaps}}.
\label{psi}
\end{align}
Thus the overlaps directly determine how the initial state evolves in time
\begin{align}
|\Psi_i(t)\ra \equiv e^{-iH(c_f)t} |\Psi_i\rangle  = \sum_{n=0}^\infty e^{-\mathrm{i}E(\{\lambda\}^{(n)}_N)t}  |\{\lambda\}^{(n)}_N\ra \la \{\lambda\}^{(n)}_N | \Psi_i\ra.
\label{psit}
\end{align}
Analytically computing the overlaps is a formidable task, even with the toolbox
of integrability at hand. Indeed, it is generally not known how to perform such
a calculation, with analytical overlaps having only been obtained in a handful
of tractable
cases~\cite{pozsgay2014overlaps,denardis2014solution,brockmann2014gaudinlike,brockmann2014overlaps,brockmann2014neelxxz,piroli2014recursive,deleeuw2015onepoint,foda2016overlaps,buhlmortensen2016onepoint,deleeuw2016adscft,deleeuw2017onepoint,bertini2017quantum,piroli2017what,piroli2017exact,de_leeuw_scalar_2018,de_leeuw_spin_2020,linardopoulos_solving_2020}.

%%%%%%%%%%%%%
\subsubsection{Formulation in terms of a perturbed Hamiltonian}
\label{Sec:formulation}
%%%%%%%%%%%%%

The time evolved state, Eq.~\ref{psit}, requires knowledge of how the initial
state (the ground state of $H(c_i)$, the initial Hamiltonian) is expressed in
terms of eigenstates of the final Hamiltonian, $H(c_f)$. If we can construct the
initial Hamiltonian directly in the basis of eigenstates of the final
Hamiltonian, diagonalization would yield the overlaps directly. In practice, we
are dealing with a continuum bosonic model, so one must truncate the constructed
Hamiltonian to obtain a finite matrix that one can diagonalize. This is a
so-called truncated spectrum approach or approximation.

The manner in which we are formulating this problem, working directly with
strongly correlated basis states, is different to previous applications of
truncated spectrum approaches to non-equilibrium
dynamics~\cite{rakovszky2016hamiltonian,kormos2017realtime,hodsagi2018quench,james2018nonthermal,robinson2018signatures,horvath2017overlaps,kukuljan2018correlation,horvath2018overlap}.
In these cases, a computational basis of \textit{non-interacting fermions/bosons}
was used, with both the initial state and final eigenbasis being constructed
from these computational states. 

For the case at hand, we are able to construct the initial Hamiltonian in the
final basis through exact knowledge of eigenstates and matrix elements from
integrability of the model~\cite{pozsgay2011local,piroli2015exact}. We begin by
writing the Hamiltonian in the form
\begin{align}
H(c_i) = H(c_f) + (c_i - c_f) \int_0^R \rd x \, \Psi\dg(x)\Psi\dg(x)\Psi(x)\Psi(x).  \label{Hpert}
\end{align}
In this manner, we have written the initial Hamiltonian as a `perturbation' of
the final Hamiltonian.\footnote{For the approach that is being discussed, the
strength of this `perturbation' $(c_i - c_f)$ does not need to be small. This
will be explicitly demonstrated in the results that follow.}  In the ground
state (zero momentum) sector with fixed particle number $N$, the matrix elements
of the initial Hamiltonian can then be written as 
\begin{align}
\la \{\lambda\}^{(m)}_N | H(c_i) | \{\lambda\}^{(n)}_N \ra = \delta_{n,m} E\big(\{\lambda\}^{(n)}_N \big) + (c_i - c_f) R\, \la \{\lambda\}^{(m)}_N |  \big(\Psi\dg(0)\big)^2 \big( \Psi(0) \big)^2 |\{\lambda\}^{(n)}_N\ra. \label{Hmatrixelm}
\end{align}
Here, as above, $|\{\lambda\}^{(n)}_N\rangle$ are $N$ particle eigenstates of
the \textit{final Hamiltonian} $H(c_f)$. We will often call these
``computational basis states''.  We see from~\eqref{Hmatrixelm} that we require
matrix elements of the operator $g_2(0) = \big( \Psi\dg(0)\big)^2
\big(\Psi(0)\big)^2$ between computational basis states. Known results for these
are recapitulated in Appendix~\ref{sec:ME}.

%%%%%%%%%%%%%%%%%%%%%%%%
\section{Developing a high overlap states truncation scheme}
\label{Sec:Overlaps}
%%%%%%%%%%%%%%%%%%%%%%%%

%%%%
\subsection{The ideal truncation scheme}
\label{Sec:Ideal}
%%%%

At the heart of the problem under study is how to truncate the initial
Hamiltonian, constructed in the computational basis, to obtain optimal
convergence of physical quantities. The time evolved wave function~\eqref{psit}
clearly points the way. Consider organizing the computational basis by the
magnitude of the overlap $w^{(n)} = | \langle \{\lambda\}^{(n)} |
\Psi_i\rangle|$. Truncation to the $N_\text{tot}$ computational states with
highest overlaps
\begin{equation}
  |\Psi(t) \rangle_{\text{approx}} = \sum_{n=0}^{N_\text{tot}} e^{-iE(\{\lambda\}^{(n)})t} |\{\lambda\}^{(n)}\rangle \langle \{\lambda\}^{(n)} | \Psi_i\rangle, \label{ideal}
\end{equation}
will give bounded errors for (bounded) physical observables. That is, saturating
the norm of the state $|\Psi_i\rangle$ to 
\begin{equation}
  s(N_{tot}) = 1 - \sum_{n=0}^{N_\text{tot}} \left\vert \langle \{\lambda\}^{(n)} | \Psi_i\rangle \right\vert^2,
\end{equation}
the maximal error $\epsilon_{\text{max}}[\cdot]$ on the time evolution of a bounded operator $A$ is
\begin{equation}
  \epsilon_{\text{max}}\Big[ A(t) - A(t)_{\text{approx}}\Big] = s(N_\text{tot})
  \text{max}_{m,n}(A_{m,n}). 
\end{equation}
Here $A(t)_{\text{approx}}$ is the operator evaluated within the time evolved
approximate state $|\Psi(t)\rangle_{\text{approx}}$, $A_{m,n} = \langle
\{\lambda\}^{(m)}| A | \{\lambda\}^{(n)}\rangle$ are the matrix element of
operator $A$ in the computational basis, and $\text{max}_{m,n}(A_{m,n})$ denotes
its maximal value. Thus if $s(N_\text{tot})$ is sufficiently small, for any
bounded operator the errors are small.

An expansion such as Eq.~\eqref{ideal} is all well and good, but we do not
\textit{a priori} know the overlaps. Thus we are unable to order the
computational basis according to the overlaps, and we must develop an approach
that mimics this. This is the subject of the remainder of this section, where we
develop a ``high overlap states truncation scheme''. We do so in a sequence of
steps, drawing inspiration from conventional truncated spectrum methods,
adaptations, and their failures, to eventually arrive at an efficient high
overlap states truncation scheme.

%%%%
\subsection{The truncated spectrum approach}
\label{sec:tsa}
%%%%

The Lieb-Liniger model is a continuum field theory of interacting bosons. The
Hilbert space is spanned by infinitely many states, and the Hamiltonian is thus
a matrix with infinite dimensions. In the formulation of our problem as a
perturbed Hamiltonian, Sec.~\ref{Sec:formulation}, the perturbed Hamiltonian is
a dense matrix in the computational basis. To proceed, we have to truncate the
Hilbert space in some manner to obtain a finite matrix, which can then be
diagonalized to obtain the eigenstates (and their energies) and hence the
overlaps.

As a starting point, we take inspiration from standard truncated spectrum
methods~\cite{yurov1990truncated,yurov1991truncated,james2017nonperturbative}.
If the perturbing operator in Eq.~\eqref{Hpert} is renormalization group
relevant, it will:
\begin{enumerate}
\item Flow to strong coupling as the renormalization group is taken to the low
  energy limit, leading to a strong mixing between low-energy states in the
  computational basis $|\{\lambda\}^{(n)}\ra$.  
\item Flow to weak coupling in the ultraviolet (high energy), meaning that high
  energy states $|\{\lambda\}^{(n)}\ra$ are approximate eigenstates of the
  perturbed Hamiltonian too.
\item As a corollary to the above points, the operator cannot strongly couple
  low-energy and high-energy states in the computational basis.  
\end{enumerate}   
In our scenario, in the non-interacting limit the perturbing operator has
scaling dimension `zero'.\footnote{That is, the two-point function of the free
bosonic field is logarithmic in form. In the conformal field theory
context~\cite{difrancesco1997conformal}, this reflects the fact that $\Psi$ is
not a primary field. See, e.g., Ref.~\cite{anand2017rg} for a detailed
discussion of the analogous case in the scalar $\phi^4$ model.} This is similar
to the scenario encountered in the $1+1$D $\phi^4$ theory, which has been
studied extensively with truncated spectrum
methods~\cite{lee2001diagonalization,lee2001introduction,rychkov2015hamiltonian,rychkov2016hamiltonian,eliasmiro2016renormalized,bajnok2016truncated,christensen2016diagonalizing,katz2016conformal,anand2017rg,eliasmiro2017nlo,eliasmiro2017highprecision}.
This suggests that perhaps the same method may achieve success here.

\begin{figure}[t]
\begin{center}
\includegraphics[width=0.7\textwidth]{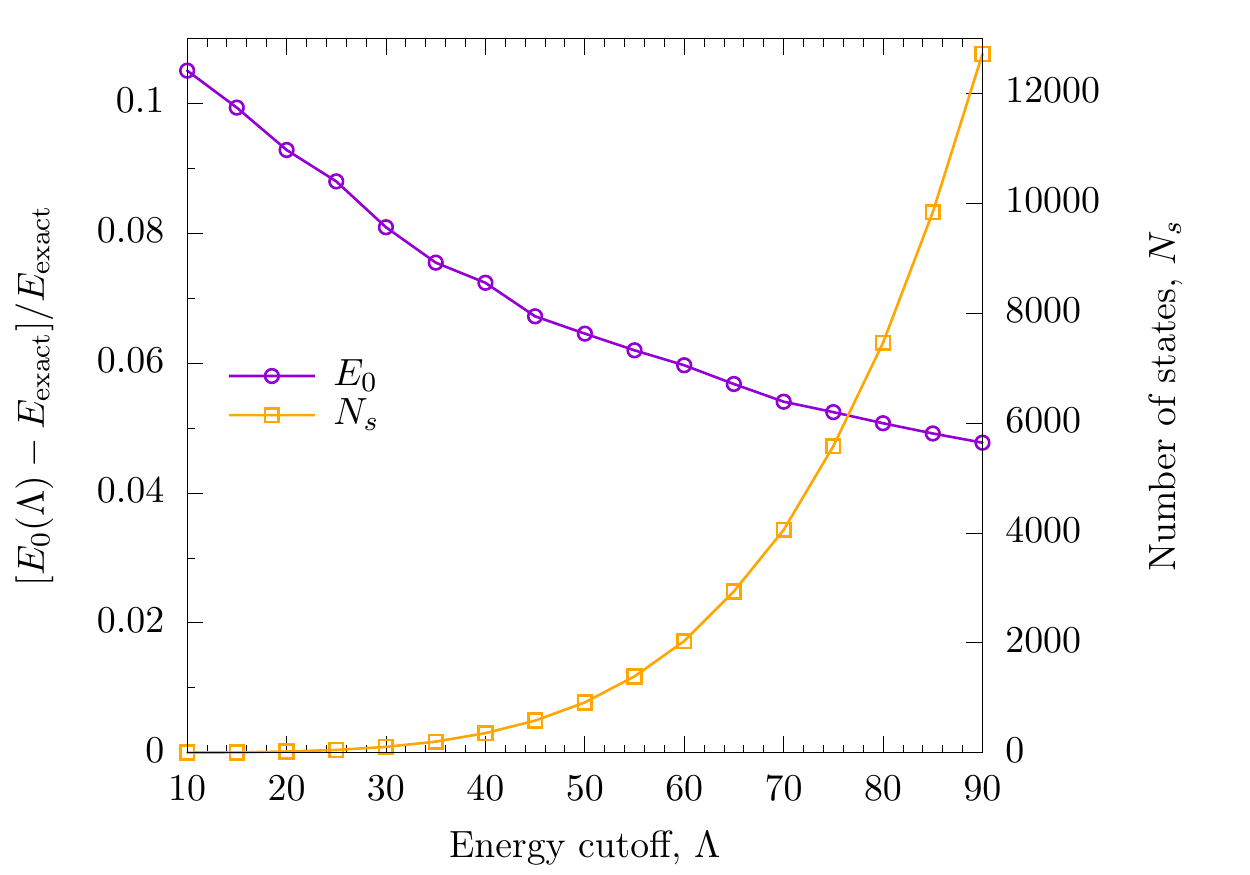}
\end{center}
\vspace{-5mm}
\caption{The ground state energy $E_0$ (compared to the exact result
$E_\text{exact}$) and the number of basis states $N_s$ as a function of the
energy cutoff $\Lambda$. The ground state of the Hamiltonian~\eqref{H} with $c_i
= 20$ is constructed in terms of eigenstates for $c_f = 10$, with $N=10$
particles at unit density, using the truncated spectrum approach.}
\label{fig:tsa}
\end{figure}

The simplest possible truncation, motivated by the `decoupling' of low- and
high-energy computational basis states, is to introduce an energy cutoff
$\Lambda$ and consider all computational states with energy below the cutoff.
This is the truncation originally envisaged by Yurov and Zamolodchikov in the
context of perturbed conformal field
theories~\cite{yurov1990truncated,yurov1991truncated}. Convergence of the ground
state energy (for example) can then be checked as a function of the cutoff
energy $\Lambda$. As a first example, we show an example of this for the
$c_i=20$ ground state of ten particles (constructed in terms of $c_f=10$
computational basis states) in Fig.~\ref{fig:tsa}. The convergence of the ground
state energy with $\Lambda$ is consistent with an exponential fit (although, we
note, that we do not have many decades of data to fit over).

For many models (see the review article~\cite{james2017nonperturbative}) it has
been found that convergence can be slow, requiring energy cutoffs far beyond
those one can treat with exact diagonalization. This can be seen clearly in
Fig.~\ref{fig:tsa}: to get the ground state energy to within just $1\%$ of the
exact value, we would expect to have to include many hundreds of thousands of
states. Various techniques have been developed to counter this, as discussed
in~\cite{james2017nonperturbative}, ameliorating the effects of the cutoff. In
the following section we discuss and implement one such approach: a numerical
renormalization group extension.  

%%%%
\subsection{Numerical renormalization group extension}
\label{sec:nrg}
%%%%

To combat slow convergence of the eigenstates and eigenvalues, we supplement the
truncated spectrum procedure with a numerical renormalization group extension.
The numerical renormalization group was first introduced by Wilson to tackle the
Kondo problem~\cite{wilson1975renormalization} and since then has become a vital
tool for tackling impurity problems~\cite{bulla2008numerical}, including in the
context of dynamical mean field theory (see, e.g.,
Ref.~\cite{stadler2015dynamical}). Its application to truncated spectrum methods
was first suggested by Konik and Adamov in 2007~\cite{konik2007numerical}, and
has since been applied to tackle a number of problems beyond the reach of the
plain truncated spectrum
approach~\cite{brandino2010energy,beria2013truncated,coser2014truncated,brandino2015glimmers,konik2015studying,azaria2016particle}.  

The numerical renormalization group procedure for the truncated spectrum
approach is formulated as follows:
\begin{enumerate}
\item Construct the computational basis $\{ |\{\lambda\}^{(j)}\ra\}$ and order
  by energy $E(\{\lambda\}^{(j)})$. 
\item Construct a truncated Hamiltonian from the first $N_s+\Delta N_s$
  computational basis states, \\$\big\{ |\{\lambda\}^{(1)}\ra, \ldots,
  |\{\lambda\}^{(N_s+\Delta N_s)}\ra\big\}$ and diagonalize it to obtain
  approximate energies and eigenstates, $\big\{ |E^{(1)}\ra , \ldots ,
  |E^{(N_s+\Delta N_s)}\ra \big\}$.
\item Discard the highest $\Delta N_s$ approximate eigenstates
  $\Big\{|E^{(N_s+1)}\ra,\ldots,|E^{(N_s+\Delta N_s)}\ra\Big\}$, from the
  truncated Hamiltonian. 
\item Construct a new basis of $N_s + \Delta N_s$ from the remaining $N_s$
  approximate eigenstates and the next $\Delta N_s$ states in the computational
  basis. 
\item Construct the Hamiltonian in this new basis, and diagonalize it to obtain
  new approximations to the eigenstates and their energies.
\item Return to the third step.  
\end{enumerate}
This process is continued, obtaining new approximate eigenstates after each
cycle of steps 3 to 5, until the required convergence of the ground state
energy/eigenstate is reached or the computational basis is exhausted. With such
a procedure, it is possible to construct the ground state of a perturbed
Hamiltonian in terms of many hundreds of thousands or millions of the
computational basis states~\cite{brandino2010energy,konik2011exciton}.

\begin{figure}[t]
  \begin{center}
      \includegraphics[width=0.49\textwidth]{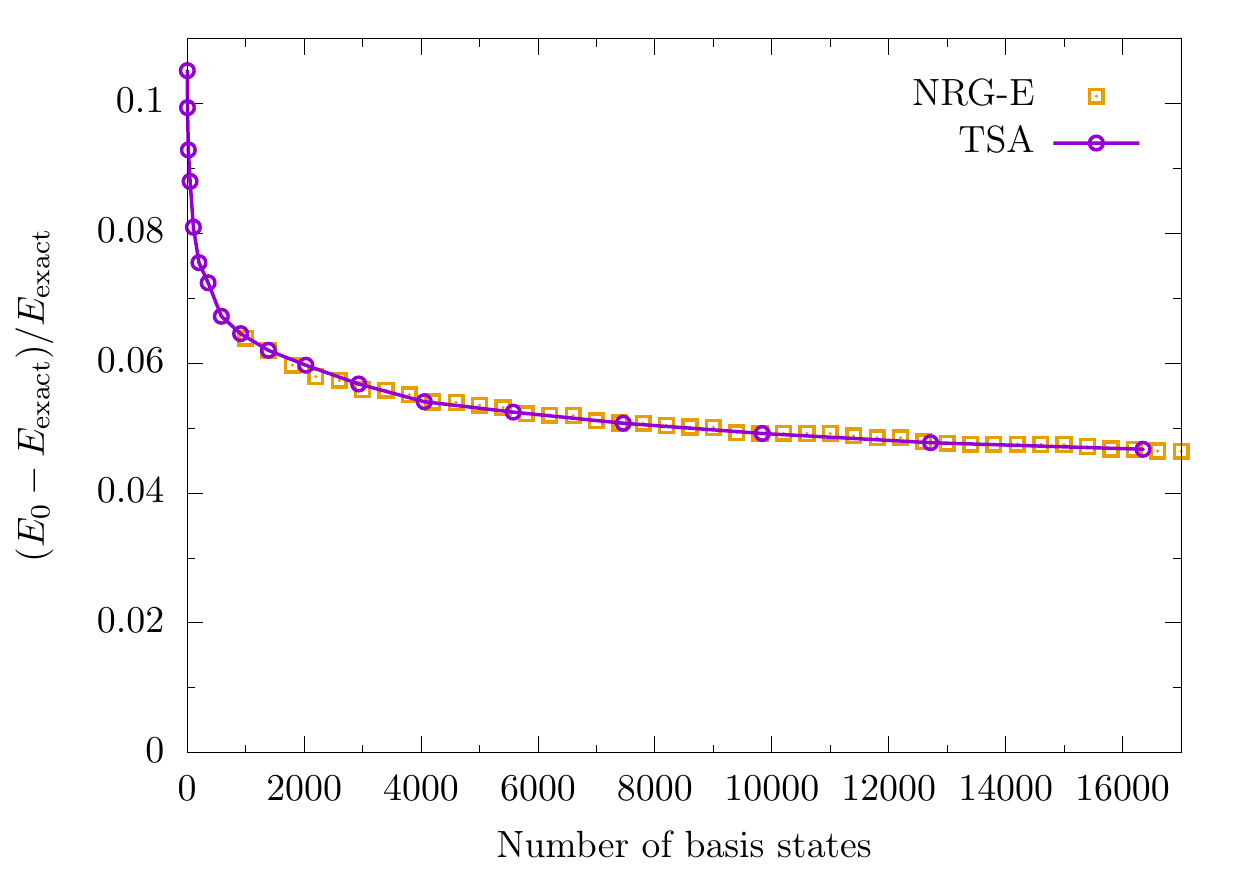} 
      \includegraphics[width=0.49\textwidth]{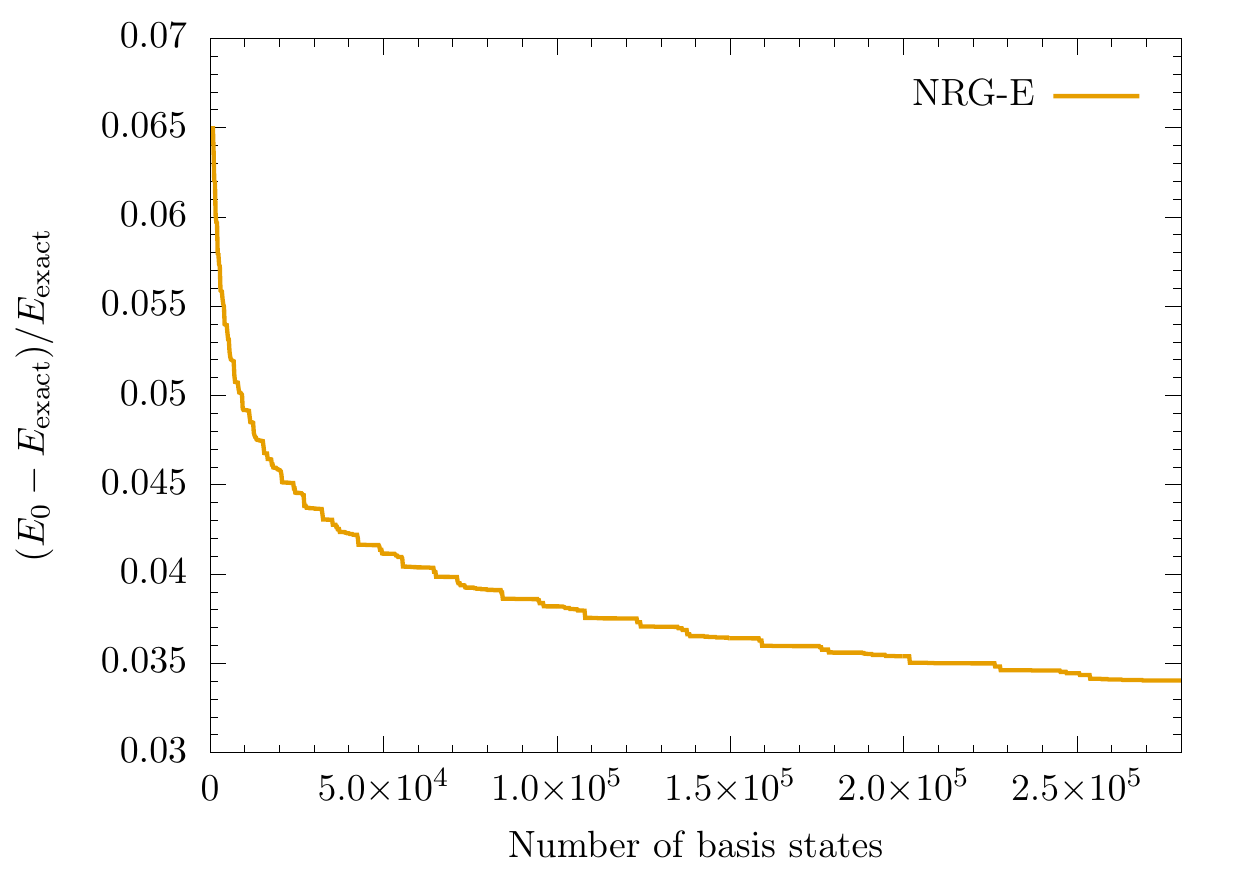}
    \end{center}
    \vspace{-5mm}
\caption{Convergence of the approximate ground state energy $E_0$ with the
number of basis states, as computed using the truncated spectrum procedure (TSA)
and its numerical renormalization group extension (NRG-E). The $c_i=20$ ground
state of the Hamiltonian~\eqref{H} is constructed in terms of $c_f = 10$
eigenstates for $10$ particles. NRG-E is performed with $N_s=600$ and $\Delta
N_s = 200$ (corresponding to an initial energy cutoff of $\Lambda \approx 50$).
We plot every other NRG-E step; excellent agreement between the results of the
numerical renormalization group and the truncated spectrum approach is seen
(left panel). The numerical renormalization group procedure can access number of
basis states far beyond those accessible to full diagonalization (right panel).}
\label{fig:nrg}
\end{figure}

As an illustration, in Fig.~\ref{fig:nrg} we present the convergence of the
ground state energy $E_0$ with $c_i = 20$ as a function of the number of
computational basis states considered in the numerical renormalization group
procedure. The computational basis is formed from eigenstates of the Hamiltonian
with $c_f=10$. As a first check, the numerical renormalization group procedure
(performed with $N_s = 600$ and $\Delta N_s = 200$, corresponding to an energy
cutoff at the first step of the procedure of $\Lambda \approx 50$) is compared
to full diagonalization in Fig.~\ref{fig:nrg}(a). Despite the small size of the
numerical renormalization group Hamiltonian (of total dimension $N_s + \Delta
N_s = 800$), we see that the obtained results accurately reproduce the full
truncated spectrum results of Fig.~\ref{fig:tsa}. The numerical renormalization
group does, however, allow us to consider many more basis states than can be
tackled with with full exact diagonalization in a time and memory efficient
manner. This is illustrated in Fig.~\ref{fig:nrg}(b), where we consider
$280,000$ computational basis states in our numerical renormalization group
procedure. This allows us to converge energies to below $3.5\%$\footnote{
This corresponds to approximately $11\%$ with respect to the Fermi energy.} at
the end of the procedure, which is significantly smaller than the level
spacing $E_1-E_0$ for the parameters under consideration.

\subsection{Ordering by an alternative metric}
%%y

As can be seen in Fig.~\ref{fig:nrg}(b), the convergence of the ground state
energy $E_0$ with the number of basis states is slow: To reach a precision of
$2\%$ it is likely that one will need to consider more than $10^6$ basis states.
It is also evidence that the convergence obtained within the numerical
renormalization group procedure has a lot of structure: there are steps of the
procedure where the ground state energy is approximately constant, whilst at
other steps it rapidly drops. One can then ask: is there an alternative ordering
of the computational basis states that prioritizes the ``important'' states,
where these large drops occur, and so improve the convergence? 

\begin{figure}[h!]
  \begin{center}
    \includegraphics[width=0.6\textwidth]{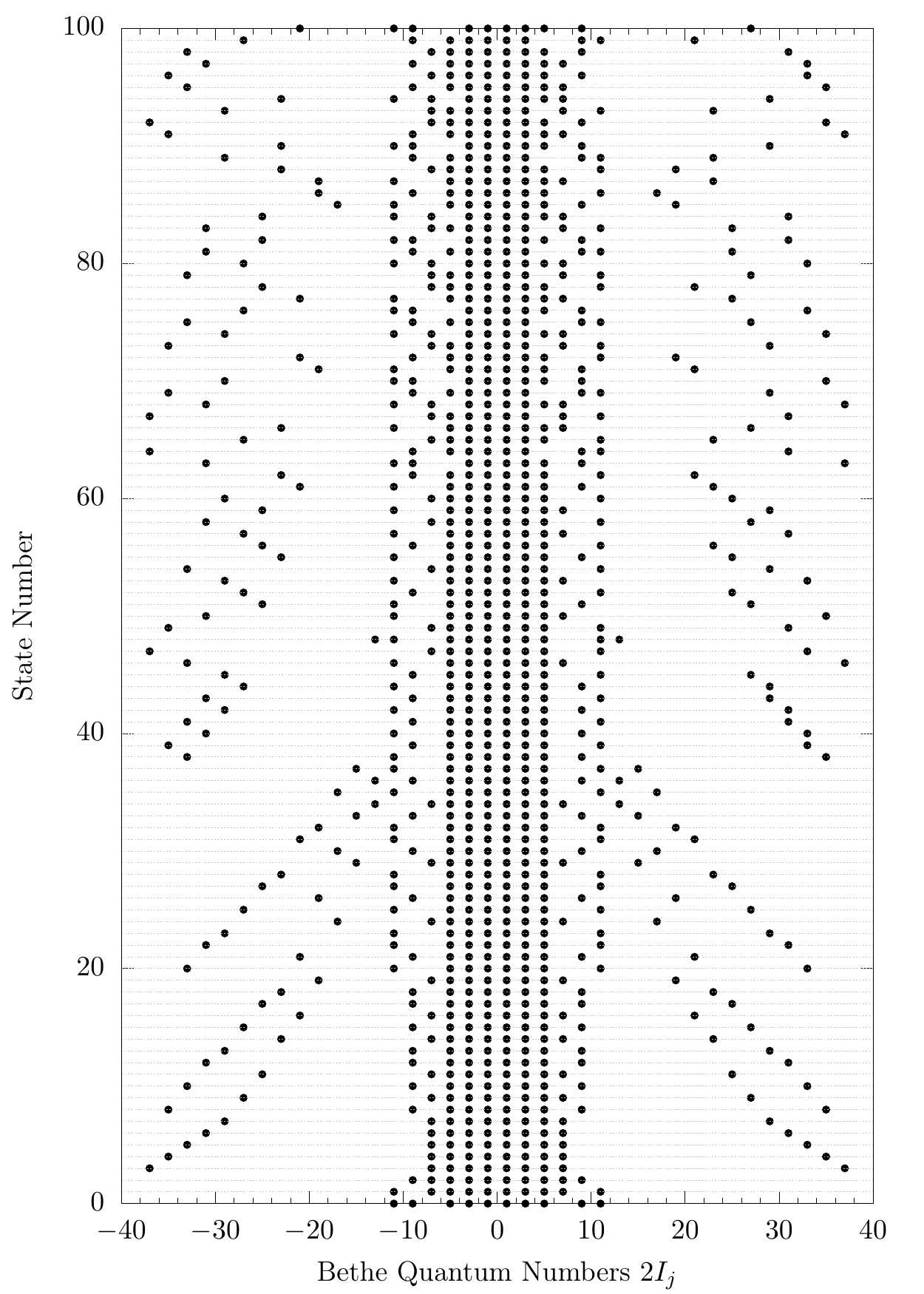}
  \end{center}
  \vspace{-5mm}
  \caption{The configurations of quantum numbers $\{I_j\}$, see
  Eq.~\eqref{eq:logbae}, characterizing the first 100 basis states ordered via
the matrix element metric, Eq.~\eqref{def:me_metric} for the $c_i = 20$ to
$c_f=10$ quench with $N=10$ particles. The total basis was formed from 273,358
states generated by the {\sc abacus} scanning routine. The highest weight
states, according to the metric, have lowest `state number'. Note that some of
the highest weight states contain high momentum (large quantum number $2I_j$)
excitations. }
  \label{fig:integers}
\end{figure}

To begin tackling this problem of modifying the ordering metric, we follow the
suggestions of
Refs.~\cite{konik2011exciton,caux2012constructing,brandino2015glimmers,konik2015predicting}
(see also the discussion in~\cite{james2017nonperturbative}) and take a
pragmatic approach. We order the computational basis states according to the
values of the matrix elements
\begin{align}
\left| \la \{\lambda\}^{(n)}_N | g_2(0) | \tilde E_j\ra \right|, \quad j =0,1,2.  \label{def:me_metric}
\end{align}
Here $|\tilde E_j\ra$ are the three lowest energy eigenstates of the initial
Hamiltonian, i.e. those states we are trying to construct. In practice, $|\tilde
E_j\ra$ are first constructed via the truncated spectrum approach with a small
energy cutoff (corresponding to circa two thousand states) and these approximate
eigenstates are then used to construct the matrix
elements~\eqref{def:me_metric}. This procedure attempts to capture those states
$|\{\lambda\}_N^{(n)}\ra$ that hybridize with and contribute most strongly to
the low energy states.

To get some understanding of how this change of metric,
Eq.~\eqref{def:me_metric}, modifies the states being considered within the
numerical renormalization group procedure, we present the configurations of
quantum numbers $\{I_j\}$ (recall Eq.~\eqref{def:betheintegers}) that
characterize the one hundred highest weight states according to this metric. We
show these in Fig.~\ref{fig:integers}. There is clearly a significant change in
ordering of the states as compared to energy ordering. Most of the highest
weight states under the metric~\eqref{def:me_metric} describe pairs of ``highly
excited quantum numbers'' that have moved away from the ``Fermi sea'' of quantum
numbers centered on zero, leaving behind holes. We see that in a family of
states with fixed configuration of quantum numbers close to zero, states
containing the most excited quantum numbers generally have highest weight.  It
is also apparent that one of the highest weight states is the ground state of
the final Hamiltonian. 

\begin{figure}[t]
  \begin{center}
    \includegraphics[width=0.7\textwidth]{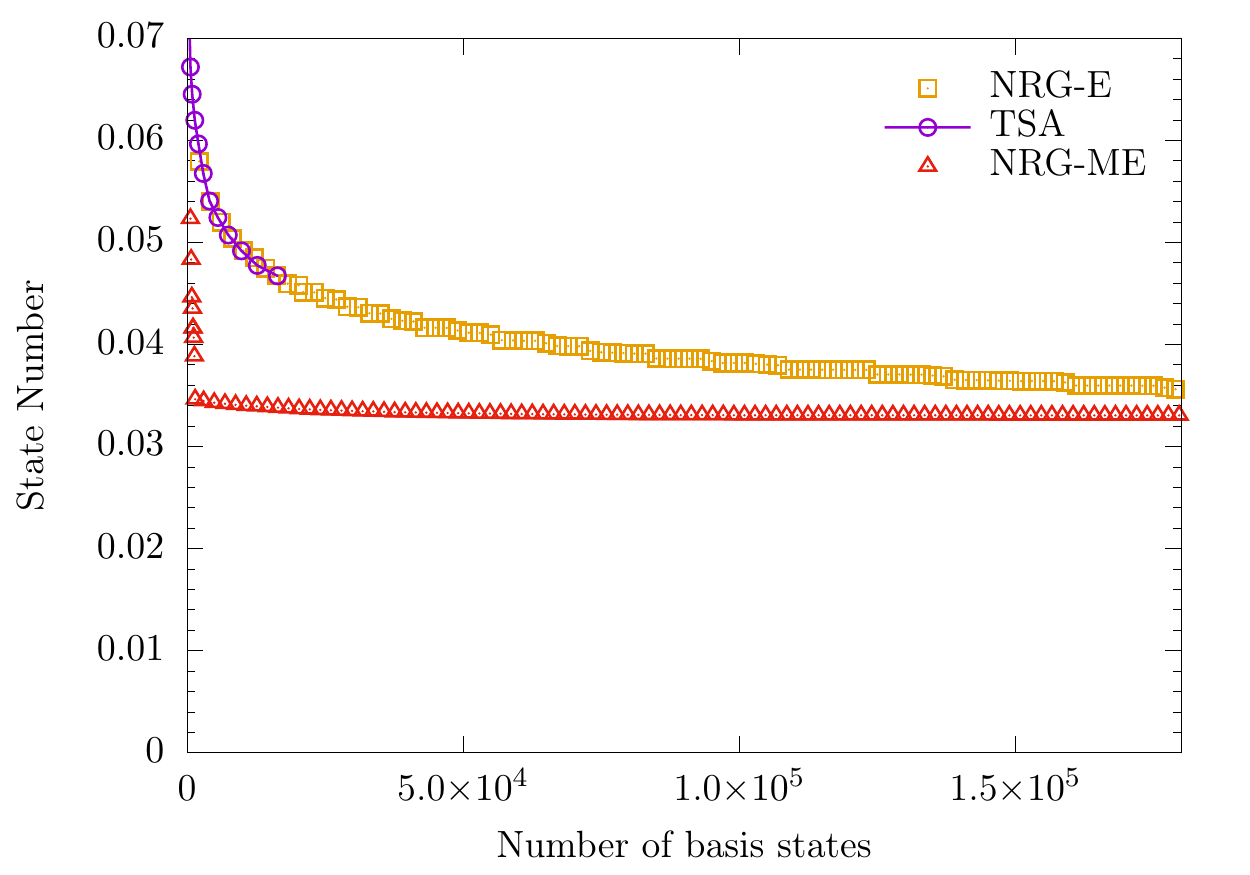}
  \end{center}
  \vspace{-5mm}
  \caption{A comparison between the numerical renormalization group (NRG-E)
  results of Fig.~\ref{fig:nrg}(b) and the modified numerical renormalization
group with matrix element ordering (NRG-ME). We see that the alternative
ordering leads to massive improvement in the convergence of the ground state
energy (and similar improvement is seen in low-lying excited states) for fixed
number of basis states. NRG-ME was performed with $N_s = 600$, $\Delta N_s =
120$ with a total of $280,000$ basis states.}
  \label{fig:nrgalt}
\end{figure}

\subsubsection{Convergence of the ground state energy with matrix element metric}

The reordering presented in Fig.~\ref{fig:integers} seems a little surprising,
but leads to considerable improvement in the convergence of the numerical
renormalization group results. This is shown in Fig.~\ref{fig:nrgalt}, where
after only seven numerical renormalization group steps, the convergence of the
ground state energy is already lower than that obtained with over $10^5$ steps
of the energy-ordered numerical renormalization group procedure. For the
computational basis considered, we have essentially saturated our approximate
representation of the initial state. With this significant improvement in
convergence, we work with alternative (non-energy ordered) metrics in the
remainder of this work.

The problem with the procedure as laid out, at the moment, is there is still a
need to generate a very large computational basis, compute the weight according
to the metric~\eqref{def:me_metric}, reorder and then perform the numerical
renormalization group procedure. The total size of this computational basis
essentially introduces an energy cutoff and this limits the extent to which we
can saturate the approximate representation of the state.

Crucially, insights from the following two sections will allow us, in
Sec.~\ref{Sec:Abacus}, to throw off the shackles of needing to generate a large
computational basis to matrix element order, and instead we will realize a way
to \textit{preferentially generate the high overlap states}.

%%%%%%%%%%%%%%%%%%%%%%%%%%%
\subsubsection{The overlaps: Convergence and structure}
\label{Sec:Structure}
%%%%%%%%%%%%%%%%%%%%%%%%%%%

Beyond examining the convergence of the energy, other checks are critical in
ascertaining the validity of results obtained within the truncated spectrum and
its numerical renormalization group extensions. Above, we are able to compute
exactly (from the Bethe ansatz) the energy to which our obtained state should be
approaching, giving us a quantitative measure of convergence. We have seen that
convergence can be obtained provided a sufficiently large number of eigenstates
are included in the computational basis. 

\begin{figure}[t]
  \begin{center}
    \includegraphics[width=0.7\textwidth]{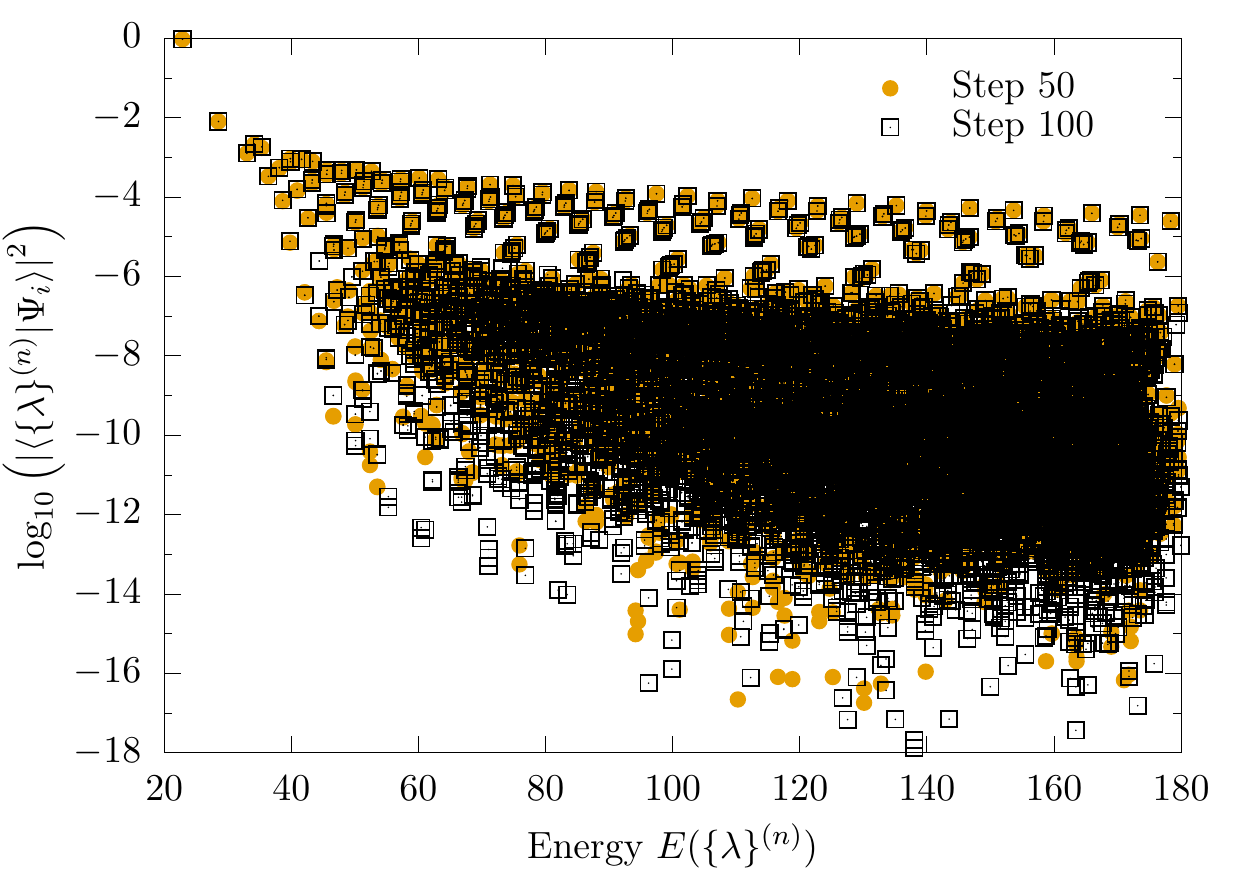}
  \end{center}
  \vspace{-5mm}
  \caption{The overlaps $\la \{\lambda\}^{(n)} | \Psi_i \ra$ between the ground
  state of the Lieb-Liniger model with $c_i = 20$ and eigenstates
$|\{\lambda\}^{(n)}\ra$ of the $c_f=10$ Lieb-Liniger model with energy
$E(\{\lambda\}^{(n)})$. The ground state $|\Psi_i\rangle$ is constructed in
terms via the NRG-ME procedure, and we present results after 50 and 100 steps,
showing that the dominant overlaps are well-converged. This is not surprising in
light of Fig.~\ref{fig:nrgalt}.}
  \label{fig:overlaps}
\end{figure}

Having constructed an approximation to our initial state, we directly have the
overlaps at hand. These are essential for computing non-equilibrium dynamics and
the long time steady state following a quench. It is worthwhile, at this point,
to examine that the overlaps themselves are well converged (which should follow
directly from the convergence of the energy). In Fig.~\ref{fig:overlaps} we plot
the square of the overlaps $|\la \Psi_i \vert \{\lambda\}^{(n)}\ra|^2$ as a
function of the  energy $E(\{\lambda\}^{(n)})$ of computational basis state
$|\{\lambda\}^{(n)}\rangle$.\footnote{We note that these overlaps are related to
  the statistics of work done~\cite{silva2008statistics}: 
\begin{align}
P(W) = \sum_n \delta\Big(W - E(\{\lambda\}^{(n)}_N) \Big) \left\vert \big\la \Psi_i \big\vert \{\lambda\}^{(n)}_N  \big\ra \right\vert^2, \nonumber
\end{align}
giving the coefficients of the delta functions. This quantity has been studied,
for example, in the Ising field theory in
Refs.~\cite{rakovszky2016hamiltonian,hodsagi2018quench}.}
Results are presented at two well-separated steps of the numerical
renormalization group procedure (conducted with metric~\eqref{def:me_metric}).
We see clearly that large overlap computational basis states large have
well-converged overlaps, being almost identical at the two different steps of
the procedure. The computational states with very small overlaps are physically
unimportant (recall Sec.~\ref{Sec:Ideal}) and clearly subject to floating point
errors of the numerical implementation.

A significant message to take from Fig.~\ref{fig:overlaps} (and implicitly from
Fig.~\ref{fig:nrgalt}) is that computational basis states with large energies
can contribute significantly to the initial state. Indeed, in
Fig.~\ref{fig:overlaps} we see a band of ``high overlap states'' with square
overlaps $\sim10^{-4}-10^{-5}$ extending out to high energies. We note that the
results of Fig.~\ref{fig:overlaps} are well-converged with numerical
renormalization group step, but cannot be well-converged with regards to
increasing the size of the computational basis. This is evident from the fact
that our procedure produces states with $\langle \Psi_i |
\Psi_i\rangle_{\text{approx}} = 1$, and there is no reason to believe high
overlap states stop at energy $E(\{\lambda\}^{(n)}) = 180$, the effective energy
cutoff of our computational basis. Note that this does not imply that physical
quantities are not well converged with the size of the basis (we indeed observe
that physical quantities \textit{are} well converged).

\begin{figure}[t!]
  \begin{center}
    \includegraphics[width=0.6\textwidth]{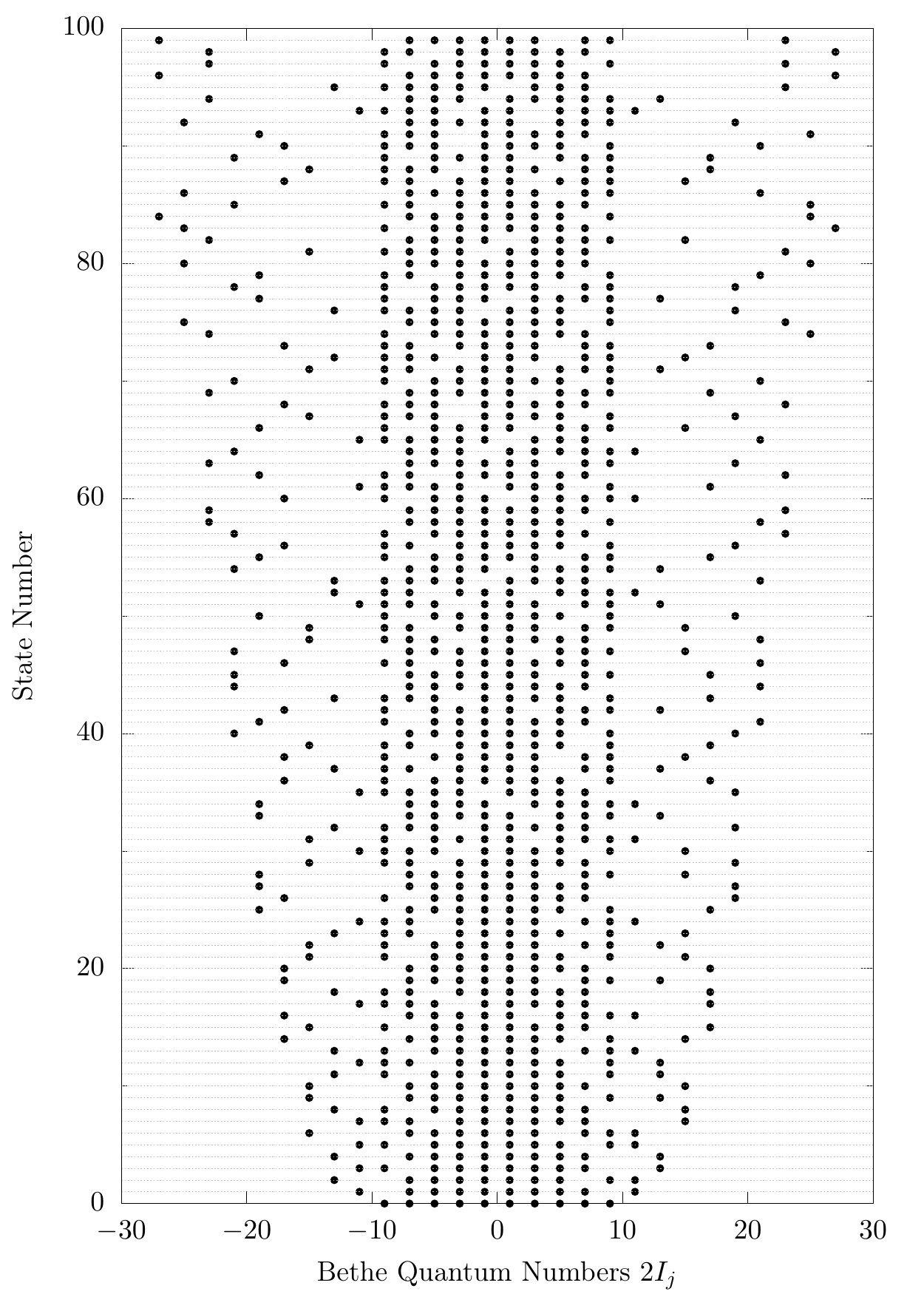}
  \end{center}
  \caption{
    The configurations of quantum numbers $\{I_j\}$, see
  Eq.~\eqref{def:betheintegers}, characterizing the 100 computational basis
states with highest overlaps in the $c_i=20$ ground state constructed in terms
of $c_f=10$ eigenstates. These were obtained from a numerical renormalization
group procedure with the basis ordered according to the
metric~\eqref{def:me_metric}. States with larger overlaps appear towards the
bottom of the figure.}
    \label{fig:highoverlapints}
\end{figure}

With convergence at fixed computational basis size confirmed for the high
overlap states, let us now illustrate the structure of these high overlap
states. From the point of view of analytical calculations, there are only a
handful of examples where overlaps can be
computed~\cite{denardis2014solution,brockmann2014overlaps,brockmann2014neelxxz,piroli2016multiparticle,piroli2016quantum,bucciantini2016stationary},
so numerical routines have significant potential when brought to bear on such a
problem.

For the $c_i=20 \to c_f = 10$ quench, the configurations of the quantum numbers
in the highest 100 overlap states are shown in Fig.~\ref{fig:highoverlapints}.
The states are organized from highest weight (bottom of the plot) to lowest
weight (top of the plot). We see that the highest weight overlap is with the
ground state of the final Hamiltonian in this case. As we proceed up the plot,
we see a pair of excited integers move away from the Fermi sea about the origin,
and the left behind holes moving around within the Fermi sea.

At first glance, it may be a little surprising that the high overlap state
integers shown in Fig.~\ref{fig:highoverlapints} are ordered so differently to
those of matrix element metric, Fig.~\ref{fig:integers}. In some sense, this
tells us that we are dealing with an ``easy quench'' where the overlaps rapidly
converge converge if we get approximately the correct ordering metric. In the
next section we will construct an alternative metric, taking some inspiration
from the information in Fig.~\ref{fig:highoverlapints} and combining it with
other knowledge, that more accurately reproduces the optimal ordering. 

%%%%%%%%%%%%%%%%%%%%%%%%
\subsection{Efficient generation of high overlap states}
\label{Sec:Abacus}
%%%%%%%%%%%%%%%%%%%%%%%

In the previous sections we have established the efficacy of ordering the
computational basis based upon information about the perturbing operator (recall
Sec.~\ref{Sec:formulation}). However, as we have already highlighted, there is a
clear issue with the procedure that has been discussed. This is best illustrated
by Figs.~\ref{fig:nrgalt} and~\ref{fig:overlaps}: The saturation of the error in
the ground state energy in this modified numerical renormalization group
procedure is ultimately set by the energy cutoff of the truncated basis on which
we perform the new ordering. The behavior of the overlaps as a function of
energy, Fig.~\ref{fig:overlaps}, clearly shows that high energy computational
basis states can contribute significantly to the initial state. Indeed, we see
in Fig.~\ref{fig:overlaps} that computational basis states with $E\sim
10^2-10^3$ can have square overlaps as large as $\sim 10^{-4}$.

Following this procedure, if we wish to achieve precision of sub-1\% in the
energy of the state for ten particles, we will need to first generate a large
basis of size $\sim 10^6$, then order according to the
metric~\eqref{def:me_metric}, and then perform a truncated spectrum or numerical
renormalization group procedure. The ordering step requires computing matrix
elements for \textit{all computational basis states}. This is computationally
costly, even if much more efficient than working with an energy-ordered
numerical renormalization group procedure. Instead, it would be much better if
we could preferentially generate the high overlap states necessary for our
algorithms.

In this section we will formulate such a preferential state generation
procedure. This will be based upon the philosophy of the {\sc abacus} ({\bf
A}lgebraic {\bf B}ethe {\bf A}nsatz-based {\bf C}omputation of {\bf U}niversal
{\bf S}tructure factors) Hilbert space scanning algorithm. A general overview of
this approach, developed to tackle the computation of equilibrium dynamical
correlation functions, can be found in Ref.~\cite{caux2009correlation}.

The essential insight for applying {\sc abacus}-inspired methods to the
non-equilibrium problem at hand is contained within
Figs.~\ref{fig:highoverlapints}. There one can see that the largest overlap is
with the ground state of the final Hamiltonian, herein denoted
$|\{\lambda\}^{(0)}\rangle$. The matrix element metric,
Eq.~\eqref{def:me_metric}, then approximately orders the states that hybridize
most strongly with $|\{\lambda\}^{(0)}\rangle$ via the perturbing operator
$g_2(x)$, i.e. the $|\{\lambda\}^{(n)}\rangle$ that maximize
\begin{align}
\int \rd x\, \langle \{\lambda\}^{(0)} | g_2(x) |\{\lambda\}^{(n)}\ra \propto \delta(P_0-P_n)\, \la \{\lambda\}^{(0)} | g_2(0) | \{\lambda\}^{(n)}\ra. \label{abacusMotivation}
\end{align}
Here we use the short hand $P_n = P(\{\lambda\}^{(n)})$. Formulated in this
manner, the case for applying an {\sc abacus}-like algorithm is clear. Consider
computing the \textit{equilibrium} dynamical correlation function of $g_2(x)$:
\begin{align}
  S_{g_2}(k,\omega) \propto \int_{-\infty}^\infty \int_{-\infty}^{\infty} \text{d}x\,\text{d}t\ e^{i(kx+\omega t)} \la\{\lambda\}^{(0)} | g_2(x,t) g_2(0) | \{\lambda\}^{(0)}\rangle. 
\end{align}
Here $g_2(x,t) = e^{iHt} g_2(x) e^{-iHt}$ is the time evolved $g_2$ operator.
The dynamical correlation function $S(k,\omega)$ can be evaluated by inserting
the resolution of identity between the two operator and summing the resulting
Lehmann spectral representation
\begin{align}
  S_{g_2}(k,\omega) \propto \sum_n \delta(\omega - E_n + E_0) \delta(k - P_n + P_0) \Big\vert \langle \{\lambda\}^{(0)} | g_2(0) | \{\lambda\}^{(n)}\rangle\Big\vert^2. \label{eq:dycorrfun}
\end{align}
Once again, we use a short hand notation $E_n = E(\{\lambda\}^{(n)})$. Thus the
states with highest weight under the metric~\eqref{def:me_metric} are
(approximately) those that contribute most strongly to Eq.~\eqref{eq:dycorrfun}
-- the problem that {\sc abacus} was designed to tackle.\footnote{If the largest
overlap was not a ground state, the same procedure could be implemented with
$|\{\lambda\}^{(0)}\rangle$ replaced with the largest overlap state, with
obvious modifications to Eqs.~\eqref{eq:perttheory} and~\eqref{abacusMetric}.} 

\begin{figure}
  \begin{center}
    \includegraphics[width=0.85\textwidth]{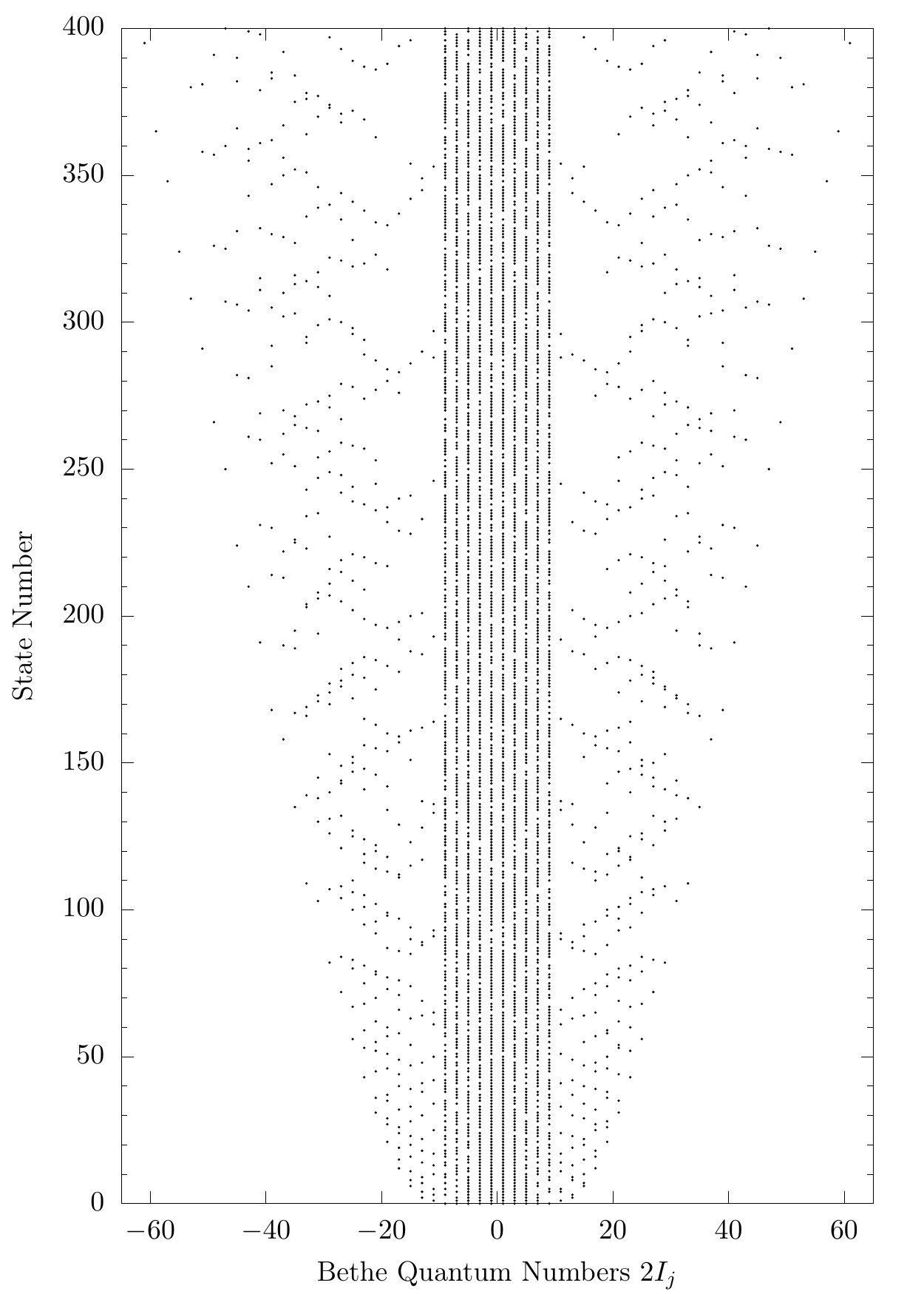}    
  \end{center}
  \caption{The configurations of integers $\{I_j\}$ in the first 400 states
  generated via preferential scanning and ordered according to the
metric~\eqref{abacusMetric} for the $c_i=20\to c_f=10$ quench. (Highest weights
correspond to lowest state numbers.) Notice the similarity with
Fig.~\ref{fig:highoverlapints}, the output of the NRG-ME procedure; the
preferential scanning algorithm efficiently generates those basis states with
largest overlaps.}
  \label{fig:intsprefscan}
\end{figure}

We are, of course, not aiming to compute equilibrium correlation functions here,
but instead non-equilibrium dynamics. By approximating the initial state in the
metric~\eqref{def:me_metric} by $|\{\lambda\}^{(0)}\rangle$ we can motivate an
{\sc abacus}-like scheme to generate the states with high weight on this metric.
We can, however, draw some inspiration from perturbation theory to construct a
better metric. If we have the state with highest overlap with the state we are
trying to construct (here this state is $|\{\lambda\}^{(0)}\rangle$), the
perturbation theory tells us the first order term in the expansion for the
approximation state should be
\begin{align}
  |\Psi_i\rangle_{\text{approx}} = |\{\lambda\}^{(0)}\rangle + (c_i-c_f)R \sum_{m\neq0} \frac{ \langle \{\lambda\}^{(m)}| g_2(0) | \{\lambda\}^{(0)}\rangle}{E_0 - E_m} |\{\lambda\}^{(m)}\rangle + O(g^2).
  \label{eq:perttheory}
\end{align}
Thus a better metric is obviously apparent: we should organize our computational
basis states according to the their weights
\begin{equation}
  w\left(|\{\lambda\}^{(n)}\rangle\right) =  \left\vert \frac{\langle \{\lambda\}^{(n)}| g_2(0) | \{\lambda\}^{(0)}\rangle}{ E_n - E_0 + \epsilon}\right\vert. 
  \label{abacusMetric}
\end{equation}
Here $\epsilon$ is a simple numerical factor introduced to avoid a divergence
for the case of $(n)=(0)$ (we take $\epsilon = 0.1$). States generated in an
{\sc abacus}-like scanning according to their matrix elements weights can easily
be post-sorted according to~\eqref{abacusMetric}.\footnote{We note that {\sc
abacus}-like scanning routines do not generally produce states contributing to
Eq.~\eqref{eq:dycorrfun} in a monotonically decreasing
manner~\cite{caux2009correlation}. Thus post-sorting for a consistent truncation
scheme is a necessity in any case.}

Implementing this procedure, the 400 highest weight states for the $c_i=20\to
c_f=10$ quench for ten particles are shown in Fig.~\ref{fig:intsprefscan}. This
is clearly rather different to the pure matrix element ordering,
Fig.~\ref{fig:integers}, and is much more in keeping with the results shown in
Fig.~\ref{fig:highoverlapints}. In the next section we will see that this
ordering leads to excellent convergence of the initial state energy, and the
ordering may be close to optimal in some scenarios (we will later see an example
where this does not appear to be the case). In light of the data presented in
Fig.~\ref{fig:intsprefscan}, it is hardly surprising that energy-ordering the
basis fails to give good convergence: Even within the first 400 state there are
states with highly excited quantum numbers, i.e. states with high energies. As
we are now able to generate computational basis states without an implicit (or
explicit) energy cutoff, we expect to be able to saturate the energy of the
initial state to a much larger extent (recall Fig.~\ref{fig:nrgalt}).

We call this procedure a ``high overlap states truncation scheme.'' We note that
we used both {\sc abacus} and an independently written Hilbert scanning routine
within this manuscript. This has helped provide independent checks of the
preferential state generation for all our results. 

%%%%%%%%%%%%%%%%%%
\subsection{Checking convergence within the high overlap states truncation scheme}
%%%%%%%%%%%%%%%%%%

With a high overlap states truncation scheme at hand, there is no longer an
energy cutoff within the computational basis. The presence of an energy cutoff
ultimately governed the value to which the energy of the initial state could
saturate in the previous sections (for example, in Fig.~\ref{fig:nrgalt} the
maximum saturation to within $\sim0.035E_\text{exact}$). This means that now we
can saturate agreement to much less than 1\%, while doing so at a significantly
decreased computational burden.

Before discussing this in more detail, it is worth first re-evaluating how we
assess convergence within the high overlap states truncation scheme. So far, we
have checked how the energy of the state varies with the number of computational
basis states, but it is not entirely clear how one and if one can extrapolate
these results to understand the exact one. There is, for example, no obvious
scaling law for the energy as a function of number of basis states.

Within our truncation scheme, a central role is played by the weights of the
computational basis states under the metric~\eqref{abacusMetric}. One potential
option for assessing convergence of computed quantities is to plot against this
weight (i.e., plotting quantities as a function of the smallest considered
weight~\eqref{abacusMetric}). We will see that this leads to a reasonable
extrapolation scheme, as compared to the number of considered computational
basis states.\footnote{We note that as we study a continuum model, we cannot
guarantee that we generate all states above a given weight of the
metric~\eqref{abacusMetric}, especially when this becomes small. Additional
checks, such as convergence of results with the number of states generated and
ordered according to the metric~\eqref{abacusMetric} must also be performed.}

%%%%%%%%%%%%%%%%%%
\subsubsection{Convergence of the energy}
%%%%%%%%%%%%%%%%%%

\begin{figure}[t]
  \begin{center}
    \begin{tabular}{ll}
      (a) & (b) \\
      \includegraphics[width=.48\textwidth]{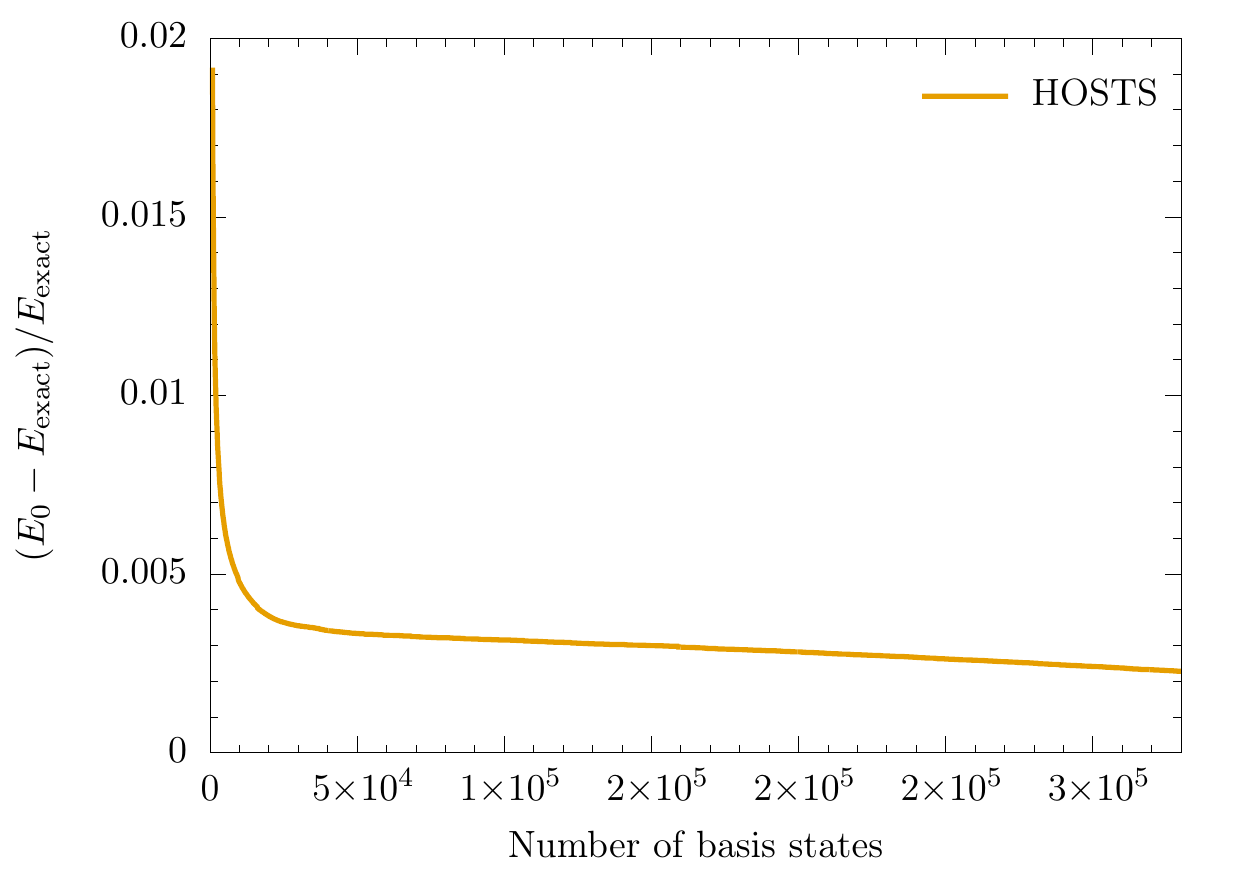} 
          & \includegraphics[width=.48\textwidth]{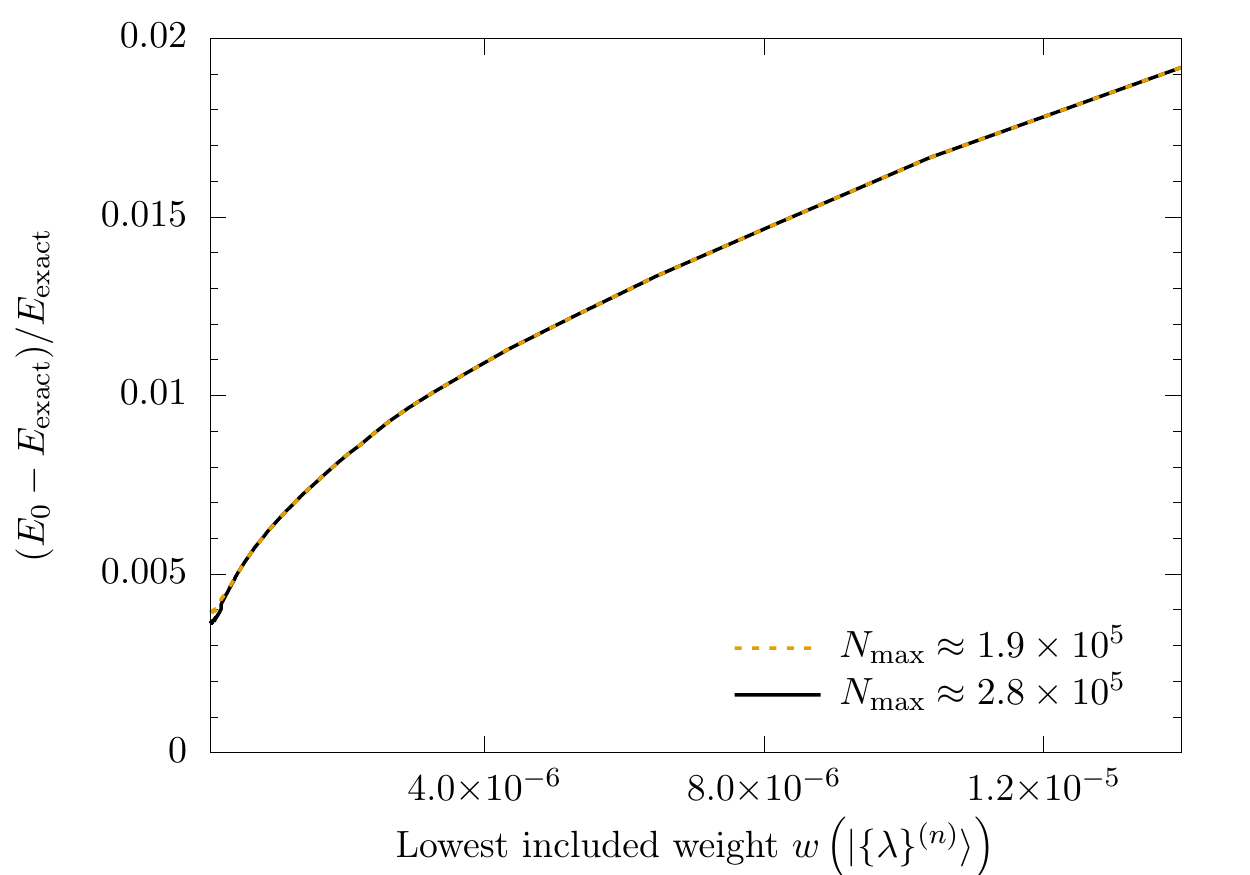}
    \end{tabular}
  \end{center}
  \vspace{-5mm}
  \caption{The convergence of the energy $E_0$ of the $c_i=20$ ground state
  constructed in terms of $c_f = 10$ eigenstates via the numerical
renormalization group within the high overlap states truncation scheme (HOSTS).
(a) $E_0$ as a function of number of basis states; $N_{\text{max}} \approx
3.9\times10^5$ states are generated via {\sc abacus} and ordered according to
their weights~\eqref{abacusMetric}. (b) $E_0$ as a function of the lowest
included weight in the first two hundred steps of the numerical renormalization
group procedure for two different total basis sizes, $N_{\mathrm{max}}$. In both
(a) and (b) the numerical renormalization group procedure is performed with
$N_\text{s} + \Delta N_s = 800$ and $\Delta N_s = 160$. Convergence of $E_0$ to
under 1\% is achieved with only a few thousand basis states, cf.
Fig.~\ref{fig:nrgalt}. The convergence w.r.t. the Fermi energy is also smaller
than $1\%$ towards the end of the procedure.} 
  \label{fig:abacusConv}
\end{figure}

Let us now examine the convergence of the energy of the initial state
constructed with the basis of high overlap states via the numerical
renormalization group. This is shown, as a function of the number of basis
states in Fig.~\ref{fig:abacusConv}(a). The computational basis states are
generated preferentially by running the {\sc abacus} algorithm for 30 seconds
and then reordering the generated states according to the
metric~\eqref{abacusMetric}. This yields an ordered basis of 220,743 states, on
which we subsequently perform the numerical renormalization group procedure (in
fact, we see that excellent convergence is achieved for basis sizes accessible
to full diagonalization). We see very rapid convergence of the initial state
energy, requiring only a few thousand computational states to obtain a
convergence of under 1\%. This should be contrasted to the traditional energy
ordering, see Figs.~\ref{fig:tsa} and~\ref{fig:nrg}, where we would likely
require $> 10^6$ computational basis states to reach the same level of
convergence.

In Fig.~\ref{fig:abacusConv}(b) we also present the convergence of the initial
state energy $E_0$ as a function of the lowest weight~\eqref{abacusMetric}
included in each iteration of the numerical renormalization group procedure. We
show data for two different total basis sizes $N_{\text{tot}}$, which shows that
at very small included weights there is some dependence on $N_{\text{tot}}$.
This implies that our preferential state generation routine has not generated
all the computational basis states with weights above a given small value.  

%%%%%%%%%%%%%
\subsubsection{Convergence of the overlaps}
%%%%%%%%%%%%%

\begin{figure}[t]
  \begin{center}
    \begin{tabular}{ll}
      (a) & (b) \\
      \includegraphics[width=.48\textwidth]{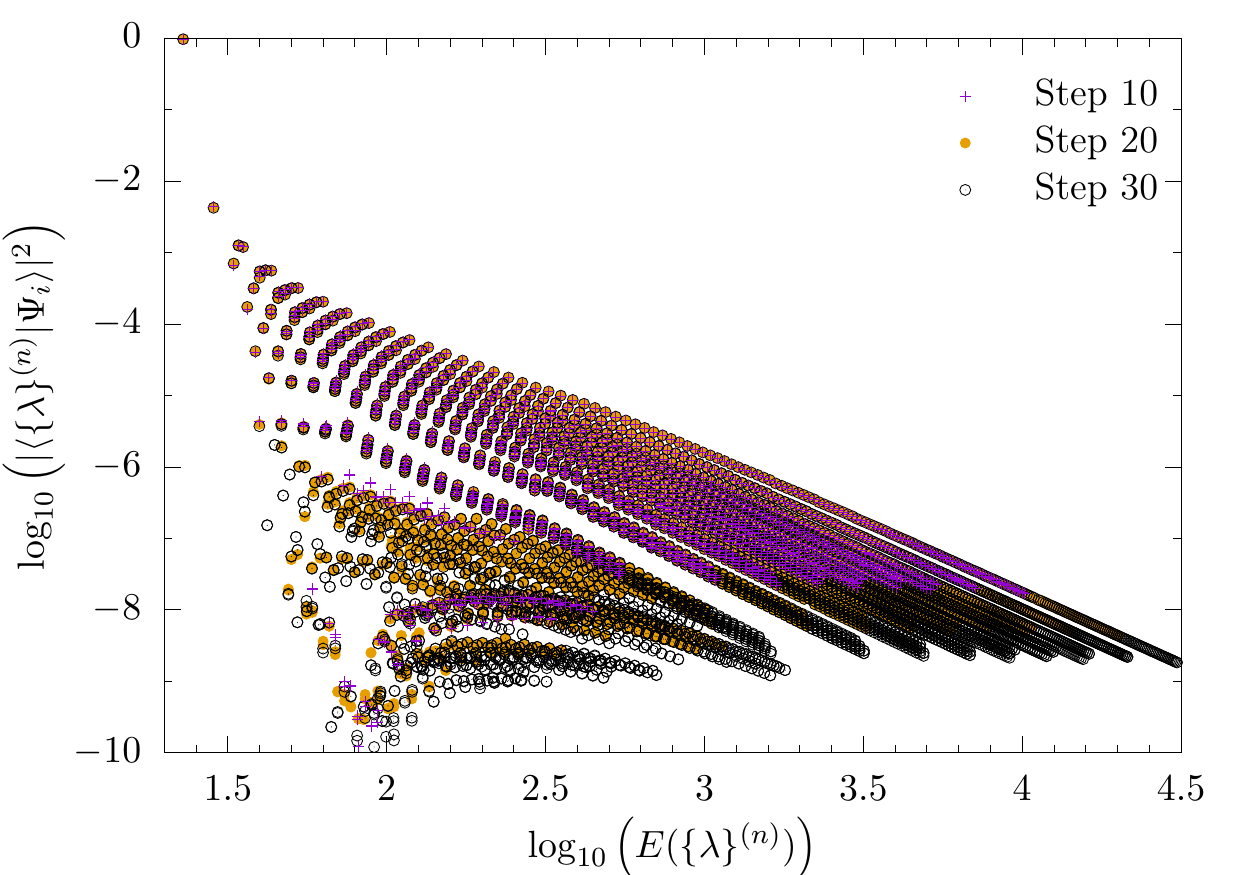}
          & \includegraphics[width=.48\textwidth]{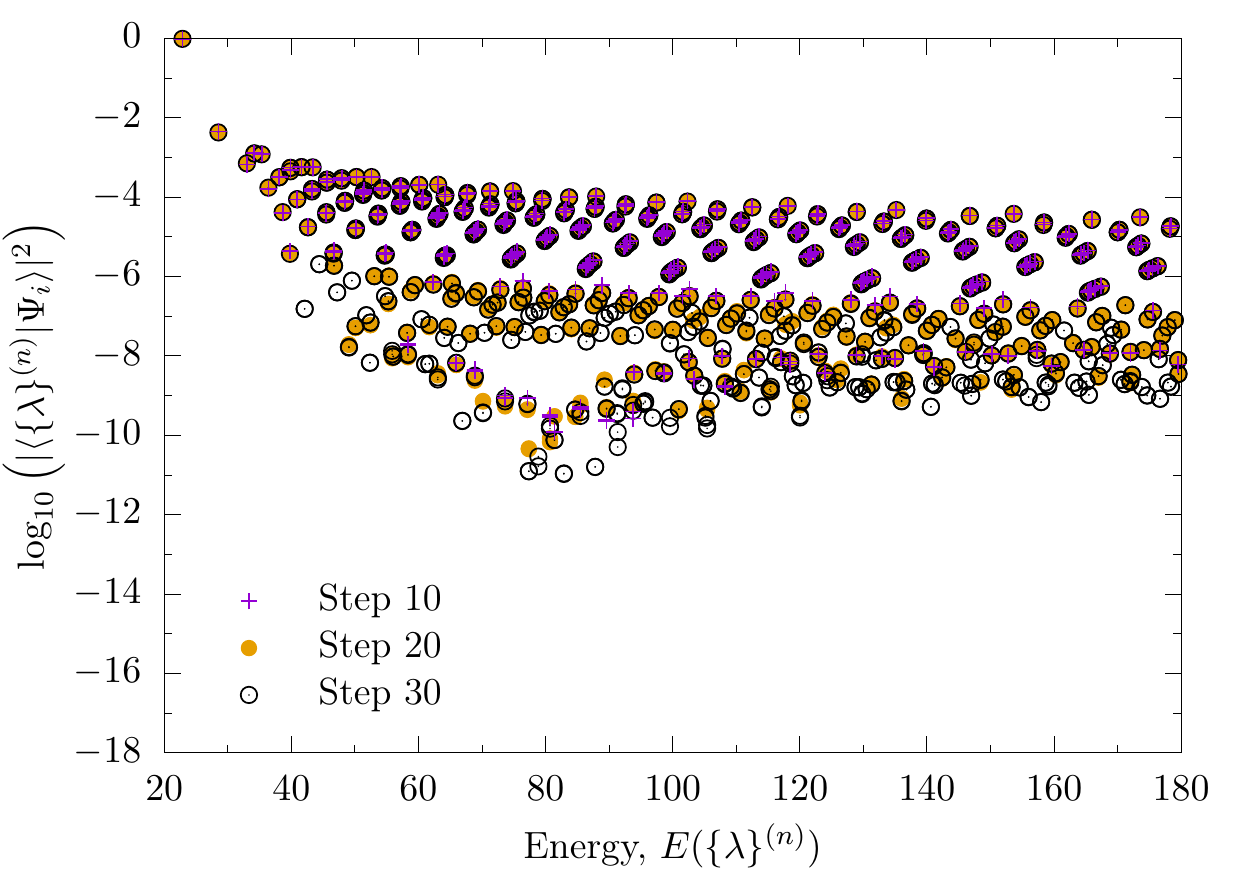}
    \end{tabular}
  \end{center}
  \vspace{-5mm}
  \caption{(a) The convergence of the overlaps at different steps of the
  numerical renormalization group procedure implemented within the high overlap
states truncation scheme. The $c_i=20$ ground state constructed is constructed
in terms of $c_f = 10$ eigenstates. Parameters of the procedure are as in
Fig.~\ref{fig:abacusConv}. (b) A focused region of the plot, to be compared
directly  with Fig.~\ref{fig:overlaps}. }
  \label{fig:abacusConvWfn}
\end{figure}

We have just seen that the high overlap states truncation scheme yields
excellent convergence of the initial state energy. Let us now turn attention to
the overlaps themselves, and how these converge. We present example data in
Fig.~\ref{fig:abacusConvWfn} for the $c_i = 20 \to c_f = 10$ quench with ten
particles. In Fig.~\ref{fig:abacusConvWfn}(a) we present the overlaps (as a
function of energy of the computational basis state) at three steps of the
numerical renormalization group procedure. From this figure, we can make a few
observations. 

Firstly, we observe that the high overlap states truncation scheme is indeed
preferentially targeting high overlap states. The overlaps generated at early
stages of the numerical renormalization group procedure are larger and remain
well converged at later steps. Secondly, we see that the quench generates very
high energy states: by just the thirtieth step of the procedure, we are probing
states with energies $E(\{\lambda\}^{(n)}) \gg 10^4$ (for reference, the ground
state energy is $E_\text{exact} = 26.9684027\ldots$). Thirdly, there appear to
be clear ``families of states'' within this plot, whose overlaps at high
energies can quite easily be predicted by extrapolation. 

We can directly compare the results of our computation to those in
Fig.~\ref{fig:overlaps}, obtained from the original matrix element ordering
without preferential generation of high overlap states. As compared to
Fig.~\ref{fig:abacusConvWfn}(b), we see that the high overlap states truncation
scheme avoids dealing with the large number of low overlap states (as it was
constructed to do), targeting instead the few high overlap states within the
energy window of Fig.~\ref{fig:overlaps}. The truncation scheme is clearly
working as desired.  

%%%%%%%%%%%%%
\subsubsection{Convergence of local expectation values}
%%%%%%%%%%%%%

We have focused thus far on obtaining the energy of the initial state to high
precision. One may ask is this convergence criteria is indeed the same as
correctly constructing the initial state? In this section we turn our attention
towards local properties of the constructed state. In particular, we consider
the behavior of expectation values of local operators within both the exact
initial state and the approximation initial state. This will allow us to
establish that we are correctly reproducing local observables within the state,
not only its energy. 

\begin{figure}[t]
  \begin{center}
    \begin{tabular}{ll}
      (a) & (b) \\ 
      \includegraphics[width=0.48\textwidth]{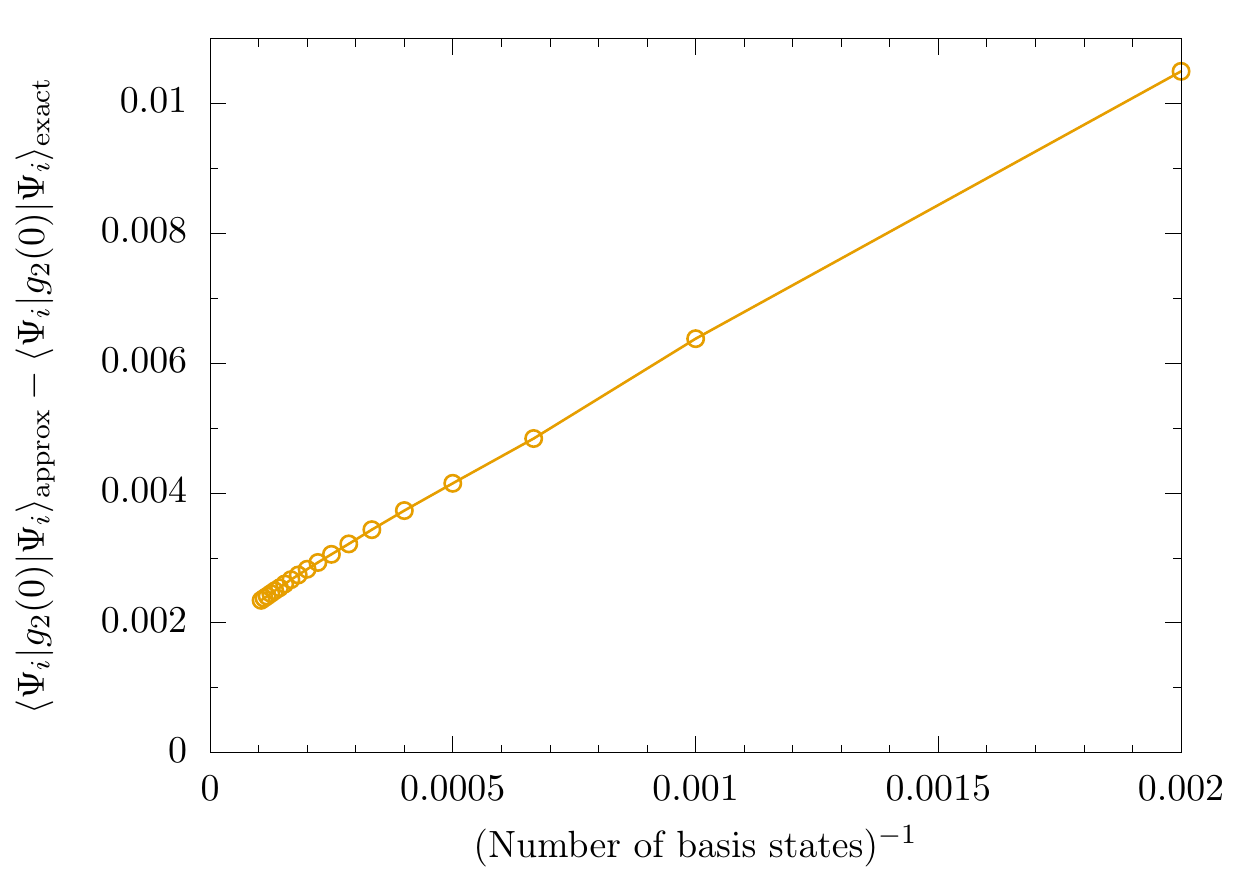}
          & \includegraphics[width=0.48\textwidth]{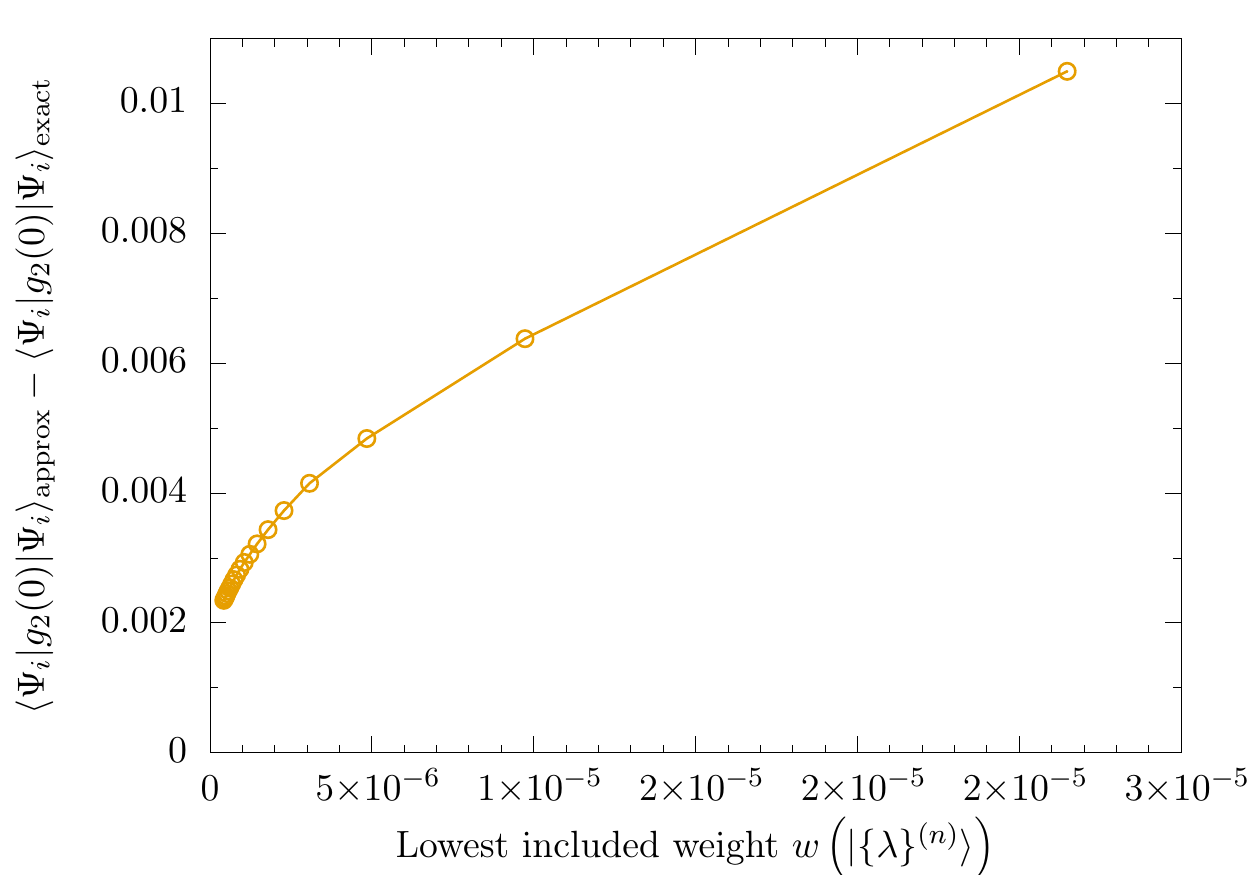}
    \end{tabular}
  \end{center}
  \vspace{-5mm}
  \caption{The difference between the expectation value of $g_2(0)$ in the
  constructed $c_i=20$ ground state (expressed in terms of $c_f=10$ eigenstates)
and the exact value. Data is presented for $N=10$ particles, where the exact
result is $\langle g_2(0)\rangle_\text{exact} = 0.0238263$. (a) Convergence with
the number of basis states; (b) convergence with lowest included weight,
according to the metric~\eqref{abacusMetric}.}
  \label{fig:g2}
\end{figure}

To start with our study of local correlations, let us note a trivial point.
Particle number $N$ is conserved within the Hamiltonian, which when combined
with translational invariance ensures that the expectation value of the local
density within all eigenstates (and the approximate initial state) satisfies
$\langle \{\lambda\}_N |\Psi^\dagger(x) \Psi(x) |\{\lambda\}_N \rangle = N/R$ by
construction. Thus our state of course satisfies this restriction, by
construction. 

\begin{figure}[t]
  \begin{center}
    \includegraphics[width=0.7\textwidth]{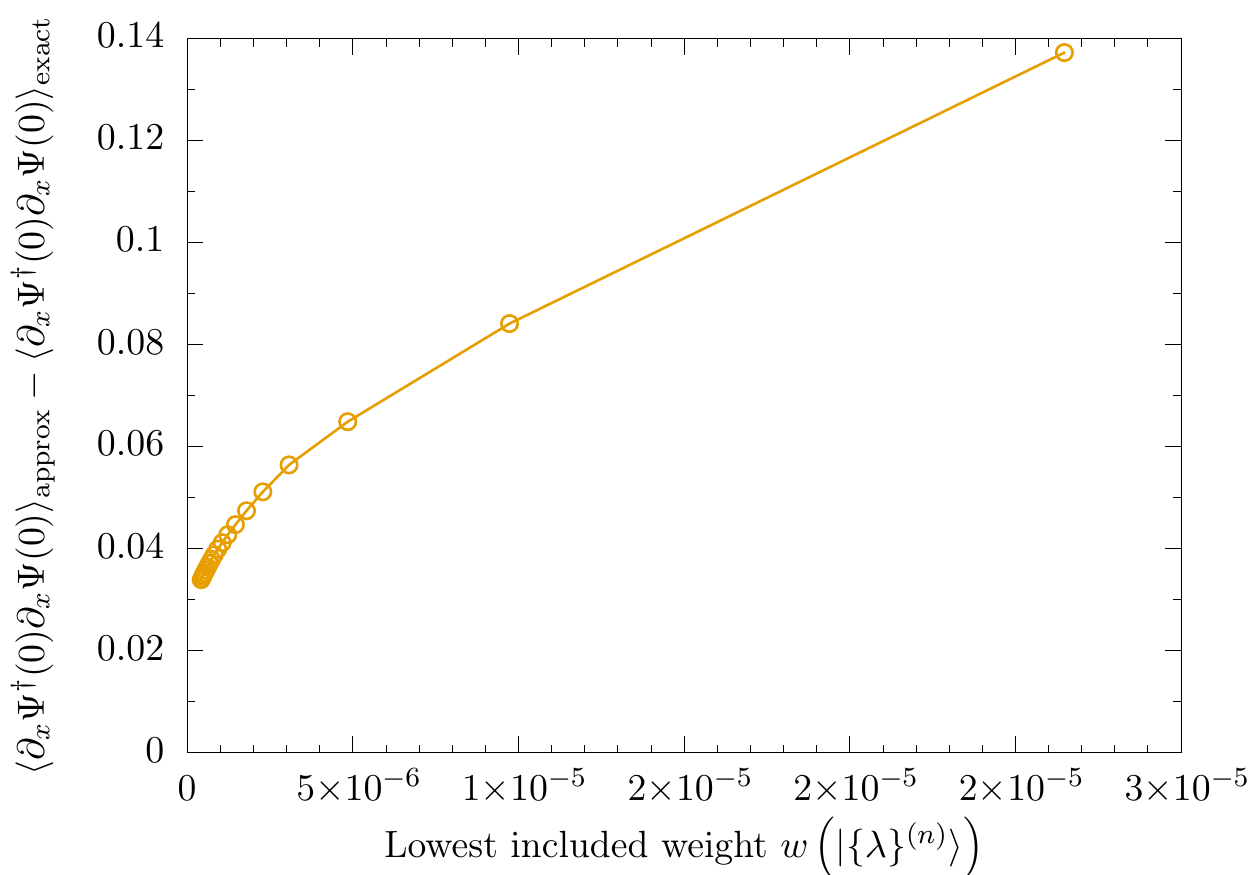}
  \end{center}
  \vspace{-5mm}
  \caption{The difference between the expectation value of $\p_x \Psi^\dagger(0)
  \p_x \Psi(0)$ in the constructed $c_i = 20$ ground state (expressed in terms
of $c_f = 10$ eigenstates) and its exact value. Data is presented for $N=10$
particles, where $\langle \p_x \Psi^\dagger(0)\p_x\Psi(0)\rangle_{\text{exact}}
= 2.22032$.}
  \label{fig:pxpx}
\end{figure}

Convergence of the energy implies that local expectation values of the operators
appearing within the Hamiltonian should also be converging. We confirm this in
Figs.~\ref{fig:g2} and~\ref{fig:pxpx}, where we present the difference between
the constructed and exact values of expectation values of $g_2(0)$ and $\p_x
\Psi^\dagger(0)\p_x \Psi(0)$, respectively. We see that the former operator,
$g_2(0)$, is not quite so well converged as $\p_x \Psi^\dagger(0)\p_x\Psi(0)$.
This makes some sense: we construct the state to ensure the energy is well
converged, and in the large $c$ limit of the Lieb-Liniger model it is the
kinetic energy that dominates the interaction energy (this is particularly
apparent in the $c=\infty$ limit, where the model maps to non-interacting
fermions). Nonetheless, we do see that we are correctly capturing expectation
values of local operators within the constructed states and, when plotted as
function of lowest included weight, the convergence to the exact value seems
reasonable.    

%%%%%%%%%%%%%%%%%%%%%%%%%%% 
\section{Non-equilibrium dynamics from the high overlap states truncation scheme}
\label{Sec:RealTime}
%%%%%%%%%%%%%%%%%%%%%%%%%%%

Having developed the high overlap states truncation scheme, we have so far used
it to construct an initial state (motivated by non-equilibrium dynamics) and we
have studied the properties of this approximate state. In this section we turn
our attention to computing the non-equilibrium dynamics following the $c=c_i \to
c_f$ sudden quantum quench. The time evolved state can easily be obtained from
Eq.~\eqref{psit}, truncated to include $N_\text{tot}$ terms via the high overlap
states truncation scheme, as in Eq.~\eqref{ideal}. We use such a representation
to first examine the time evolved wave function via the return amplitude and the
fidelity, before turning our attention to the time evolution of local
observables. 

%%%%%%%%%
\subsection{The return amplitude and the fidelity}
%%%%%%%%%

\begin{figure}[t]
  \begin{center}
      \includegraphics[width=0.48\textwidth]{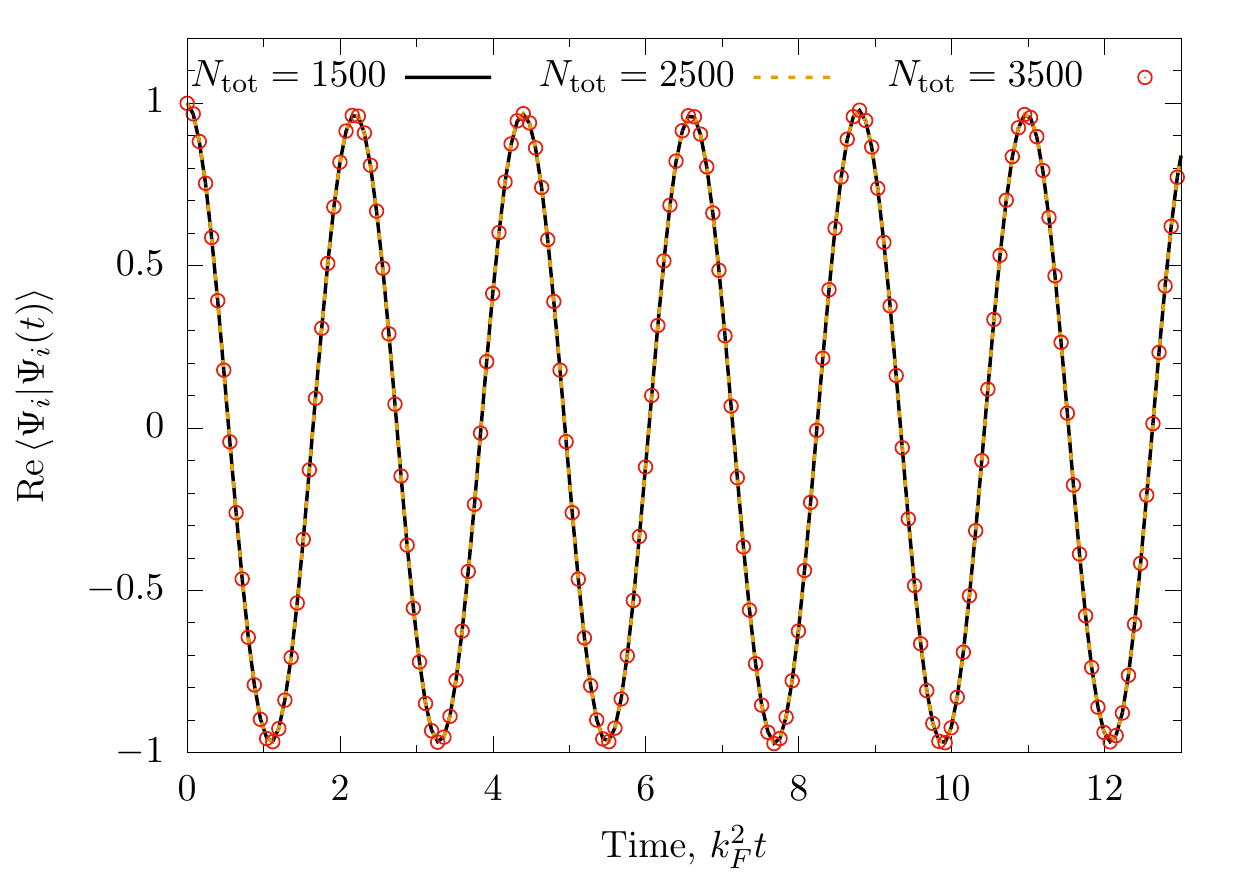}
      \includegraphics[width=0.48\textwidth]{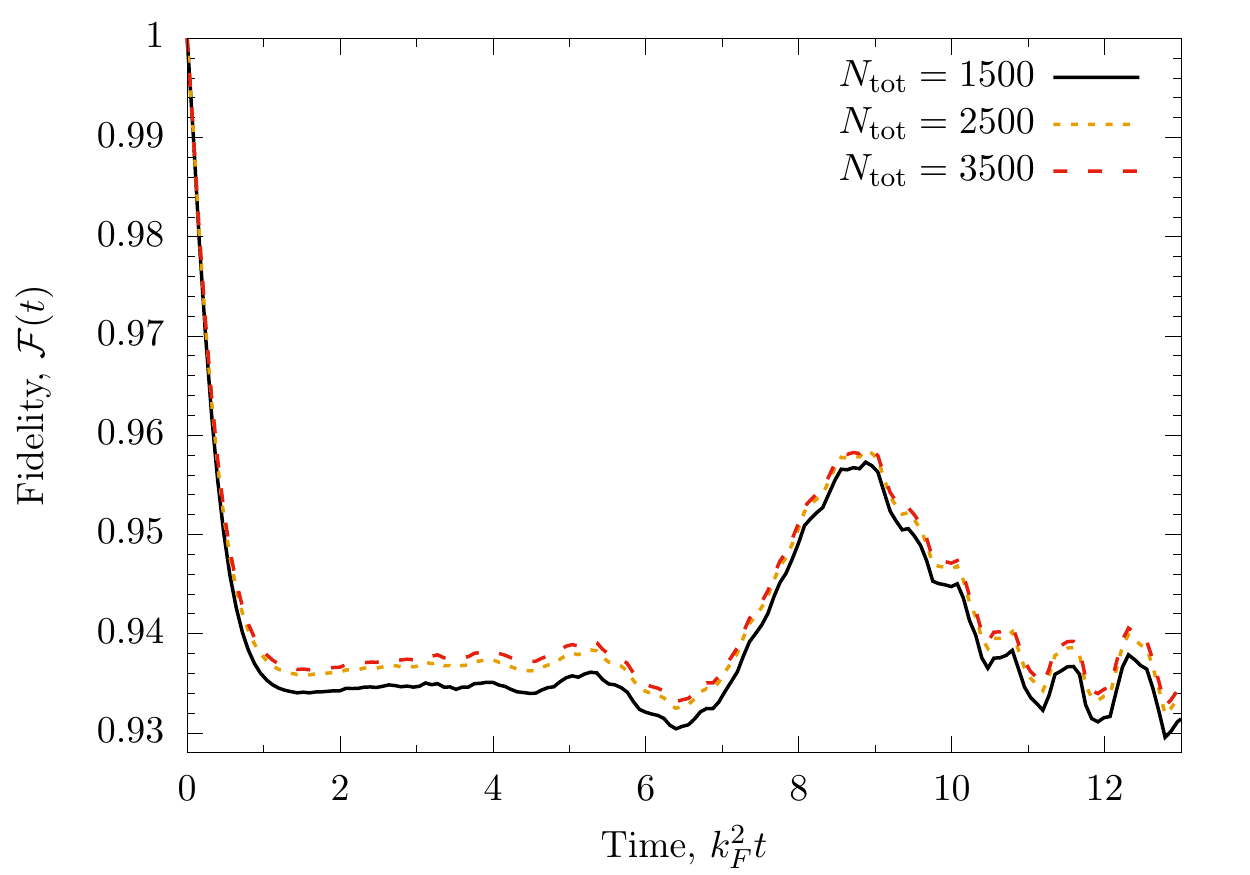}
    \end{center}
    \vspace{-5mm}
  \caption{The real part of the return amplitude~\eqref{returnAmpli} (left) and
  the fidelity~\eqref{fidelity} (right) following the quench $c_i = 20 \to c_f =
10$ in the Lieb-Liniger model, starting from the $c_i$ ground state with $N=10$
particles at unit density, $N/R=1$. ($k_F = \pi (N-1)/R$ is the Fermi wave
vector in the $c=\infty$ limit.) In both cases, we show the time evolution for
three sizes of the truncated Hamiltonian (a matrix of size $N_{\text{tot}}\times
N_{\text{tot}}$). Both these quantities rapidly converge with the number of
states, see the convergence of the initial state energy for comparison in
Fig.~\ref{fig:abacusConv}.}
  \label{fig:returnAmpli}
\end{figure}

To begin, we consider a particularly simple quantity to evaluate: the return
amplitude
\begin{align}
  \langle \Psi_i |\Psi_i(t)\rangle \approx \sum_{n=0}^{N_{\text{tot}}} e^{-\mathrm{i}E(\{\lambda\}^{(n)}t} \left\vert \langle \{\lambda\}^{(n)} | \Psi_i \rangle \right\vert^2. \label{returnAmpli}
\end{align}
This return amplitude has received significant attention in the context of
quantum quenches, where it was realized that
\begin{align}
  f(t) = -\lim_{R\to\infty} \frac{1}{R} \log\, \langle \Psi_i | \Psi_i(t)\rangle
\end{align}
can display non-analytic behavior, related to dynamical quantum phase
transitions (see, e.g.,~\cite{heyl2018dynamical} for a review
and~\cite{jurcevic2017direct} for an example experiment). The absolute value
squared of the return amplitude,
\begin{align}
  {\cal F}(t) = |\langle\Psi_i|\Psi_i(t)\rangle|^2, \label{fidelity}
\end{align}
is known as the fidelity.

Here the return amplitude and the fidelity will serve as useful test-beds for
understanding the effect truncation of the Hilbert space has on non-equilibrium
quantities. This may, in principle, be quite different to the behavior shown in
the convergence of the initial state energy studied above. This is because the
energies of the eigenstates $|\{\lambda\}^{(n)}\rangle$ entering
Eq.~\eqref{ideal} are \textit{unbounded}, while each term appearing within the
return amplitude (and the fidelity) is bounded. Indeed, we see precisely this
difference in Fig.~\ref{fig:returnAmpli}, where we show the time evolution of
the return amplitude and the fidelity at short times. For small numbers of
states in the truncated Hilbert space, $N_{\text{tot}}$, both of these
quantities are well converged, unlike the initial state energy for the same
number of states, see Fig.~\ref{fig:abacusConv}. This is particularly
convenient, as we can achieve excellent convergence of time evolved physical
quantities for (very) small number numbers of states.

We note that bench marking convergence of time evolution with the return
amplitude, or the fidelity, is also convenient as it involves evaluating only a
single sum over the final eigenstates. It can, thus, be evaluated very
efficiently even if one requires $N_{\text{tot}}$ large. In the next subsection,
we consider time evolution of local observables, which requires evaluating a
double sum over the truncated Hilbert space. 

%%%%%%%%%
\subsection{Time evolution of local observables}
%%%%%%%%%

\begin{figure}[t]
  \begin{center}
    \includegraphics[width=0.7\textwidth]{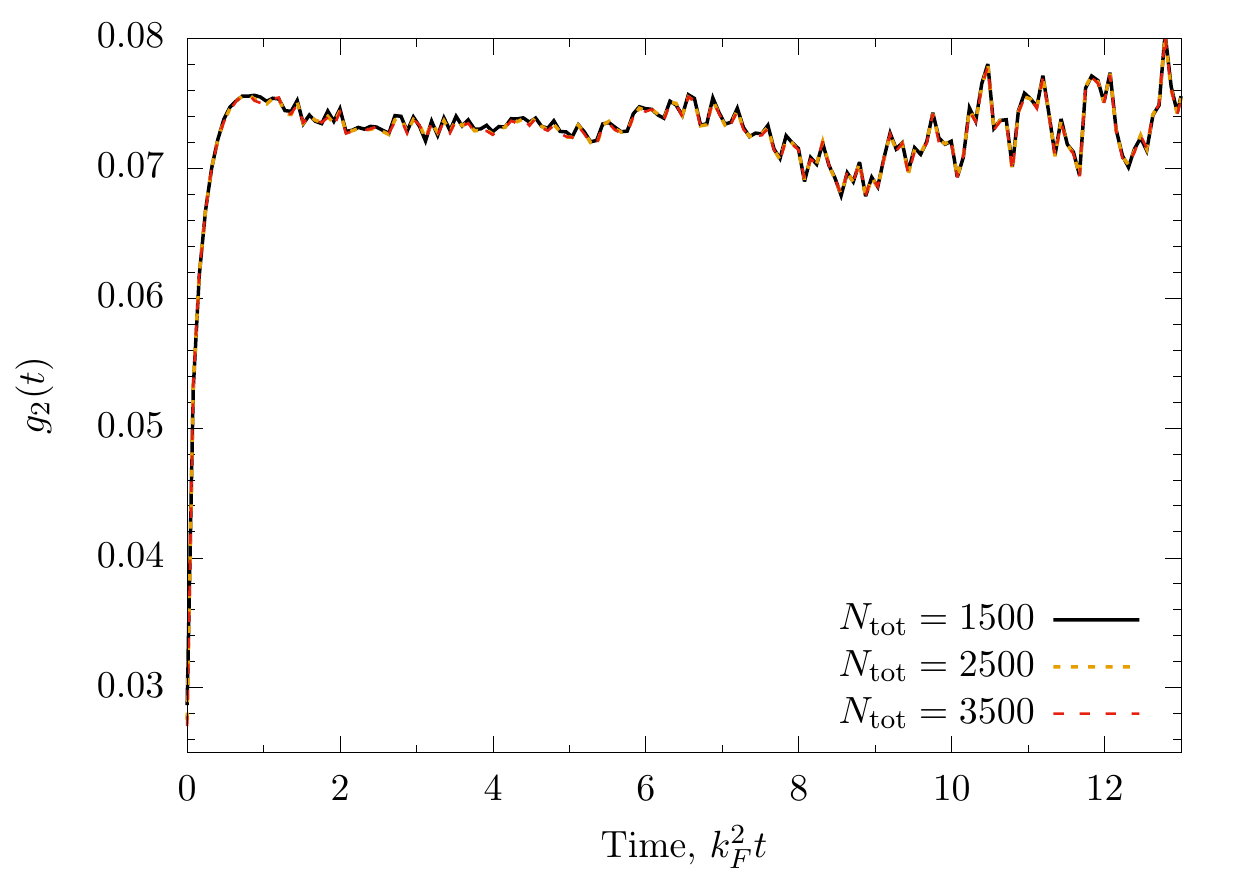}
  \end{center}
  \vspace{-5mm}
  \caption{The time evolution of the local observable $g_2 = \left(
  \Psi^\dagger(0)\right)^2 \left( \Psi(0) \right)^2$ following a quench $c_i=20
\to c_f=10$ in the Lieb-Liniger model, starting from the ground state at $c_i$
with $N=10$ particles at unit density, $N/R=1$. ($k_F = \pi (N-1)/R$ is the
Fermi wave vector at $c=\infty$.) Results are shown for a number of different
truncated basis sizes, $N_{\text{tot}}$, which illustrates the excellent
convergence of $g_2(t)$ for small numbers of states.}
  \label{fig:g2t}
\end{figure}

Having examined the return amplitude, which depends solely on the time evolved
wave function, we turn our attention to the non-equilibrium behavior of local
observables $O$. These are computed by evaluating the double sum over the
truncated Hilbert space
\begin{align}
\la O(t)\ra_i &\equiv \la \Psi_i(t) | O |\Psi_i(t)\ra, \nonumber\\
& = \sum_{n,m=0}^{N_\text{tot}} e^{-\mathrm{i}t[E(\{\lambda\}^{(n)}) - E(\{\lambda\}^{(m)})]} \la \Psi_i | \{ \lambda \}^{(m)} \ra \la  \{ \lambda \}^{(m)}  | O |\{\lambda\}^{(n)} \ra \la \{\lambda\}^{(n)} | \Psi_i\ra. 
\label{overlaptevo}
\end{align}
Clearly for any observable of interest $O$ we require knowledge of the matrix
elements between the Bethe eigenstates, $ \la  \{ \mu \}^{(m)}  | O
|\{\lambda\}^{(n)} \ra$.

Having understood how the truncation of the wave function affects the return
amplitude, we check whether the same excellent convergence occurs in the
time evolution of local quantities. We focus on the operator $O = g_2(0)$, whose
matrix elements are given in Appendix~\ref{sec:ME}. Its time evolution, $g_2(t)
= \langle \Psi(t) | g_2(0) | \Psi(t)\rangle$, is shown in Fig.~\ref{fig:g2t} for
the same quench as previously. We observe convergence properties similar to the
return amplitude and the fidelity, see Fig.~\ref{fig:returnAmpli}, i.e.
excellent convergence for small numbers of states within the truncated Hilbert
space.

The results of this subsection, taken with those of the previous one, are
strongly suggestive that we will be able to efficiently generate the
time evolution of observables for relatively large numbers of particles with
modest computational resources. We show an example of this in
Sec.~\ref{Sec:MoreParticles}.

\subsubsection{Comparison to the coordinate Bethe ansatz}

\begin{figure}[t]
  \begin{center}
    \includegraphics[width=0.7\textwidth]{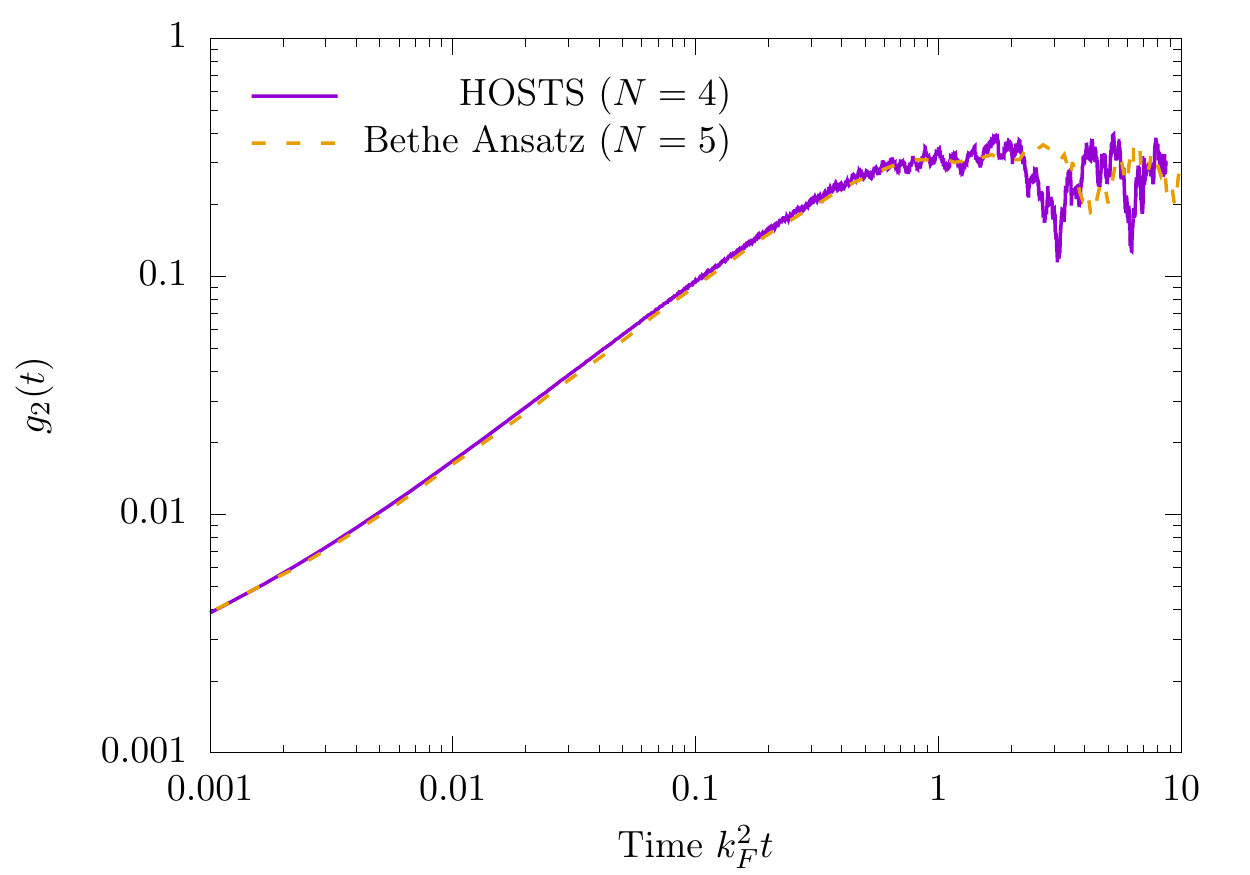}
  \end{center}
  \vspace{-5mm}
  \caption{The time evolution of the local observable $g_2(0)$ following the
  $c_i=100 \to c_f = 3.7660$ in the Lieb-Liniger model starting from the ground
state at $c_i$ with unit density. Exact data (dashed line) computed via the
coordinate Bethe ansatz with $N=5$ particles (from
Ref.~\cite{zill2016coordinate}) is compared to high overlap states truncation
scheme (HOSTS) computations with $N=4$ particles. HOSTS results for higher
numbers of particles are discussed in Sec.~\ref{Sec:MERG}.}
  \label{fig:coordBA}
\end{figure}

As a check of our results, we turn attention to results within the literature.
In particular, in Ref.~\cite{zill2016coordinate} a quench of the interaction
parameter $c_i = 100 \to c_f = 3.7660$ was considered via the coordinate Bethe
ansatz. This is a large, challenging quench where the interaction parameter
changes drastically. The energy density between the initial $c_i$ ground state
and the final ground state $c_f$ is significant and presumably many excitations
are generated in the quench. This is a challenging scenario for any numerical
approach, and it will allow us to assess the precision of our results. (We note
that calculations of non-equilibrium time evolution at finite $c_i$ and $c_f$
are very limited in the literature.)

The coordinate Bethe ansatz calculations of Ref.~\cite{zill2016coordinate} are
computational intensive, and scale very poorly with particle number. Data is
limited to cases with just small numbers of particles, $N=5$ in the case at
hand~\cite{zill2016coordinate}. Our algorithm is currently limited to even
numbers of particles, to avoid technical issues in dealing with coinciding
rapidities that often occur with $N$ odd. Thus we will compare the $N=5$ results
of Ref.~\cite{zill2016coordinate} to $N=4$ data obtained within our high overlap
states truncation scheme (we will discuss $N=6$ later). 

Our comparison to results of the coordinate Bethe ansatz is shown in
Fig.~\ref{fig:coordBA}, where the data from Fig.~4 of~\cite{zill2016coordinate}
was extracted directly from the image. For $N=4$ particles, our data is
generated with $N_{\text{tot}} = 5000$ states via full diagonalization (i.e. we
do not need to use the numerical renormalization group). We see excellent
agreement up to the finite size revival time. This is promising, as the
computational effort within our scheme is rather modest in such a scenario.
These results further confirm that the high overlap states truncation scheme is
correctly capturing all of the physics within the problem.

\subsubsection{The long time limit: The diagonal ensemble}
\label{Sec:DE}

\begin{figure}[t]
  \begin{center}
    \includegraphics[width=0.48\textwidth]{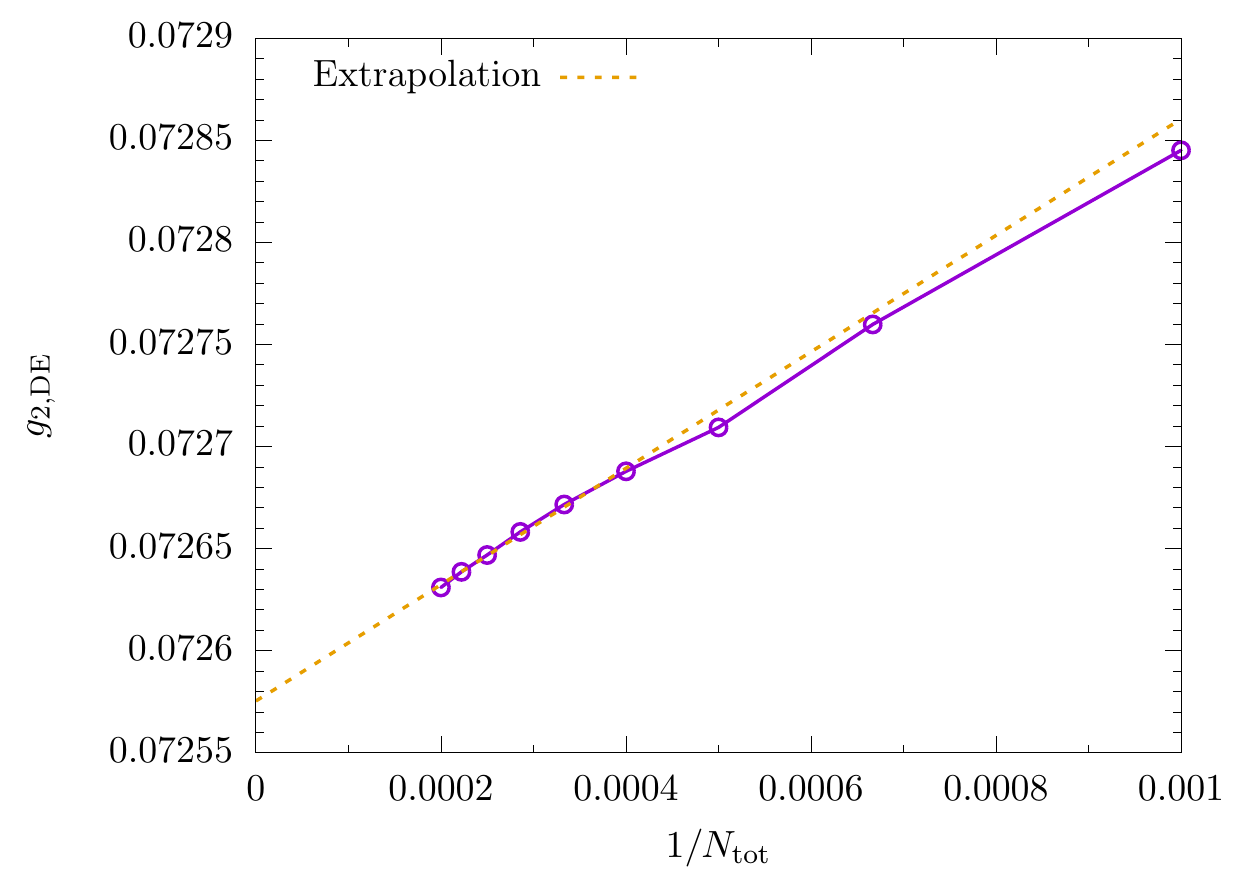}
    \includegraphics[width=0.48\textwidth]{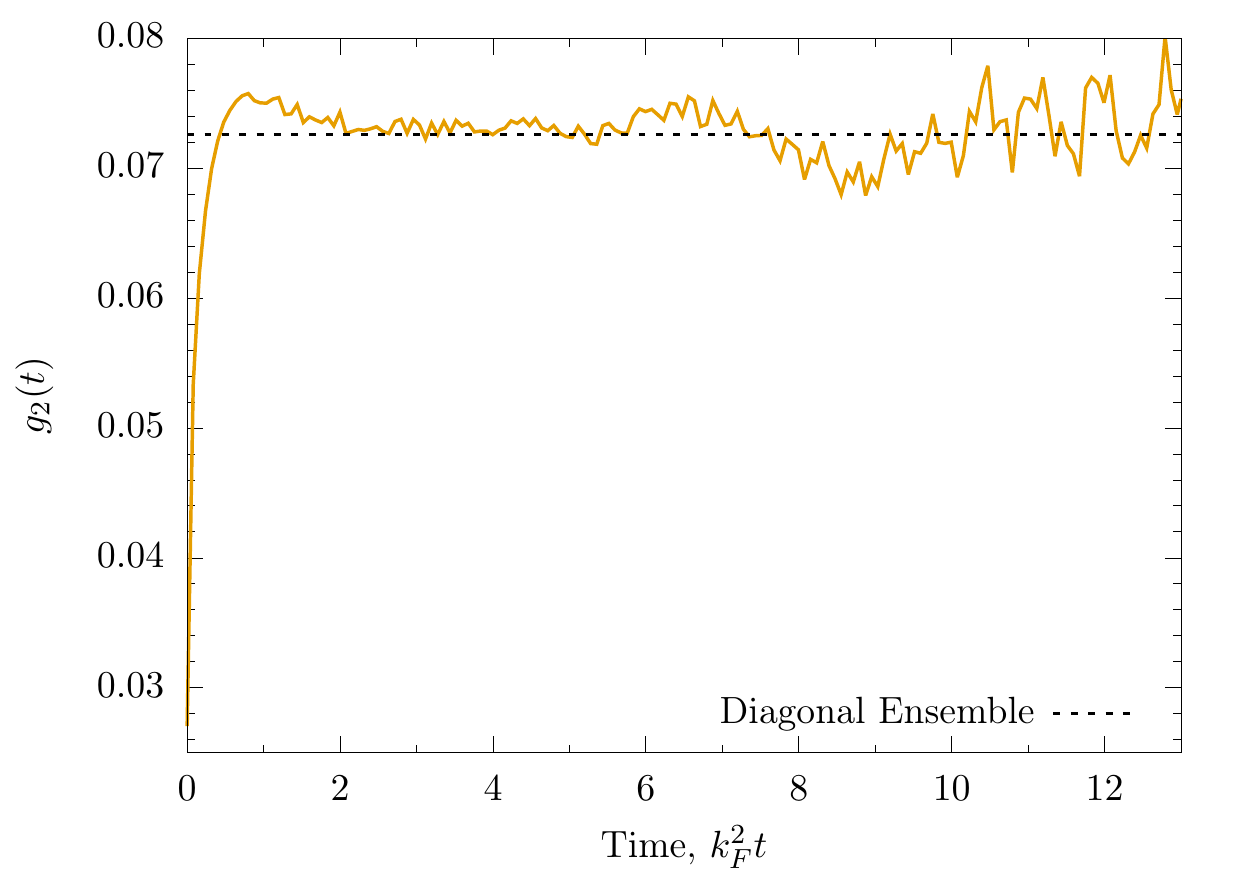}
  \end{center}
  \vspace{-5mm}
  \caption{The scaling with $N_{\text{tot}}$ of the diagonal ensemble result
  (left panel) for $c_i =20 \to c_f = 10$ quench with $N=10$ particles. The
dotted line shows a linear extrapolation to $N_{\text{tot}} = \infty$. A
comparison of the $N_{\text{tot}} = 3500$ data of Fig.~\ref{fig:g2t} with the
extrapolated diagonal ensemble result for the long time limit (right panel).}
  \label{fig:DE}
\end{figure}

Beyond accessing finite time dynamics of observables, we can also access the
long-time limit, $t\to\infty$. Here a number of simplifications occur, as
detailed in many works (see, e.g.,~\cite{rigol2008thermalization}); for example,
with the overlaps at hand, we can compute the diagonal ensemble result for the
long-time limit~\cite{rigol2008thermalization}
\begin{align}
\la O \ra_\text{DE} = \sum_{j} \la \{\lambda\}^{(j)} \vert O \vert \{ \lambda\}^{(j)} \ra  \,\Big\vert  \la \{\lambda\}^{(j)} | \Psi_i\ra \Big\vert^2. 
\end{align}
This describes the infinite time limit of the time-averaged observables in the
large system size limit (see, e.g., the discussion in the appendix
of~\cite{zill2016coordinate})
\begin{align}
\lim_{T\to\infty} \frac{1}{T} \int_0^T \rd t \la \Psi_i(t)|O|\Psi_i(t)\ra \to \la O \ra_\text{DE}.
\end{align}
If the observable $O$ relaxes to a stationary value in a sufficiently fast
manner then that long time limit of the expectation value (without time
averaging) will be reproduced
\begin{align}
\lim_{t\to\infty} \la \Psi(t) | O | \Psi(t)\ra \to \la O \ra_\text{DE}. 
\end{align}
see, for example, Ref.~\cite{piroli2017correlations}. As such, the diagonal
ensemble can be a computationally convenient way to access the infinite time
limit if overlaps are known.

We illustrate that our truncation scheme can evaluate the diagonal ensemble in
Figs.~\ref{fig:DE}. We first study the convergence of the diagonal ensemble
result as a function of the truncated Hilbert space dimension (left panel),
before comparing the extrapolated $N_{\text{tot}}\to\infty$ result to the
real-time dynamics shown in Fig.~\ref{fig:g2t} (right panel). We see
well-behaved and rapid convergence of the diagonal ensemble value to its
``non-truncated'' limit, and that it captures well the values to which local
observables relax at intermediate times. We note that fluctuations about the
diagonal ensemble value within the real time dynamics are large for the system
sizes considered.

We note that expectation values of local operators, long after a quench in an
integrable model, are expected to be described by a generalized Gibbs
ensemble~\cite{vidmar2016generalized}. In the case of the Lieb-Liniger model,
one can run into issues constructing this ensemble because of the asymptotic
behavior (in rapidity space) of the steady state root distribution, which leads
to diverging expectation values of ultra-local charges. This can, in principle,
be remedied by working with a different set of charges (see, e.g.,
Ref.~\cite{palmai2017quasilocal}), but we do not pursue that approach here.

\subsubsection{An example with larger numbers of particles}
\label{Sec:MoreParticles}

\begin{figure}[t]
  \begin{center}
    \begin{tabular}{ll}
      (a) & (b) \\
      \includegraphics[width=0.48\textwidth]{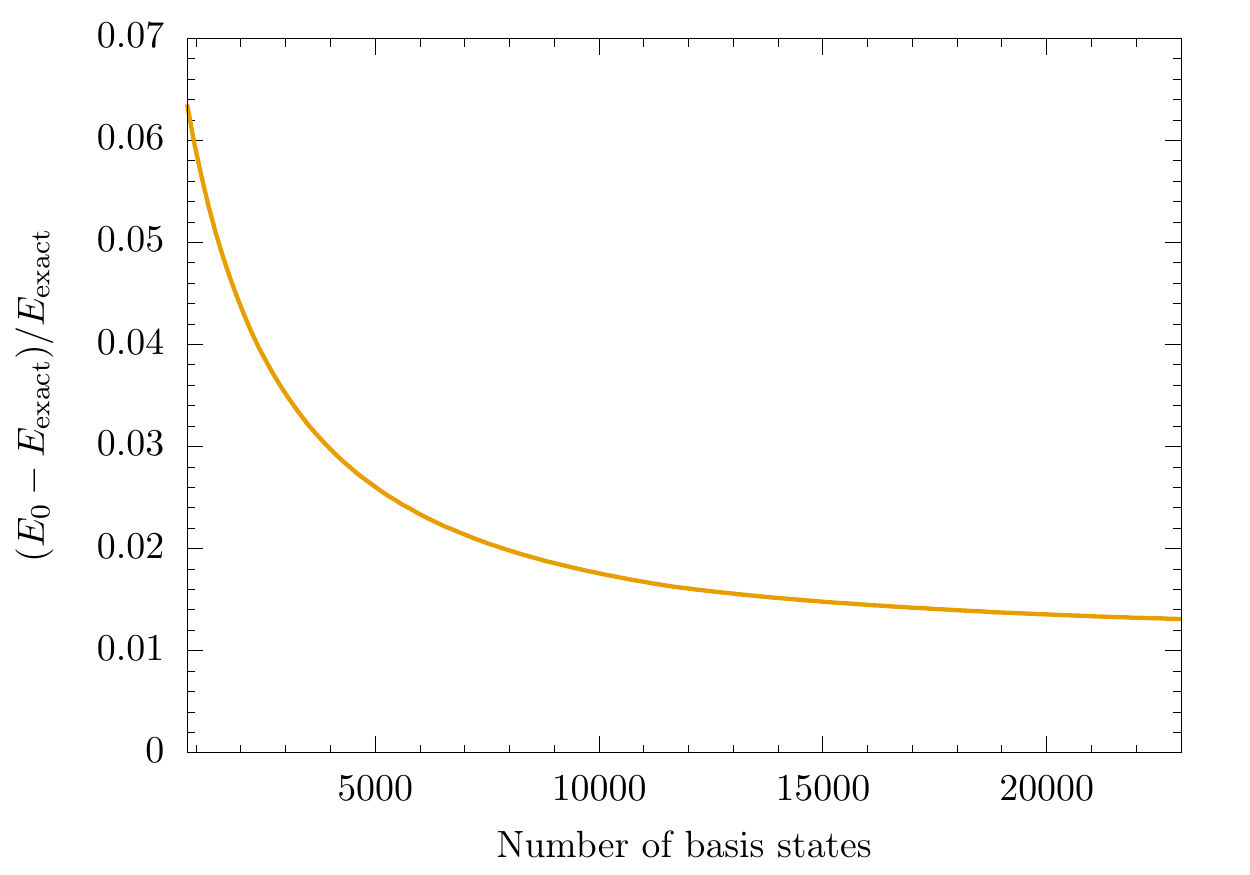}
          & \includegraphics[width=.48\textwidth]{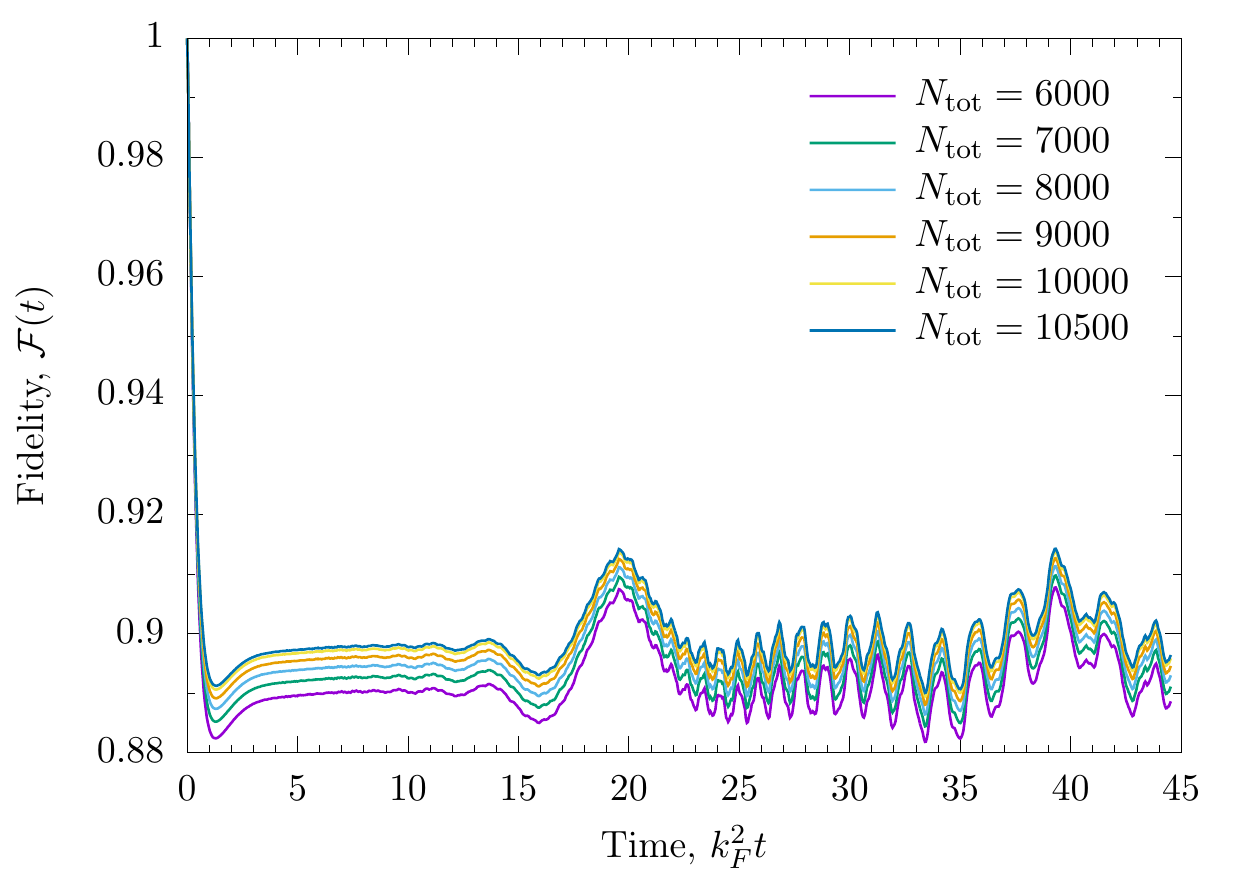} \\
    \end{tabular}
  \end{center}
  \vspace{-5mm}
  \caption{Example calculation showing (a) the energy convergence of the initial state; (b) the time evolution of the fidelity for $N=20$ particles, for the quench $c_i = 20 \to c_f = 10$.} 
  \label{fig:moreParticles}
\end{figure}

So far, we have examined quenches with relatively small numbers of particles,
$N=4,6,10$. It is worth emphasizing that even for these numbers of particles,
exact calculations via the coordinate Bethe ansatz are computationally
expensive, and exact calculations with $N=10$ corresponding to summing $\propto
(N!)^2 \sim 1.3\times10^{13}$ terms. To even contemplate exact evaluation for
$N=20$ particles, $\propto 5.9\times 10^{36}$ terms, seems futile. Instead the
high overlap states truncation scheme, in combination with the numerical
renormalization group, gives one a handle on such problems.

Here, we consider the $c_i = 20 \to c_f = 10$ quench for a larger numbers of
particles ($N=20$) as an illustrative example. We leave detailed study, both of
larger numbers of particles and different quenches (as well as expanding on
results presented above), to future works~\cite{TBP}. Results for the energy
convergence and the time evolution of the fidelity, ${\cal F}(t)$, are shown in
Fig.~\ref{fig:moreParticles}.   

%%%%%%%%%%%%%%%%%%%%%%%%%%%
\section{Introducing the Matrix Element Renormalisation Group}
\label{Sec:MERG}
%%%%%%%%%%%%%%%%%%%%%%%%%%%

In this section, we consider a perturbing operator $(c_i-c_f)R g_2(0)$ whose
matrix elements, with respect to some of the computational basis states, are
large compared to the energy difference between these states and the unperturbed
ground state. We call quenches for which the perturbing operator satisfies this
property ``strongly non-perturbative''. In such cases relying solely on the
metric~\eqref{abacusMetric} discussed in the previous section, which was
motivated by leading order perturbation theory, is no longer justified. Higher
order terms are expected to be relevant and, as a result, we need to modify the
way we select and order states. Not only this, but we need to re-examine the
assumptions behind the numerical renormalization group procedure discussed in
Sec.~\ref{sec:nrg}, and modify these accordingly.  In the final part of this
section we show how the resulting procedure allows us to not only treat quenches
where the initial state is a ground state, but also those where it is an excited
state.

As matrix elements of the perturbing operator become large, contributions of a
given computational basis state to the ground state mediated via other
intermediate computational basis states can become relevant. This corresponds to
the second order terms in Eq.~\eqref{eq:perttheory} no longer being negligible
compared to the first order terms. However, such contributions are not
considered in the standard numerical group procedure as discussed in
Sec.~\ref{sec:nrg}. As a result, these contributions are missed when states are
not by chance included in the same step of the renormalization group procedure.
We will see that these contributions can play an important role for strongly
non-perturbative quenches, so they need to be taken into account.

\begin{figure}[t] \begin{center}
\includegraphics[width=0.7\textwidth]{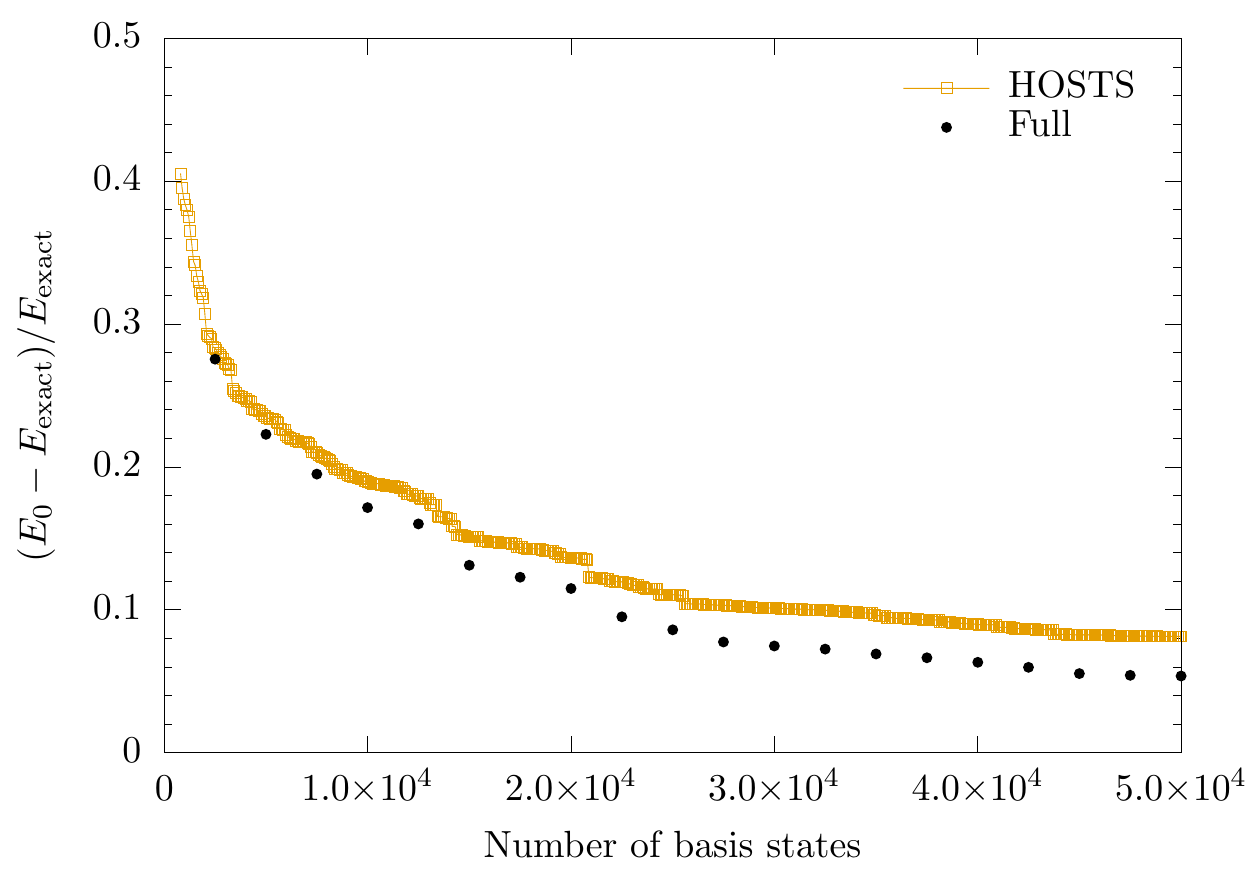} 
\end{center}
\vspace{-5mm} 
\caption{An example calculation showing that for strongly non-perturbative
  quenches there can be a large discrepancy between the results from full
  diagonalisation and the NRG-extension of HOSTS.  We consider the $c_i = 100
  \to c_f = 3.7660$ quench for $N=6$ particles, cf.  Fig.~\ref{fig:coordBA}.
  $N+\Delta N_s = 800$ is fixed within each data set.}
  \label{Fig:nrgGoneWrong} 
\end{figure}

An illlustration of how naively applying the algorithms developed thus far can
lead to inaccurate results for strongly non-perturbative quenches is shown in
Fig.~\ref{Fig:nrgGoneWrong}. The initial steps still correspond to the results
obtained from diagonalising the full truncated Hamiltonian, but as the number of
iterations increases, the discrepancy becomes larger. 
  % TODO: Check for different step sizes and say something about that!  
  % TODO: Check for more particles and see if the problem becomes larger

%%%%%%%%%%%
\subsection{The matrix element renormalization group algorithm for the ground
state}
\label{sec:MERG algorithm}
%%%%%%%%%%%

To deal with the problem for strongly non-perturbative quenches discussed in the
previous section, we develop a reworking of the numerical renormalization
group procedure that we refer to as the matrix element renormalization group.
The main differences between the matrix element renormalization group and the
conventional renormalization group algorithm are:
\begin{enumerate}
\item The approximate eigenstates obtained at each step of the algorithm (from
  diagonalization of a truncated Hamiltonian) are kept, unlike in the
  conventional case where one discards $\Delta N_s$ states at each iteration. 
\item When introducing new computational basis states, we select which of the
  previously obtained approximate eigenstates to include in the Hamiltonian
  using a weighing function based on the quadratic terms in the
  perturbation-series expansion of the wave function \eqref{eq:perttheory}
  instead of including the approximate eigenstates with the lowest energies.
\end{enumerate}

The matrix element renormalization group takes seriously the idea that matrix
elements of the perturbing operator, rather than energies, are the important
quantity when operators are not strongly renormalization group relevant. The
central idea is that computational basis states $|\{\lambda\}^{(j)}\rangle$
included at a given step can mediate strong coupling between the approximate
ground state and approximate ``excited states'' obtained at an earlier
iteration. These ``approximate states'' must then be included in the truncated
Hamiltonian at this diagonalization step to ensure an accurate description of
the ground state. So, instead of blindly removing the high energy approximate
``excited states'' at each step of algorithm (as in the conventional numerical
renormalization group), we keep all approximate eigenvectors, and
at each iteration include the states most important for mediating the coupling
between the approximate ground state and the newly added states from the
computational basis. 

Let $|\Omega \rangle$ be the ground state of the final Hamiltonian,
then the steps of the matrix element renormalization group are as follows:
\begin{enumerate}

\item Generate the computational basis via preferential state generation from
  the ground state $|\Omega \rangle$. Order the states in the computational basis
  according to the metric in Eq.~\eqref{abacusMetric}, to obtain $\{
  |\{\lambda\}^{(j)}\rangle \}$. 

\item Construct a truncated Hamiltonian from the first $N_s+\Delta N_s$
  computational basis states, $\big\{ |\{\lambda\}^{(1)}\ra, \ldots,
  |\{\lambda\}^{(N_s+\Delta N_s)}\ra\big\}$ and diagonalize this Hamiltonian to
  obtain the first approximate eigenstates $\big\{|E_1\ra, \ldots , |E_{N_s+\Delta
  N_s}\ra\big\}$ with energies $\{ E_1, \ldots, E_{N_s + \Delta N_s} \}$. These
  approximate eigenstates replace the first $N_s + \Delta N_s$ states in the
  computational basis, and are ordered such that $E_1 < \dots < E_{N_s + \Delta
  N_s}$.

\item Define a new basis of $N_s + \Delta N_s$ eigenstates for the truncated
  Hamiltonian by adding the next $\Delta N_s$ states from the computational
  basis $\big\{ |\{\lambda\}^{(j)}\rangle \big\}$ to the approximate eigenvector
  with the lowest energy $|E_1\rangle$ as well as the $N_s - 1$ approximate
  eigenstates $\{|E_i\rangle\}_{i>1}$ whose ``second order weight'', given by  
\begin{align}
	w_2\left( |E_i \ra \right) = \sum_{j} \frac{ \langle E_i | \delta H
  |\{\lambda\}^{(j)} \rangle \langle \{\lambda\}^{(j)} | \delta H | E_1 \rangle}{( E_1 - E_{\{\lambda\}}^{(j)}) (E_1 - E_i)}. 
\end{align}
  is largest. Here $\delta H = cR\times g_2(0)$ is the perturbing operator (see
  Sec.~\ref{Sec:formulation}), the sum ranges over all $\Delta N_s$ newly added
  computational basis states, and $E_{\{\lambda\}}^{(j)} = E(\{\lambda\}^{(j)})$
  are the energies of the newly added computational basis states.

\item Construct the truncated Hamiltonian in this new basis and diagonalize it
  to obtain $N_s + \Delta N_s$ new approximate eigenstates. These newly
  constructed approximate eigenstates replace the states in the
  computational basis used to construct the truncated Hamiltonian.

\item Return to the third step.  

\end{enumerate}
This process is continued, obtaining new approximate eigenstates after each
cycle of steps 3 to 5, until the required convergence of the ground state
energy/eigenstate is reached or the computational basis is exhausted. 

The matrix element renormalization group has some slight disadvantages when
compared to the conventional numerical renormalization group.  Firstly, it is
more memory intensive: a complete set of approximate eigenstates must be
retained in the procedure, while in the conventional routine we only need keep
track of $N_s$ such approximate eigenstates.\footnote{In practice, since we need
  at most $N_s$ approximate eigenvectors at any one time, these eigenvectors do
  not have to be stored in memory.} 
Secondly, the matrix element renormalization group has a higher computational
burden since it requires the computation of the second order weight for all
approximate eigenvectors at the start of each iteration. However, we have seen
that the conventional numerical renormalization group fails to produce accurate
results for strongly non-perturbative quenches (see Fig.~\ref{Fig:nrgGoneWrong})
so these savings in memory and computations compared to the matrix element
renormalization group are moot.

\begin{figure}[h!]
  \begin{center}
    \begin{tabular}{ll}
      (a) & (b) \\
      \includegraphics[width=0.48\textwidth]{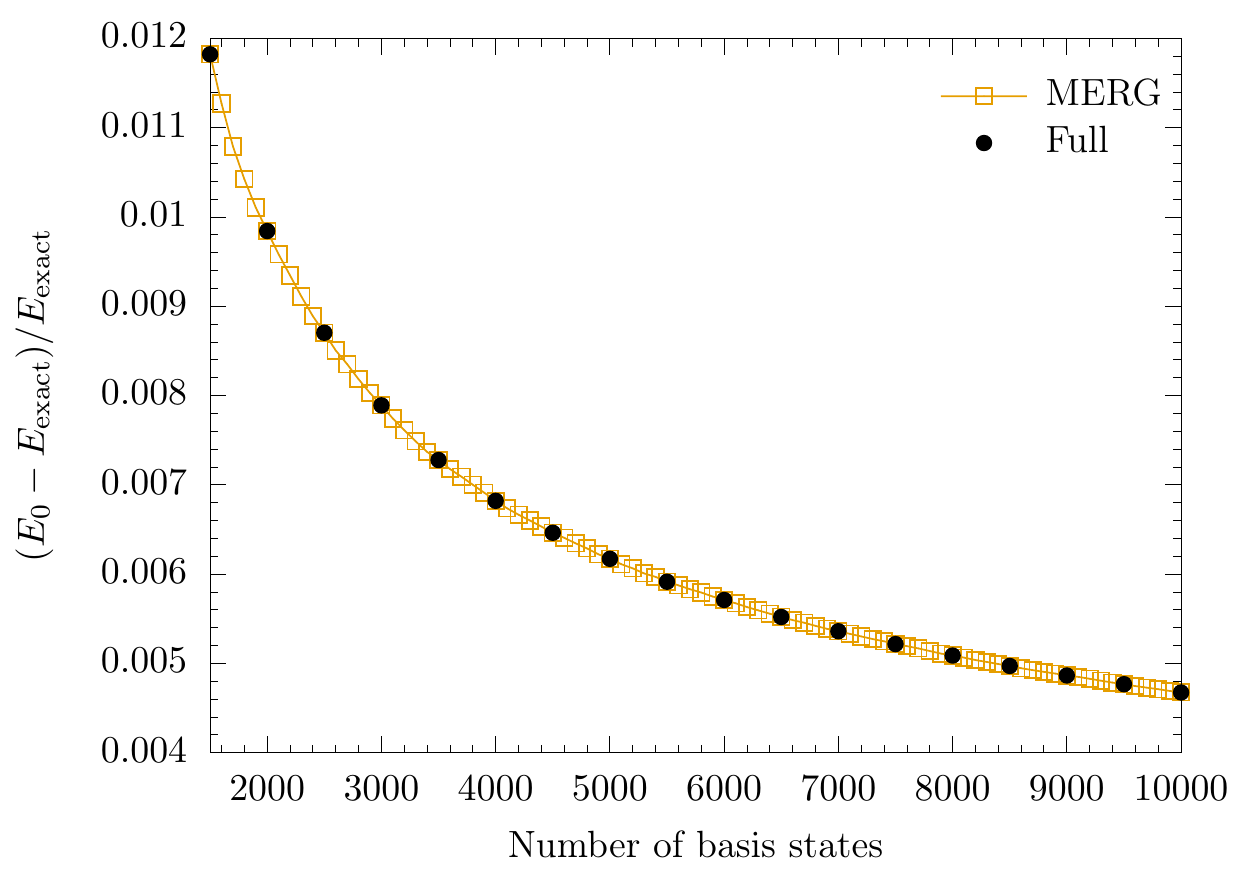}
          & \includegraphics[width=0.48\textwidth]{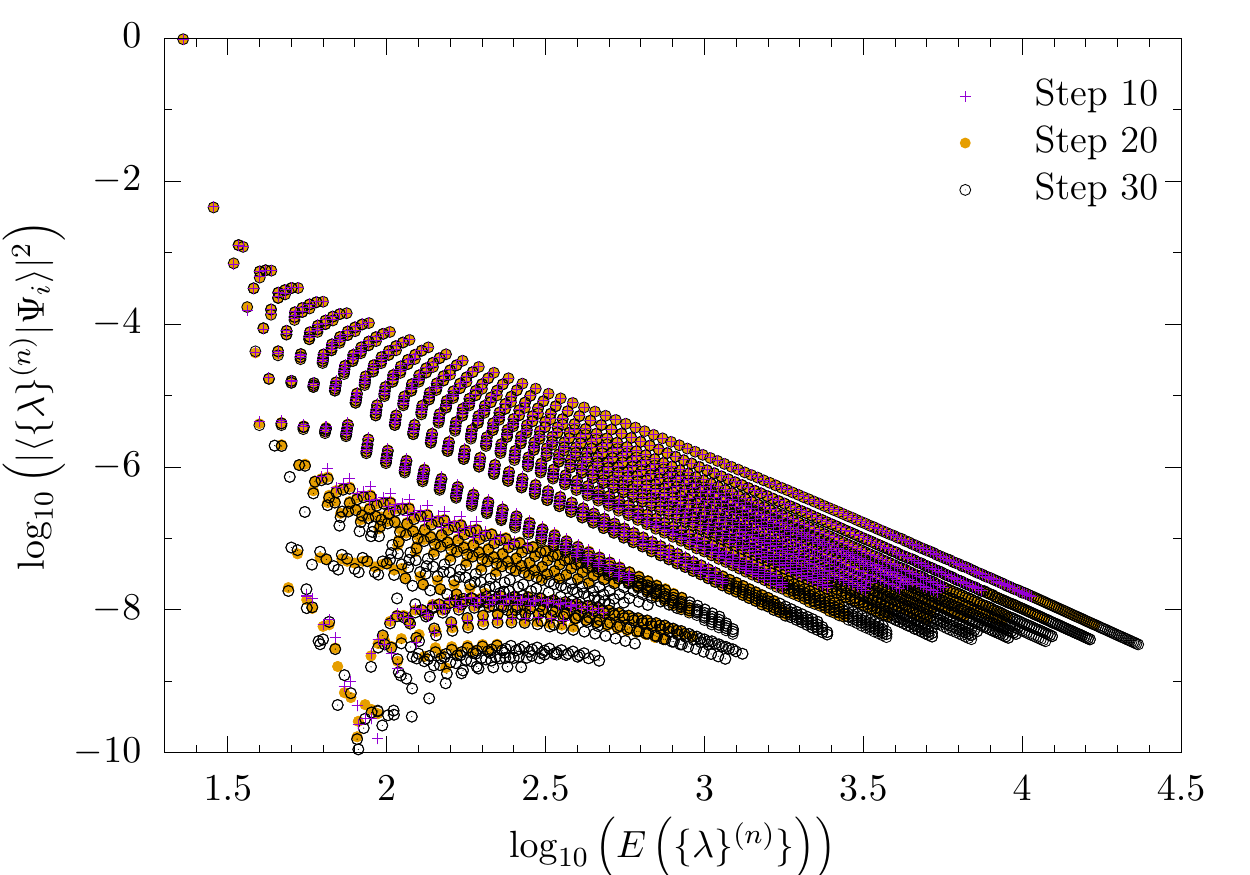} \\
      (c) & (d) \\
      \includegraphics[width=0.48\textwidth]{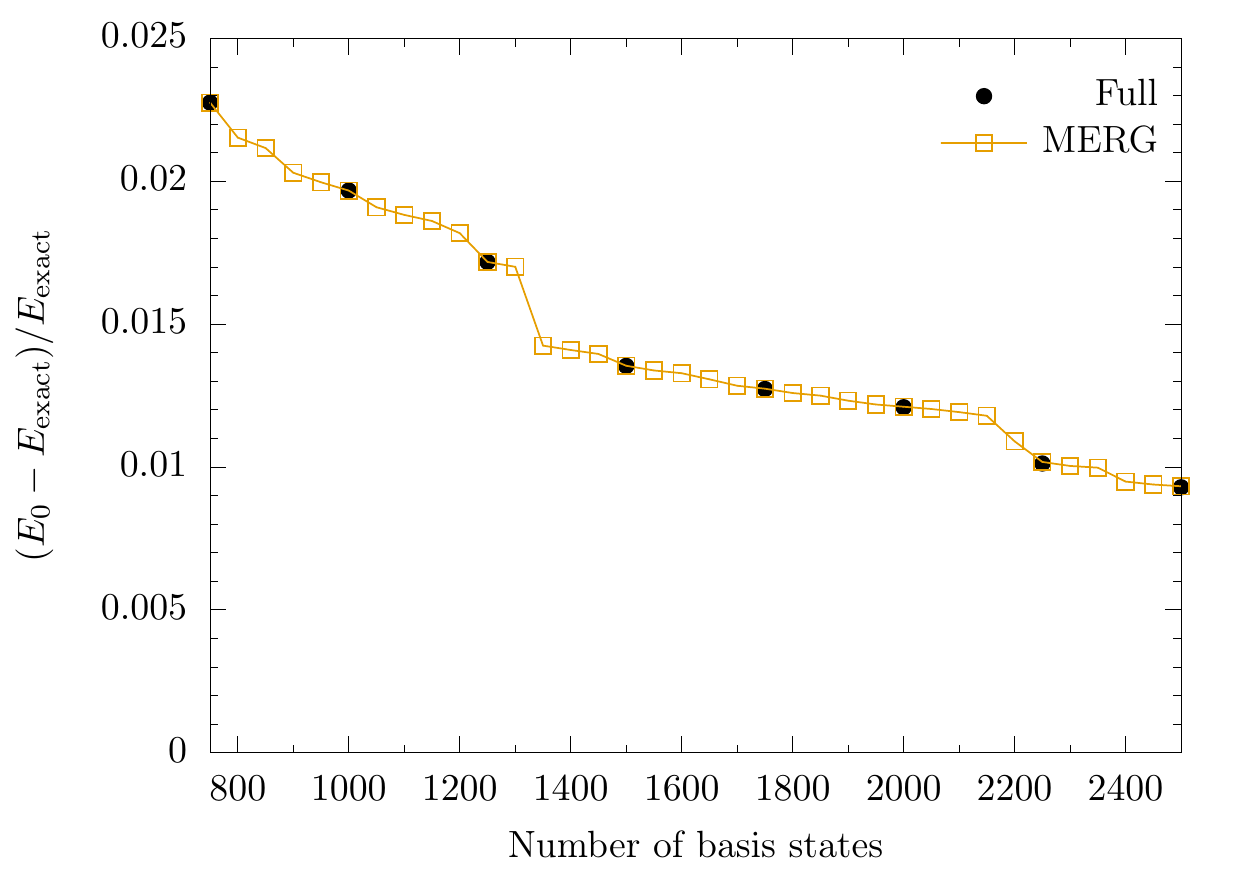}
          & \includegraphics[width=0.48\textwidth]{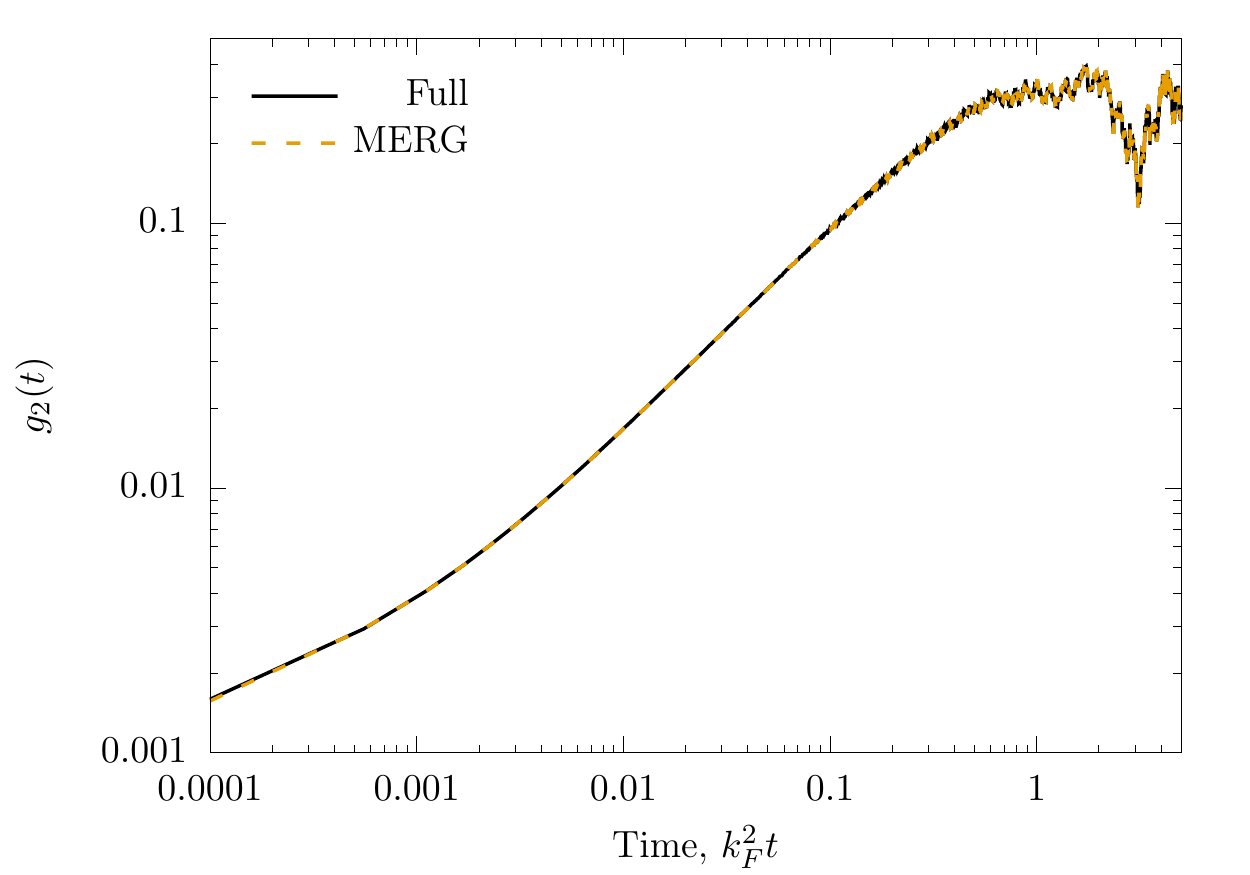} \\
    \end{tabular}
  \end{center}
  \vspace{-5mm}
  \caption{{\textit{Top row:}} Matrix element renormalization group (MERG) and
  numerical renormalization group (NRG) for the $c_i =20 \to c_f = 10$ quench
for $N=10$ particles, computed within the high overlap states truncation scheme.
(a) The convergence of the initial state energy as a function of number of basis
states; (b) the convergence of the overlaps at different steps of the MERG
procedure (cf. Fig.~\ref{fig:abacusConvWfn}). \textit{Bottom row:} MERG and full
diagonalization results for the $N=4$ particle quench $c_i = 100 \to c_f =
3.7660$: (c) the convergence of the energy of the initial state with the number
of basis states; (d) the time evolution of $g_2(t)$ with $2500$ basis states.} 
  \label{fig:mergTest}
\end{figure}

There are a couple of alternative, complementary, schemes that could be used to
construct the initial state. Firstly, there exist ``sweeping'' improvements of
the conventional numerical renormalization group (see, e.g., their discussion
in~\cite{james2017nonperturbative}). If succesful however, this additional would
certainly come at a higher computational cost than directly using the matrix
element renormalization group. Secondly, one could invoke iterative
diagonalization (via, e.g., Lanczos or Davidson) within a given truncated basis.
In such a procedure, one would have to check convergence of results with basis
size, but one can (in principle) deal with very large bases.  How quickly such
iterative diagonalization converges, with our matrix being dense, is not clear.
We have yet to explore this avenue, but it is an interesting direction for
future works. 

%%%%%%%%%%%%
\subsection{Results from the matrix element renormalization group for the ground
state}
\label{Results from the matrix element renormalization group for the ground
state.}
%%%%%%%%%%%%

With the matrix element renormalization group algorithm in place, we can employ
it to tackle problems that are inaccessible to the conventional numerical
renormalization group.

However, in the first case, we check that the matrix element renormalization
group correctly reproduces results in cases where the conventional numerical
renormalization group approach works. This is a basis sanity check: can we
reproduce the initial state and its dynamics in these simpler cases. Our first
example is $c_i = 20 \to c_f = 10$ quench studied earlier in this work. We
present the convergence of the energy and the overlaps in
Figs.~\ref{fig:mergTest}(a)--(b). In particular, Fig.~\ref{fig:mergTest}(b)
should be compared to Fig.~\ref{fig:abacusConvWfn} obtained previously. We see
excellent agreement between the conventional numerical renormalization group and
the matrix element renormalization group in this scenario.

As a second check, we turn our attention to the harder quench considered in the
previous section for $N=4$ particles, $c_i = 100 \to c_f = 3.7660$. Here we
check against full diagonalization of the truncated Hamiltonian (as the required
number of states for excellent convergence is rather small), as shown in
Figs.~\ref{fig:mergTest}(c)--(d). The matrix element renormalization group gives
results in excellent agreement with full diagonalization of the same basis, both
in terms of energy of the initial state, Fig.~\ref{fig:mergTest}(c), and the
non-equilibrium dynamics of observables, Fig.~\ref{fig:mergTest}(d). 

\begin{figure}[t]
  \begin{center}
      \includegraphics[width=0.7\textwidth]{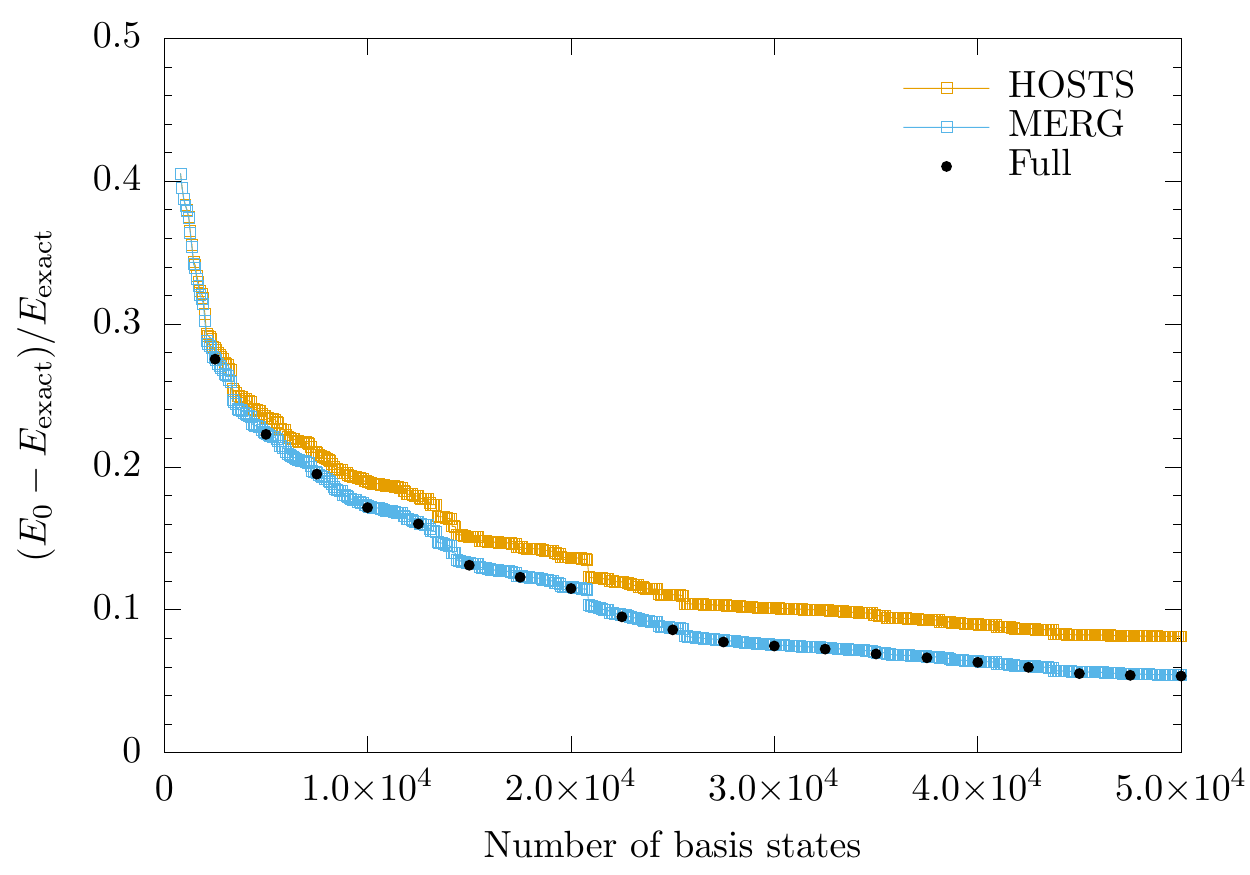}
  \end{center}
  \caption{ The convergence of the initial state energy $E_0$ for $N=6$ particle
  quench $c_i = 100 \to c_f = 3.7660$ obtained with the matrix element
renormalization group (MERG). MERG is performed with $N_s = 720$ and $\Delta N_s
= 80$. Conventional numerical renormalization group approaches breakdown in this
scenario (for the same $N_s,\Delta N_s$) as shown in
Fig.~\ref{Fig:nrgGoneWrong}. Full diagonalization results, for the same number
of basis stats, are shown for comparison.}
  \label{fig:mergSavesTheDay}
\end{figure}

With the matrix element renormalization group correctly reproducing both full
diagonalization (in small bases) and conventional numerical renormalization
group (in large bases) results, we examine the problematic scenario discussed in
the previous section. In this strongly non-perturbative quench, the matrix
element renormalization group is vital for correctly constructing the initial
state. In scenarios where the conventional numerical renormalization group fails
to produce results which agree with results obtained by diagonalization of the
full truncated Hamiltonian, such as the one illustrated in
Fig.~\ref{Fig:nrgGoneWrong}, the matrix element renormalization group continues
to produce results that agree with great accuracy as is shown in
Fig.~\ref{fig:mergSavesTheDay}.

\begin{figure}
  \begin{center}
    \includegraphics[width=0.7\textwidth]{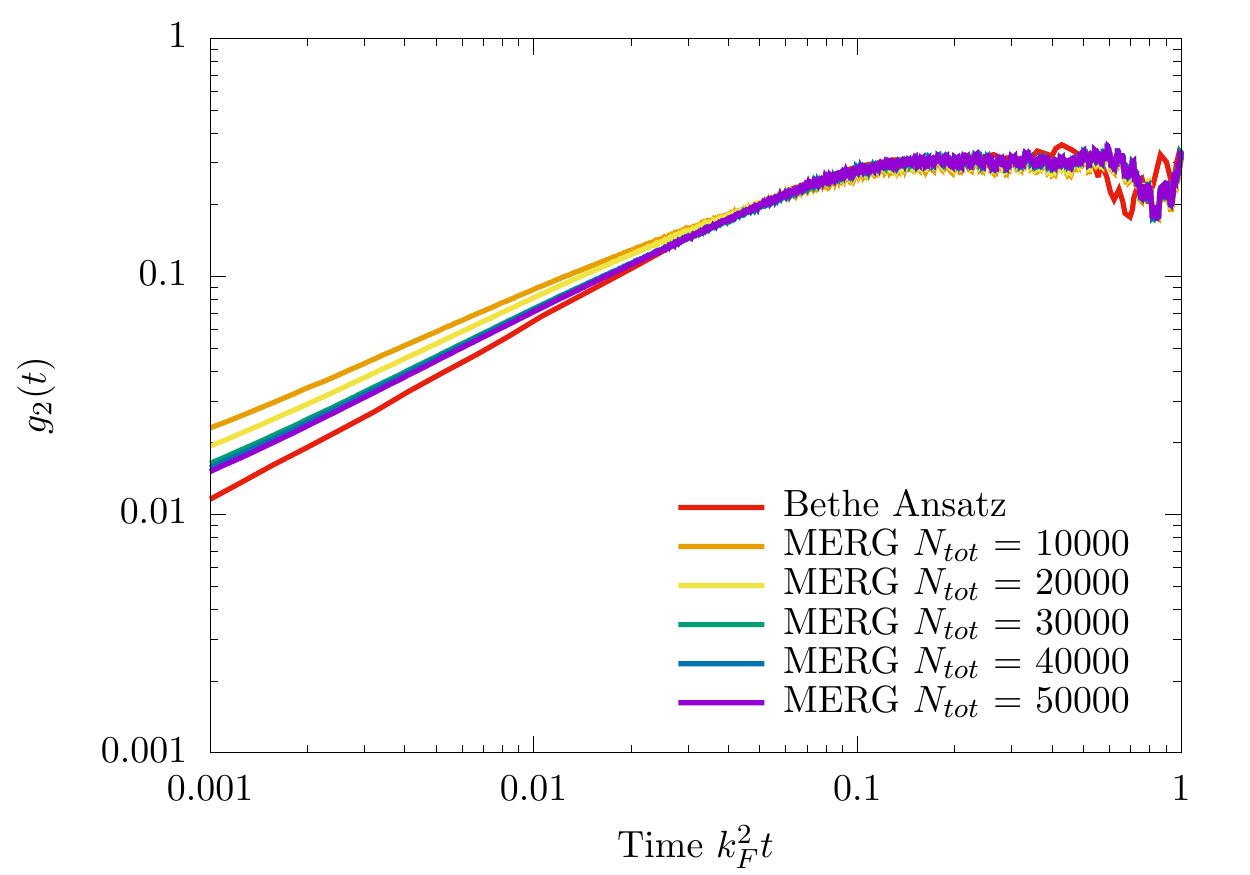}
  \end{center}
  \caption{The time evolution of the local observable $g_2(0)$ following the
    $c_i=100 \to c_f = 3.7660$ in the Lieb-Liniger model starting from the
    ground state at $c_i$. Exact data (dashed line) computed via the coordinate
    Bethe ansatz with $N=5$ particles (from Ref.~\cite{zill2016coordinate}) is
    compared to matrix element renormalization group (MERG) calculations with
  $N=6$ particles.}
  \label{fig:coordBA2}
\end{figure}

In Fig.~\ref{fig:mergSavesTheDay} we see a number of features.  Firstly, we note
that the agreement between the results obtained by full diagonalisation of the
truncated Hamiltonian and using the matrix element renormalization group are in
excellent agreement.  Secondly, regardless of the procedure, the convergence of
the initial state energy shows some plateaus and jumps, which implies that the
metric~\eqref{abacusMetric} is not the perfect one.  Understanding how to
construct the most convergent metric for a given problem is an outstanding
challenge, which requires further investigations. Thirdly, we see that for $N=6$
particles the problem is very challenging: By including $50,000$ states, we
still only achieve initial state energies correct to within $\sim 5.5\%$ ($\sim
10\%$ w.r.t. the Fermi energy). Whilst
a better ordering metric might improve this, it still seems likely that strongly
nonperturbative quenches will present a significant numerical challenge. This is
further supported by Fig.~\ref{fig:coordBA2}, where we show the time evolution
of $g_2(t)$ as compared to results from the coordinate Bethe ansatz discussed
previously. We see that even a truncated wave function~\eqref{ideal} with 50,000
states included does not accurately realize the short time dynamics of
observables. At longer times, once the steady state plateau is approached (and
high energy modes have dephased, effectively averaging to zero), the truncated
wave function does describe $g_2(t)$ well. 

%%%%%%%%%%%%
\subsection{The matrix element renormalization group algorithm for excited states}
\label{Sec:Constructing excited states}
%%%%%%%%%%%%

In contrast to conventional renormalization group techniques, which at most
target the first few excited states in addition to the ground state, the matrix
element renomalization group can also be used to target more highly excited
states. Furthermore, the algorithm can construct these states without the need
to construct all the states of lower energy making the procedure a more
efficient tool to target excited states than the methods discussed thus far. In
order to do so we have to make some changes to the algorithm described in
Sec.~\ref{sec:MERG algorithm}.

% 2, problems with replacing GS with excited state
To understand why we need to change the algorithm in Sec.~\ref{sec:MERG
algorithm} in order to consider excited states, let us consider what happens if
we replace the ground state with an excited state in the algorithm. This state
will henceforth be referred to as the seed state.  First of all, the
preferential state generation routine leads to a different computational basis,
as we now consider $|\{\lambda\}^{(0)}\rangle$ in Eq.~\eqref{abacusMetric} to be
an excited state. In particular, the ground state may not have a high weight
according to this metric, so it may not even be included in the computational
basis obtained via preferential state generation.  Second of all, the algorithm
retains the lowest energy approximate eigenstate $|E_1\rangle$ at every step,
and selects the approximate eigenstates most relevant to this approximate
eigenstate based on the second order weight.  This means that the selection
rules are still set to promote the convergence of the lowest energy eigenstate,
rather than an excited state.  Finally, it is generally unclear which eigenstate
of the perturbed Hamiltonian corresponds to which of the approximate eigenstates
obtained by the algorithm, as energies of different excited states can be very
similar and even degenerate. 

% 3, tracking the seed state
Before we discuss how we resolve these issues, note that every step of the
algorithm represents a mapping between the states used to construct the
truncated Hamiltonian and the approximate eigenstates obtained by
diagonalization. To identify which of the approximate eigenstates a given state
used in the basis for the truncated Hamiltonian is mapped to, we compute the
overlaps between this state and all newly obtained approximate eigenstates. The
approximate eigenstate with the largest overlap is then said to be its image
provided that the RG-step is ``small enough''. This allows us to track the
approximate eigenstate derived from the seed state throughout the procedure. 

% 4, how one can do this if one makes a simple assumption
The main assumption behind our method of tracking the seed state is that the
number of states added at every step of the routine is small enough, so that no
single iteration wildly changes the (image of) the seed state. What step size is
small enough for this assumption to hold depends on the quench and seed state
under consideration.\footnote{It may happen that no step size is small enough
when we consider a very strong quench and/or a highly excited state.} However,
there are some general methods by which one can check if an appropriate step
size has been chosen. Firstly, one can consider the overlaps computed at each
iteration of the routine and verify that there is only one state with a
significant overlap.  Secondly, one can rerun the routine with a smaller step
size and check that it produces the same results. With the preferential scanning
routine in place, by which the most significantly states are identified and
included first, the start of the routine is where the changes are most drastic
and therefore the procedure is most likely to break down there. As a result, the
checks proposed here need not be time-consuming.

% How tracking the seed state changes the procedure
Now that we have established how we can track the seed state, we note that we
can replace the lowest energy approximate eigenstate with the image of the seed
state in the second order metric used in Eq.~\eqref{eq:second order weight}. 
This change, together with the replacement of the ground state with an arbitrary
seed state in the preferential scanning routine results in a routine designed to
optimize the convergence of the approximate eigenstate associated to the seed
state. The resulting algorithm can be summarized as follows.

% However, the question remains if we can a priori identify which
% eigenstate of the perturbed Hamiltonian the approximate eigenstate associated to
% a seed state will correspond to. We conjecture that, at least for mild quenches,
% this is the eigenstate with the same quantum numbers. We use this assumption for
% the comparions made in Sec.~\ref{sec:results from MERG for excited states}

Let $|\Omega \rangle$ be some eigenstate of the final Hamiltonian,
which in this case can be an excited state, then the steps of the matrix element
renormalization group are as follows:
\begin{enumerate}

\item Generate the computational basis via preferential state generation from
  the seed state $|\Omega \rangle$. Order the states in the computational basis
  according to the metric in Eq.~\eqref{abacusMetric}, to obtain $\{
  |\{\lambda\}^{(j)}\rangle \}$. 

\item Construct a truncated Hamiltonian from the first $N_s+\Delta N_s$
  computational basis states, $\big\{ |\{\lambda\}^{(1)}\ra, \ldots,
  |\{\lambda\}^{(N_s+\Delta N_s)}\ra\big\}$ and diagonalize this Hamiltonian to
  obtain the first approximate eigenstates $\big\{|1\ra, \ldots , |N_s+\Delta
N_s\ra\big\}$ with energies $\{ E_1, \ldots, E_{N_s + \Delta N_s} \}$.

\item Compute the overlaps between $|\{\lambda\}^{(0)}\rangle$ and the newly
  acquired approximate eigenvectors $\{|1\rangle, \ldots, |N_s + \Delta
  N_s\rangle \}$. Then relabel the approximate eigenstates such that $|1\rangle$
  refers to the approximate eigenstate with the largest overlap.

\item Take the next $\Delta N_s$ computational basis states
  $|\{\lambda\}^{(j)}\rangle$ and compute the ``second order weight'' for each
  of the approximate eigenstates $|i \rangle$ with $i>1$ obtained in previous
  steps: 
\begin{align}
	w_2\left( |i \ra \right) = \sum_{j} \frac{ \langle i | \delta H |\{\lambda\}^{(j)} \rangle \langle \{\lambda\}^{(j)} | \delta H | 1 \rangle}{( E_1 - E_{\{\lambda\}}^{(j)}) (E_1 - E_i)}. 
  \label{eq:second order weight}
\end{align}
Here $\delta H = cR\times g_2(0)$ is the perturbing operator (see
Sec.~\ref{Sec:formulation}), the sum ranges over all the newly added
computational basis states, and $E_{\{\lambda\}}^{(j)} = E(\{\lambda\}^{(j)})$
are the energies of the newly added computational basis states.

\item Form a truncated basis consisting of $|1\rangle$, the $N_s-1$ states in
  $\{|i \rangle \big\}_{i>1}$ with the largest $w_2$-weight, and the $\Delta N_s$
  computational basis states introduced in step 4, construct the truncated
  Hamiltonian in this basis and diagonalize it to obtain $N_s + \Delta N_s$ new
  approximate eigenstates $\{|1'\rangle, \ldots, |(N_s + \Delta N_s)'
  \rangle\}$.

\item Compute the overlaps between $|1\rangle$ and the newly acquired
  approximate eigenvectors, and replace $|1\rangle$ by the approximate
  eigenvector with the largest overlap. Replace the remaining approximate
  eigenvectors used to form the truncated basis in step 5 with the remaining
  newly obtained approximate eigenvectors. 

\item Return to the fourth step.  

\end{enumerate}
This process is continued, obtaining new approximate eigenstates after each
cycle of steps 4 to 6, until the required convergence of the eigenstate is
reached or the computational basis is exhausted.  

This version of the matrix element renormalization group is not more memory
intensive than the routine presented for constructing ground states and it is
only slightly more computationally intensive. The additonal computational cost
comes from computing the overlaps at each iteration. 

%%%%%%%%%%%%
\subsection{Results from the matrix element renormalization group for excited
states}
\label{sec:results from MERG for excited states}
%%%%%%%%%%%%

As mentioned in Sec.~\ref{Sec:Constructing excited states}, one of the
subtleties that arises when considering the matrix elemenent renormalization
group for excited states is that, even though we know that we construct an
approximate eigenstate of the perturbed Hamiltonian, we do not necessarily know
a priori what eigenstate this will correspond to. For the interaction quench
considered here, the most natural eigenstate of $H(c_i)$ to end up with when
starting from an eigenstate of $H(c_f)$ is the eigenstate with the same quantum
numbers.  In this section we show some preliminary results to verify this claim,
although we do note that to assert with more certainty that this claim is true,
more properties of the eigenstates other than the energies would have to be
considered. This is left to future works.

\begin{figure}[t]
  \begin{center}
      \includegraphics[width=0.7\textwidth]{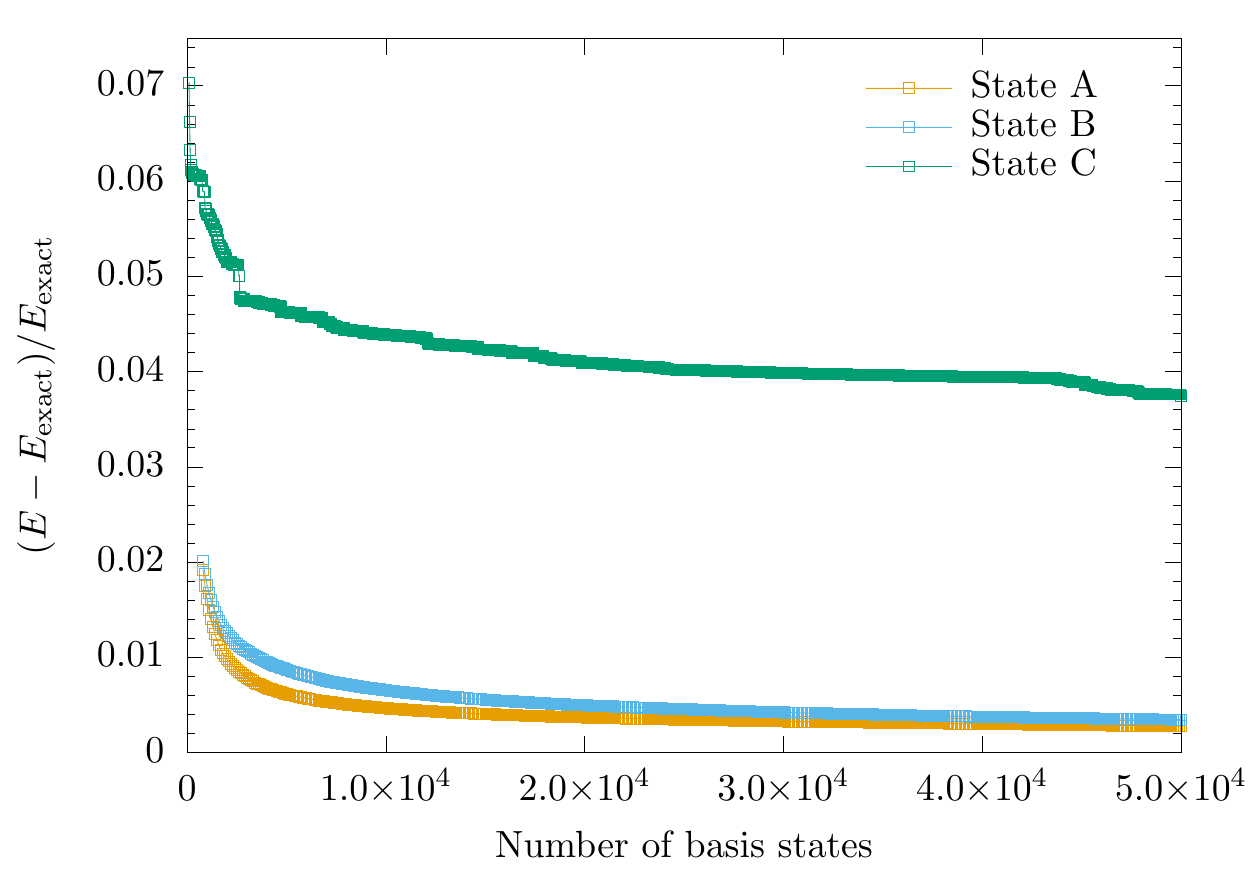}
  \end{center}
  \caption{The convergence of the states A, B, and C, for the $N = 10$ particle
    quench $c_i = 20 \to c_f = 10$ obtained using the matrix element
    renormalization group (MERG). For state A, and B, MERG is performed with
    $N_s = 700$ and $\Delta N_s = 100$, whereas for state C, MERG is performed
    with $N_s = 75$ and $\Delta N_s = 25$.
  }
\label{fig:excitedStateMERG}
\end{figure}

Consider again the quench from $c_i = 20$ to $c_f = 10$. In the following we
present the results obtained from running the algorithm three times with three
different seed states, whose doubled quantum numbers are given by
\begin{align}
  \textrm{State A:}& \textrm{ } \{ -9, -7, -5, -3, -1, 1, 3, 5, 7, 9\} \\
  \textrm{State B:}& \textrm{ } \{ -11, -7, -5, -3, -1, 1, 3, 5, 7, 11\} \\
  \textrm{State C:}& \textrm{ } \{ -17, -13, -9, -5, -1, 1, 5, 9, 13, 17\}.
\end{align}
The results for the energy convergence of the approximate eigenstates
corresponding to these seed states obtained from the matrix element
renormalization group are shown in Fig.~\ref{fig:excitedStateMERG}. 
In order to keep track of the convergence, we again consider the percentual
error of the energy only this time with respect to a different target energy of
the each of the runs. The target energy $E_{exact}$ is taken to be the energy of
the eigenstate of $H(c_i)$ with quantum numbers identical to those of the seed
state under consideration. We note that even though we only consider data for
the energy convergence here, we have still computed the expansion of the
approximate eigenstates in terms of the eigenstates of the intiial basis, so
could still compute the time evolution of operators if we please to do so.

The rate of convergence of the run seeded by the lowest excited state, state B,
is comparable to the convergence when considering the ground state, state A. 
On the other hand, when considering a run seeded by a highly excited state,
state C, the convergence shows characteristics reminding us of the strongly
non-perturbative quenches considered in section Sec.~\ref{Results from the
matrix element renormalization group for the ground state.}.
Also, in order to keep track of the right approximate eigenstate throughout the
procedure, we had to significantly alter the parameters characterizing the size
of the renormalization group steps to $N_s = 75$ and $\Delta N_s = 25$.

The fact that considering higher energy states requires a decrease in step 
size is what currently limits how high the energy of the seed states may be. To
overcome this limitation, we would have to reorder the computational basis so
that single steps of the procedure do not change the targeted approximate
eigenstate as violently. Nevertheless, even without such alterations, our
current algorithm goes beyond what one could target using the conventional
renormalization group techniques we discussed at the start of this paper,
because in that case one would have to construct all lower energy eigenstates.

%%%%%%%%%%%%%%%%%%%%%%%%%%%
{\section{Conclusions}
\label{Sec:Conclusions}
%%%%%%%%%%%%%%%%%%%%%%%%%%%

Even in the presence of integrability, the computation of non-equilibrium
dynamics following a quantum quench remains a great challenge for theory.
Well-controlled numerical approaches are vital for accessing the physics away
from analytically tractable limits, including for the cases of finite-time
dynamics of observables. Here we have presented a proof-of-principle
investigation of finite-$c$ to finite-$c$ quenches in the Lieb-Liniger model
using a high overlap states truncation scheme, in combination with full
diagonalization, the numerical renormalization group, and a new matrix element
renormalization group algorithm. We have worked with interacting computational
basis states, which intrinsically have built-in strong correlations, and we have
systematically constructed initial states in terms of high overlap states, for
quenches starting from ground states as well as excited states.
Using these, we have computed both real-time dynamics and the long-time limit of
physical observables following a quench.

In our development of a high overlap states truncation scheme, and the matrix
element renormalization group, we have highlighted the important role played by
the ordering of the computational basis. Applying the conventional metric,
energy of the computational basis states, we observe poor convergence of
properties of the initial states. This poor convergence means applying
conventional ``truncated spectrum methods'' (in their naive form) requires the
use of unfeasibly many computational basis states. By modifying the metric, to a
``matrix element'' focused one that takes into account the structure of the
operator coupled to the quench parameter, we achieve orders-of-magnitude
improvement in the convergence of properties of the initial state with truncated
Hilbert space size. This was studied in detail in Sec.~\ref{Sec:Overlaps}.
Along the way, we were able to develop a routine that preferentially generates
the states with high overlap following a quench, and this enabled efficient
convergence of the initial state energy to sub-percent precision.  

This improved convergence opened the door to computing non-trivial
non-equilibrium dynamics for numbers of particles far beyond the reach of brute
force computations. This was discussed in Sec.~\ref{Sec:RealTime}. Convergence
of real-time non-equilibrium dynamics of local observables with the number of
computational basis states was surprisingly fast: for $N=10$ particles $c_i = 20
\to c_f = 10$ quench, well-converged results for time evolution of $g_2(0)$ are
obtained with just thousands of states (some of which are of very high in
energy). The long-time limit was also shown to be efficiently accessed via the
diagonal ensemble, with results agreeing with the intermediate time dynamics, as
expected.

In the case of strongly non-perturbative quenches, we found that conventional
numerical renormalization group improvements have to be significantly modified
to achieve accurate results. This modification, the so-called matrix element
renormalization group, takes seriously that the properties of the perturbing
operator should govern the whole procedure. We found this modification to be
necessary in the ``large quench'' studied previously in the
literature~\cite{zill2016coordinate}, $c_i = 100 \to c_f = 3.7660$, when
considering more than four particles. Our results were compared to the
coordinate Bethe ansatz results of Zill \textit{et
al.}~\cite{zill2016coordinate}, and were found to be in excellent agreement. We
note, however, that such strongly non-perturbative quenches remain challenging
problems, with the quench projecting the initial state on to many states with
sizable overlaps. This makes it tough to tackle even relatively small numbers of
particles, even with our computationally efficient approach. This seems like an
insurmountable problem with introducing additional approximations, beyond the
scope of this work, or alternative Hilbert space ordering metrics.

Finally, we considered the construction of excited states of the perturbed
Hamiltonian. We wrote down a variant of the matrix element renormalization group
algorithm able to directly construct excited states of a perturbed Hamiltonian
in terms of the eigenbasis of another Hamiltonian without having to construct
all lower energy eigenstates. This allows one to consider more highly excited
states than one normally could using the truncated spectrum approach and
focusses the computational resources on this particular eigenstate rather than
it being a less well-converged side-product of trying to converge the
approximate eigenstate representing the ground state of the perturbed
Hamiltonian. 

The presented high overlap states truncation scheme, combined with full
diagonalization and renormalization group improvements, can be applied to many
other models and scenarios. Perhaps the most interesting is to consider the case
with integrability-breaking where, provided matrix elements of the
integrability-breaking terms are known, one can directly apply the same
approach. This enables, for example, non-perturbative studies of
prethermalization (see, e.g.,
Refs.~\cite{moeckel2008interaction,marcuzzi2013prethermalization,essler2014quench,bertini2015prethermalization,langen2016prethermalization})
in continuum quantum gases. Other interesting directions include extensions to
other integrable continuum models, such as two-component Bose and Fermi gases or
the sine-Gordon regime away from the ultra-relevant perturbation
limit~\cite{kukuljan2018correlation}.
Finally, we would like to point out that the method developed in this paper
provides, in principle, all the ingredients necessary to compute for example the
time evolution of the entanglement entropy. In order to come to a tractable
computation one can convert the overlaps coming from the NRG-routines to a root
distribution and then use the quasi-particle picture formulas for the
entanglement entropy, see e.g.
\cite{calabrese_evolution_2005,alba_entanglement_2018}. However, in order to
ascertain the accuracy of results obtained in this way, a careful quantitative
study of finite-size effects is required in order to determine if we can
accurately match results in the thermodynamic and scaling limits. We leave
addressing this challenge to future work.

Extending these methods to lattice models should also be possible, using
strongly correlated integrable eigenstates. Such an algorithm may complement
existing ones: being able to tackle longer times, but smaller systems, than the
time-dependent density matrix renormalization group, but larger system sizes
than exact diagonalization. It may also be interesting to implement the ideas
behind the matrix element renormalization group to lattice and impurity models,
invoking a Wilsonian numerical renormalization group-like picture with strongly
correlated basis states. These points remain for future works. 

The approach implemented within this work for simulating continuum
one-dimensional models provides an alternative, complementary approach to
continuum matrix product state
methods~\cite{verstraete2010continuous,draxler2013particles}. Utilizing the
solvability of a proximate integrable point, time evolution is easy within our
approach and can be performed to long times with high precision. This opens the
door to novel, non-perturbative studies of non-equilibrium dynamics in models of
relevance to cold atomic gases. 

\section*{Acknowledgments}

We are grateful to Stijn de Baerdemacker, Andrew James, Robert Konik, G\'abor
Tak\'acs and Matthew Walters for useful and inspiring conversations over the
past two years while we explored these ideas. Thanks also go to Matthew Davis
and Jan Zill for useful correspondence regarding previous coordinate Bethe
ansatz calculations of non-equilibrium dynamics, Ref.~\cite{zill2016coordinate},
presented in Fig.~\ref{fig:coordBA}.

\section*{Funding Information}
This work received funding from the European Union's Horizon 2020 research and
innovation programme under grant agreement No 745944 (N.J.R) and the European
Research Council under ERC Advanced grant No 743032 DYNAMINT (all authors).

\appendix

%%%%%%%%%%%%%%%%%%%%%%%%%%%%%%%%%%%%%
\section{Matrix elements}
\label{sec:ME}
%%%%%%%%%%%%%%%%%%%%%%%%%%%%%%%%%%%%%

In the main text, we have made use of many known expressions for matrix elements
of operators. In this appendix, we provide a summary of these results taken from
Refs.~\cite{piroli2015exact,pozsgay2011local,slavnov1990nonequaltime,slavnov1989calculation}. 

%%%%%
\subsection{Matrix elements of $\Psi^\dagger(0)\Psi(0)$: determinant representation}
%%%%%

Expectation values of the density operator $\Psi^\dagger(0)\Psi(0)$ are fixed by
the U(1) number conservation and translational invariance to read
\begin{align}
  \frac{\langle\{\mu\}_N | \Psi^\dagger(0)\Psi(0) | \{\mu\}_N\rangle}{\langle \{\mu\}_N | \{\mu\}_N \rangle} = \frac{N}{R}. 
\end{align}
That is, the expectation value is simply the average density.

Off-diagonal matrix elements can be expressed in terms of a single determinant,
as can easily be obtained from~\cite{slavnov1990nonequaltime}. These read: 
\begin{align}
  \la \{\mu\}_N | \Psi^\dagger(0)\Psi(0) |\{\lambda\}_N\rangle =&\ \mathrm{i}{\cal J}_1\left(\{\mu\}_N,\{\lambda\}_N\right)  \prod_{j=1}^N \left( V^+_j - V^-_j\right) \prod_{j,k=1}^N \left( \frac{\lambda_j - \lambda_k + \mathrm{i}c}{\mu_j-\lambda_k}\right)\nonumber\\
                                                                & \times\frac{\text{det}\left(\delta_{jk}+U_{jk}(\lambda_p)\right)}{V_p^+ - V_p^-},
\end{align}
where, explicitly, $\{\mu\}_N \neq \{\lambda\}_N$. In the above, we use the
following notations for functions and matrices:
\begin{align}
  {\cal J}_1 =&\ P(\{\lambda\}_N) - P(\{\mu\}_N), \\
  V^\pm_j =& \prod_{m=1}^N \frac{\mu_m - \lambda_j \pm \mathrm{i}c}{\lambda_m - \lambda_j \pm \mathrm{i}c}, \label{eq:defVpm} \\ 
  U_{jk}(\lambda_p) =&\ \frac{\mathrm{i}}{V_j^+-V_j^-}\frac{\prod_{m=1}^N (\mu_m-\lambda_j)}{\prod_{m\neq j=1}^N (\lambda_m-\lambda_j)} \Big[K(\lambda_j,\lambda_k) - K(\lambda_p,\lambda_k) \Big],
\end{align}
and $K(\lambda_j,\lambda_l)$ is given in Eq.~\eqref{eq:defK}. Notice that ${\cal
J}_1$ implies that the matrix element element of the density operator between
non-identical states within the same momentum sector vanish. 

%%%%%
\subsection{Off-diagonal matrix elements of $g_2(0)$: determinant representation }
\label{sec:g2}
%%%%

An efficient, single determinant representation for the off-diagonal matrix
elements of $g_2(0)$ is provided by Piroli and Calabrese~\cite{piroli2015exact}:
\begin{align}
\begin{split}
\la \{\mu\}_N|  g_2(0) | \{\lambda\}_N\ra =&\ 
\frac{(-1)^N}{6c} {\cal J}_2\left(\{\mu\}_N,\{\lambda\}_N\right) \prod_{j=1}^N \Big( V_j^+ - V_j^- \Big) \prod_{j,k=1}^N \left( \frac{\lambda_j - \lambda_k + \mathrm{i}c}{\lambda_j-\mu_k}\right) \\
&\times \frac{\text{det}_N(\delta_{jk}+U_{jk}(\lambda_p,\lambda_s))}{(V_p^+-V_p^-)(V^+_s - V^-_s)}.
 \end{split}
 \label{eq:offdiagME}
\end{align}
Here explicitly: (i) the sets of rapidities do not coincide ($\{\mu\}_N \neq
\{\lambda\}_N$); (ii) no individual elements of the sets coincide ($\mu_j \neq
\lambda_k$ $\forall j,k$). In Eq.~\eqref{eq:offdiagME} the following functions
and matrices are required: 
\begin{align}
  {\cal J}_2 =& \Big[ P\big( \{\lambda\}_N\big) - P\big(\{\mu\}_N\big) \Big]^4 + 3 \Big[ E\big( \{\lambda\}_N\big) - E\big(\{\mu\}_N\big) \Big]^2 \nonumber\\
& - 4\Big[ P\big( \{\lambda\}_N\big) - P\big(\{\mu\}_N\big) \Big] \Big[ Q_3\big( \{\lambda\}_N\big) - Q_3\big(\{\mu\}_N\big) \Big], \\
U_{jl}(\lambda_p,\lambda_s) =& \frac{\mathrm{i}}{V_j^+ - V^-_j} \frac{\prod_{m=1}^N (\mu_m - \lambda_j)}{\prod_{\substack{m=1,\\m\neq j}}^N (\lambda_m - \lambda_j)} \Big[ K(\lambda_j,\lambda_l) - K(\lambda_p, \lambda_l) K(\lambda_s,\lambda_j)\Big]. \label{eq:defU}
\end{align}
Furthermore $V^\pm_j$ is given in Eq.~\eqref{eq:defVpm},
$K(\lambda_j,\lambda_l)$ is defined in Eq.~\eqref{eq:defK}, $Q_3(\{\lambda\})$
is given by Eq.~\eqref{defQ}, and $\lambda_p,\lambda_s$ are \textit{arbitrary}
complex numbers.

Within the main text, we consider states within the same momentum sector, where
\begin{align}
  {\cal J}_2(\{\mu\}_N,\{\lambda\}_N)\Big\vert_{P(\{\mu\}_N) = P(\{\lambda\}_N)}  = 3 \Big[ E(\{\lambda\}_N) - E(\{\mu\}_N)\Big]^2.
\end{align}

%%%%
\subsection{Matrix elements of $g_K(0)$}
%%%%
We now recount known results for matrix elements of the operator
\begin{align}
  g_K(0) = \Big( \Psi^\dagger(0)\Big)^K \Big( \Psi(0) \Big)^K, 
\end{align}
as derived by Pozsgay~\cite{pozsgay2011local}. We have implemented these
expression both for the diagonal elements of $g_2(0)$. 

\subsubsection{Diagonal elements}
The diagonal elements of $g_K(0)$ are given by:
\begin{align}
\frac{\la \{ \lambda\}_N | g_K(0) |\{\lambda\}_N\ra}{\la \{\lambda\}_N | \{\lambda\}_N \ra}  = (K!)^2 \sum_{\substack{\{\lambda^+\}\cup\{\lambda^-\}\\ |\{\lambda^+\}| = K}}\bigg[ \prod_{j>l}\frac{\lambda_j^+ - \lambda_l^+}{(\lambda^+_j - \lambda^+_l)^2 + c^2} \bigg] \times \frac{\text{det} {\cal M}}{\text{det}{\cal N}},
\label{eq:diagME}
\end{align}
where ${\cal M}$ is an $N\times N$ matrix with elements
\begin{align}
{\cal M}_{jl} = \left\{ 
\begin{array}{ccc} 
(\lambda_j)^{l-1}, & \quad & \text{for }l=1,\ldots,K,\\
{\cal N}_{jl} & \quad & \text{for }l=K+1,\ldots,N.
\end{array}\right. ,
\end{align}
and ${\cal N}$ is the Gaudin matrix, see Eq.~\eqref{eq:defGaudin}. Here it
should be understood that the rapidities are ordered as $\{\lambda\} = \big\{
\{\lambda^+\},\{\lambda^-\} \big\}$. Whilst this is a sum of determinants, so
not as computationally efficient to evaluate as the previous single determinant
representation, it is still relatively easy to compute numerically.

\subsubsection{Off-diagonal elements}
The off-diagonal matrix elements for $g_K(0)$ read:
\begin{align}
  \langle \{\lambda\}_N | g_K(0) |\{\mu\}_N\rangle  =
  c^K (K!)^2&\sum_{\substack{\{\lambda^+\}\cup\{\lambda^-\}\\|\{\lambda^+\}|=K}}\left(\prod_{o,\ell}  \frac{\lambda_o^--\lambda_\ell^+ + \mathrm{i}c}{\lambda_o^--\lambda_\ell^+}\right) \nonumber \\
  & \times \frac{\prod_{i,j} (\lambda_i-\lambda_j^- + \mathrm{i}c)}{\prod_{m<n} (\mu_m-\mu_n)\prod_{r<s}(\lambda^-_r-\lambda^-_s)}\ \mathrm{det}\ {\cal W}, \label{gk}
 \end{align}
where ${\cal W}$ is an $N\times N$ matrix with elements
\begin{align}
  {\cal W}_{j,l} =& (\mu_j)^{l-1}, \qquad \text{for~}l=1,\ldots,K, \\
  {\cal W}_{j,K+l} =& \frac{\mathrm{i}c}{(\mu_j-\lambda^-_l)(\mu_j-\lambda^-_l+\mathrm{i}c)}\nonumber\\
  & + \frac{\mathrm{i}c}{(\lambda^-_l - \mu_j)(\lambda^-_l - \mu_j + \mathrm{i}c)} \prod_{o=1}^N \frac{(\lambda_l^--\mu_o+\mathrm{i}c)(\lambda_l^- - \lambda_0 - \mathrm{i}c)}{(\lambda_l^- - \mu_o -\mathrm{i}c)(\lambda^-_l - \lambda_o+\mathrm{i}c)},
\end{align}
and the Bethe states are normalized as in Eq.~\eqref{eq:defNorm}.

\bibliography{bib}

\end{document}